\documentclass[12pt,letterpaper]{article}
\pdfoutput=1
\usepackage[colorlinks=false,
   linkcolor=red, 
   citecolor=blue,
    filecolor=red,
    urlcolor=red,
    linktoc=all, %page 
    pdfstartview=FitV,
    bookmarksopen=true]{hyperref}
\usepackage[left=2cm,top=1cm,right=3cm,nohead]{geometry}
\usepackage[utf8]{inputenc} 
\usepackage{epsfig,latexsym,amsfonts,amsmath,amsthm,amssymb,amsbsy,multirow,slashed,color,mathrsfs,wasysym,textcomp,subfigure,wrapfig,comment,bbold}

\usepackage{cite}
\usepackage{color,datetime,graphicx,textcomp}

\newcommand{\pink}[1]{\textcolor{\pink}{#1}}

\newtheorem*{theorem}{Theorem}

\numberwithin{equation}{section}

\graphicspath{{./Immagini/}}

\begin{document}

\thispagestyle{empty}
 \begin{flushright}
 MPP-2015-185\\
LMU-ASC 50/15
 \end{flushright}
\begin{center}
%\boxed{\texttt{Draft: \today, \currenttime }}
\end{center}
\vspace{1cm}
\begin{center}
{\LARGE \bf{The monodromy of T-folds and T-fects}}
 \vskip1.5cm 
Dieter
    L\"ust$^{a,b}$, Stefano
Massai$^a$ and Valent\'i Vall Camell$^a$\\
\vskip0.8cm
\textit{$^a$ Arnold Sommerfeld Center for Theoretical Physics,\\
Theresienstra\ss e 37, 80333 M\"unchen, Germany}\\
\vskip0.3cm
\textit{$^b$ Max-Planck-Institut f\"ur Physik\\F\"ohringer Ring 6, 80805
  M\"unchen, Germany}
\vskip0.3cm
\noindent {\small{\texttt{dieter.luest@lmu.de, stefano.massai@lmu.de, \\v.vall@physik.uni-muenchen.de}}}
\vspace{1.5cm}
\end{center}

\begin{abstract}

\noindent
We construct  a class of codimension-2 solutions in supergravity that
realize T-folds with arbitrary $O(2,2,\mathbb{Z})$ monodromy and we
develop a geometric point of view in which the monodromy is identified
with a product of Dehn twists of an auxiliary surface $\Sigma$ fibered on a
base $\mathcal{B}$.
These defects, that we call T-fects, are identified by the monodromy of
the mapping torus obtained by fibering $\Sigma$ over the boundary of a small disk
 encircling a degeneration.
We determine all possible local
 geometries by solving the corresponding Cauchy-Riemann equations,
 that imply the equations of motion for a semi-flat metric ansatz.
 We discuss the relation with the F-theoretic approach and
we consider a generalization to the T-duality group
 of the heterotic theory with a Wilson line.
\end{abstract}

\newpage

\tableofcontents

\section{Introduction}

The geometrization of duality groups has been the source of many
insights both in string and field theories. Perhaps the simplest
example is that of Montonen-Olive duality in four-dimensional
$\mathcal{N}=4$ SYM, where the discrete $SL(2,\mathbb{Z})$ group acting on
the complexified coupling $\tau$ is
interpreted as the modular group of a $T^2$. The torus is physical:
one can obtain the four-dimensional theory with parameter $\tau$ by compactifying a
six-dimensional $(2,0)$ theory on $\mathbb{R}^4\times T^2$~\cite{Witten:2009at}. A related example
is S-duality in $\mathcal{N}=2$ theories, which can be identified with
the mapping class group of an $n$-punctured genus-$g$ Riemann surface
$\Sigma_{g,n}$~\cite{Witten:1997sc,Halmagyi:2004ju,Gaiotto:2009we,Seiberg:1994rs,Klemm:1994qs,Klemm:1996bj}.
In type IIB string theory, identifying the axio-dilaton $\tau$ with
the modular group of an auxiliary torus leads to
F-theory~\cite{Vafa:1996xn}. Even if we do not have a complete twelve
dimensional picture, the torus is again physical and this is best
understood via duality to M-theory. Similar attempts have been made to
geometrize the full U-duality
group~\cite{Kumar:1996zx,Liu:1997mb}. Recently, investigations along
these lines have been carried out for the T-duality group in the
context of heterotic/F-theory duality
in~\cite{McOrist:2010jw,Malmendier:2014uka,Gu:2014ova}, following
earlier works~\cite{Greene:1989ya,Hellerman:2002ax}.

One of the most interesting outcomes of this analysis is the
description of inherently quantum bundles, namely spaces in which
different patches are glued together with T-dualities. These spaces
are usually referred to as ``T-folds''~\cite{Hull:2004in} or more
generally as ``non-geometric
backgrounds''~\cite{Hellerman:2002ax}. While string dualities may help
to understand such backgrounds, a direct study beside the most simple
realizations such as asymmetric orbifolds has proved quite
challenging. A clearly fascinating question is how the notion of
differentiable manifolds should be changed to accommodate such stringy
geometries.
One might think that T-duality covariant formalisms such as
Generalized Complex Geometry~\cite{Hitchin:2003uq,Gualtieri:2004kx}
and Double Field Theory~\cite{Siegel:1993th,Hull:2009mi} are relevant tools for such 
investigations. These approaches provide a natural framework
for the inclusion of the B-field (as well as RR-fields), but there are
also attempts to describe non-geometric backgrounds in this
context, see for example~\cite{Grana:2008yw,Andriot:2012an,Andriot:2014qla,Hassler:2014sba,Blumenhagen:2013aia,Blumenhagen:2014gva,Blumenhagen:2015zma,Andriot:2011uh,Andriot:2012wx,Blumenhagen:2012nt,Blumenhagen:2012nk}.\footnote{See also~\cite{Aldazabal:2013sca,Hohm:2013bwa} for reviews and~\cite{Hull:2014mxa,Lee:2015xga} for a recent discussion about the relation
between the two formalisms.}

In view of these motivations, it is of interest to obtain
approximate solutions in which there is a non-trivial duality symmetry
between various patches, and which provide a concrete realization of
T-folds. This will be the subject of the present paper. A useful approach is to consider string theory compactified
on a torus, and fiber the resulting T-duality group over a base
$\mathcal{B}$. We develop a geometric point of view in which the
duality group is identified with the mapping class group
$M(\Sigma)$ of a given
surface $\Sigma$ fibered over $\mathcal{B}$. We then classify all
possible torus fibrations with geometric and non-geometric twists in
terms of a classification of elements in $M(\Sigma)$. In particular, we identify
T-duality elements with Dehn twists of $\Sigma$ and we categorize them
according to the type of surface diffeomorphisms.

Most of the paper will deal with the simplest example of a two torus $T^2$, in
which case a partial geometrization of the T-duality group
$O(2,2,\mathbb{Z}) = SL(2,\mathbb{Z})_{\tau} \times
SL(2,\mathbb{Z})_{\rho}\times \mathbb{Z}_2\times \mathbb{Z}_2$ is
obtained by simply taking $\Sigma = T^2_{\tau} \cup T^2_{\rho}$,
where $\tau$ is the complex structure of $T^2$, $\rho$ is the
complexified K\"ahler form and $T^2_{\tau}$ is identified with the
compactification torus. As we will discuss shortly, in some cases
there is a physical picture for the
auxiliary $T^2_{\rho}$.

We first consider the case in which the base
$\mathcal{B}$ is a circle $S^1$. If we only vary $\tau$, the fibration is
specified by a monodromy $\phi \in M(T^2_{\tau})$ which
is used to identify the fibers after encircling the base $S^1$. The
monodromy can be factorized as a product of Dehn twists, and the total
space of the fibration is a 3-manifold $\mathcal{N}_{\phi}$, known as
the mapping torus for $\Sigma=T^2$, whose
geometry is determined by the type of torus diffeomorphism $\phi$
(which can be either reducible, periodic or pseudo-Anosov,
respectively matching parabolic, elliptic and hyperbolic
$SL(2,\mathbb{Z})$ conjugacy classes)
and is either an Euclidean space, a Nil or a Sol geometry. By replacing
$\tau$ with $\rho$ we obtain a classification of possible T-fold
metrics. We can think of the monodromy in $\rho$ as a product of Dehn
twists of the auxiliary torus $T^2_{\rho}$ fibered over $S^1$.

We then consider a fibration of the torus over a two dimensional base
$\mathcal{B}=\mathbb{P}^1$. This is the familiar situation of stringy cosmic
strings~\cite{Greene:1989ya,Hellerman:2002ax} and the parameters
$\tau$ and $\rho$ need to be holomorphic functions on the base where a
collection of points, at which the fiber torus degenerates, has been
removed. This situation can be described by two elliptic
fibrations~\cite{Hellerman:2002ax} and the corresponding
degenerations, classified by Kodaira~\cite{kodaira123}, are
associated to stringy cosmic five-branes. For each degeneration is associated
a monodromy, namely an element of the mapping class group of
$T^2_{\tau}$ or $T^2_{\rho}$. Non-trivial elements of
$M(T^2_{\rho})$, with the exception of a perturbative shift $\rho
\rightarrow \rho + 1$ which is associated to a NS5
brane~\cite{Ooguri:1995wj}, will give non-geometric solutions. The
fibration data for the  auxiliary $T^2_{\rho}$ can be determined in
the heterotic theory by a fibration of heterotic/F-theory
duality~\cite{McOrist:2010jw}, and the non-geometric backgrounds are
mapped to geometric K3-fibered Calabi-Yau compactifications on the
F-theory side.
We will generically refer to such degenerations as T-duality defects, or \emph{T-fects}. 

It is interesting to ask to what extent one can invert this
reasoning, namely if it is possible to determine, from a
classification of elements in the mapping class group, which monodromies of the fibration
 on the boundary $S^1 = \partial D^2$ of a small disk $D^2$ correspond
 to a family
 of degenerating surfaces at a point in $D^2$. As a result of an
 elaborate theorem~\cite{MatMont}, one can make this approach precise
 and basically obtain a topological classification of singular fibers.

In order to construct approximate solutions for such backgrounds we
use a semi-flat
approximation~\cite{Strominger:1996it,Vegh:2008jn,Becker:2009df}
where the fields do not depend on the fiber coordinates. In our
case this means to preserve the $U(1)\times U(1)$ isometry of the
torus. For the
geometric fibrations this is a very good approximation of the full
metric, up to exponentially suppressed terms close to the
degenerations. We then construct all possible local geometries for 
a fibration on a small disk $D^2$ with a given monodromy $\phi=[f]$ at
the boundary, for each conjugacy class $[f]$ of
$SL(2,\mathbb{Z})_{\tau}\times SL(2,\mathbb{Z})_{\rho}$, by
solving the corresponding Cauchy-Riemann equations. Some of the geometric solutions are
identified with a smeared version of known branes, such as the KK
monopole. Solutions for the parabolic conjugacy class
in $SL(2,\mathbb{Z})_{\rho}$ give, other then a smeared NS5 brane solution, a
class of non-geometric backgrounds in which patching requires
$\beta$-transformations. In particular, we recover the solution for
the $5_2^2$ exotic brane recently discussed
in~\cite{deBoer:2012ma,Hassler:2013wsa,Chatzistavrakidis:2013jqa}.\footnote{This is sometimes
  referred to as a Q-brane, since it provides a source for the
  so-called Q-flux~\cite{Hassler:2013wsa}, but we choose not to use this term to
  avoid confusion with the Q seven-branes of~\cite{Bergshoeff:2006jj}
  which are closely related to our elliptic T-fects.} We obtain a
class of solutions for elliptic $\rho$-fects which are related to an
asymmetric orbifold description~\cite{Condeescu:2013yma}. We also
construct solutions with hyperbolic monodromy, whose corresponding
mapping torus is a Sol-geometry. The volume of the fiber torus in such
backgrounds is a highly
oscillating function near the would-be degeneration point, in line with the fact that
hyperbolic monodromies cannot be obtained from colliding degenerations of
elliptic fibrations. However, such solutions can approximate a
distribution of branes. When both $\tau$ and $\rho$ fibrations are
allowed to degenerate at the same point, we obtain non-geometric
solutions which are not T-dual to geometric ones. In particular we
discuss the example of double elliptic monodromies and the relation
with the corresponding asymmetric orbifold studied
in~\cite{Condeescu:2013yma}. 
The solutions that we construct give approximate local geometries around
degenerations of a given global model. For the case of elliptic
fibrations, these local solutions will be approximated by an expansion
of the solutions for $\tau$ and $\rho$ determined by the appropriate
combination of hypergeometric functions that invert Klein's
$j$-invariant. The same result holds for the axio-dilaton profile in
F-theory.

Finally, we describe from the geometric point of view fibrations of a
genus 2 surface $\Sigma_2$, that provides a geometrization of the heterotic
T-duality group $O(2,3,\mathbb{Z})\approx Sp(4,\mathbb{Z})\subset
M(\Sigma_2)$~\cite{Malmendier:2014uka,Gu:2014ova,Mayr:1995rx}. Degenerations of
surface fibrations over $\mathbb{P}^1$ correspond to heterotic
T-fects. We describe a classification of such defects in terms of
elements in $M(\Sigma_2)$ and we give the monodromy factorization for
some of them.

Let us mention few possible applications of our results. Having explicit solutions which require patching by arbitrary elements
of the T-duality group should be useful to investigate the correct geometric
description of such backgrounds, as in the approach
of~\cite{Andriot:2014uda,Andriot:2014qla}, as well to understand general
non-geometric microstate solutions for black
holes~\cite{deBoer:2012ma,Park:2015gka}. It would be
interesting to see if solutions with the monodromies considered in
this paper can arise via brane polarization as
in~\cite{deBoer:2012ma}, as this would be an additional mechanism to
obtain a globally well defined solution, since only dipole exotic
charges would be nonzero. We also note that exotic branes should be
relevant for cosmological billiards~\cite{Damour:2002et,Brown:2004jb}
and it would be nice to clarify the relation with the T-fects described
here. We mention that a particular kind of elliptic $\rho$-fects (described in terms of a gauged linear sigma model) were
used in~\cite{Dong:2010pm} as an ingredient to ``uplift''  AdS/CFT
dual pairs to dS ones. The semi-flat solutions we are
considering incorporate the backreaction of such objects.

Lastly we note that, as mentioned in~\cite{deBoer:2012ma}, there is a
close relation between the codimension-2 T-fects that we study in this
paper and the theory of non-abelian anyons and topological quantum computation~\cite{2008RvMP...80.1083N}. The approach we take
here, by identifying the monodromy (or charge) of the T-fects in terms of Dehn twists, makes this analogy even more compelling,
since in both situations braid groups play a central role. In the
case of abelian anyons, there is also a close relation with quantum
groups~\cite{Lerda:1992yg}. It would be interesting to see if there is
a connection with the
approach we develop in this paper.

\vspace{0.5cm}

This paper is organized as follows. In section~\ref{Sec:torusbundles} we recall a number of
elementary facts about torus fibrations and the $SL(2,\mathbb{Z})$
mapping class group, in a way that makes easy the
generalization to fibrations of arbitrary genus. 
In section~\ref{sec:Tfolds} we classify torus fibrations over a circle by means of
 monodromy elements in the mapping class group and we write explicit
 metrics for the total spaces. We then apply the same technique to
 fibrations in which the K\"ahler modulus undergoes non-trivial
 monodromies, obtaining a classification of T-fold metrics. In
 section~\ref{sec:T-walls} we discuss a way to obtain explicit
 solutions for domain walls that realize T-folds backgrounds. We dub
 such solutions T-walls.
In section~\ref{sec:exoticbranes} we consider fibrations over a two-dimensional base and we
derive local solutions around a degeneration with arbitrary monodromy
in $SL(2,\mathbb{Z})_{\tau}\times SL(2,\mathbb{Z})_{\rho}$, by solving
the corresponding Cauchy-Riemann equations for $\tau$ and $\rho$. We
refer to such degenerations as T-fects. Some of
these T-fects are identified with a semi-flat approximation of
known brane solutions, while some of them are new exotic solutions.
In section~\ref{sec:hyperellipticfibrations} we briefly discuss the
generalization of our results to
fibrations of genus 2 surfaces and their relation to heterotic
T-fects. We 
present our conclusions and a list of open questions in
section~\ref{sec:conclusions}. We relegate various materials in the
appendices. In appendix~\ref{app:mcg} we recall
briefly basic results regarding mapping class groups and hyperbolic maps, in
appendix~\ref{app:braid} we discuss the relation between the
braid action on the monodromy factorization in terms of Dehn twists and the
familiar technique of moving branch cuts and the ABC factorization
used in F-theory. In appendix C and D we give some details on the
semi-flat geometries and the embedding of the mapping class groups in $O(2,2,\mathbb{Z})$.

\section{T-duality and monodromy}\label{Sec:torusbundles}

We begin by considering a $T^2$
fibration over a base $\mathcal{B}$. We will give a presentation of
the results that can be readily generalized to higher genus
fibrations. The fiber torus can be either part of the ten dimensional
space-time, or be an auxiliary space that geometrizes an
$SL(2,\mathbb{Z})$ duality group, as in F-theory~\cite{Vafa:1996xn}. 
We will be mainly interested in fibering the T-duality group
$O(2,2,\mathbb{Z})=SL(2,\mathbb{Z})_{\tau}\times
SL(2,\mathbb{Z})_{\rho}\times \mathbb{Z}_2\times \mathbb{Z}_2$ over the base
$\mathcal{B}$. The
complex structure $\tau$ and the complexified K\"ahler form $\rho$ are defined as:
\begin{equation}\label{taurhodef}
\tau= \frac{G_{12}}{G_{22}} + i \frac{\sqrt{G}}{G_{22}} \, ,\qquad
\rho = B + i \sqrt{G} \, ,
\end{equation}
where $G_{ab}$ is the metric on $T^2$, $G =G_{11}G_{22}
-G_{12}^2$ and $B$ is the component of
the $B$-field. The actions of the two
$SL(2,\mathbb{Z})$ groups on $\tau$ and $\rho$
are M\"obius transformations:
\begin{align}\label{sl2action}
\tau & \rightarrow M_{\tau}[\tau] \equiv \frac{a \tau + b}{c \tau + d} \,
,\qquad M_{\tau} = \begin{pmatrix} a & b\\ c& d\end{pmatrix} \in
SL(2,\mathbb{Z})_{\tau} \, ,\\
\rho &\rightarrow M_{\rho}[\rho] \equiv \frac{\tilde a \rho + \tilde
                                       b}{\tilde c \rho + \tilde d} \,
,\qquad M_{\rho} = \begin{pmatrix} \tilde a & \tilde b\\ \tilde c& \tilde d\end{pmatrix} \in
SL(2,\mathbb{Z})_{\rho} \, .
\end{align}
Note that the kernel of this action is $\{ \pm \mathbb{1}\}$.
The two $\mathbb{Z}_2$ factors are the mirror symmetry $(\tau,\rho)
\rightarrow (\rho,\tau)$ and a reflection $(\tau, \rho)\rightarrow
(-\bar \tau, -\bar \rho)$. A partial geometrization of the duality group is obtained by identifying the
two $SL(2,\mathbb{Z})$ factors with the group of large diffeomorphisms
of two tori, $T^2_{\tau}$ (the compactification torus) and
$T^2_{\rho}$. 

We will first consider torus bundles with $\mathcal{B}$ a circle, and later we will study fibrations over a
two dimensional base, where the circle becomes contractible. 
The first
situation has been discussed many times in the context of
Scherk-Schwarz dimensional reduction and as toy model for
non-geometric
backgrounds (see for
instance~\cite{Dabholkar:2002sy,Hull:2005hk,Wecht:2007wu,Grana:2013ila,Hohm:2013bwa}),
but our discussion will focus more on the role of monodromy. The case
$\mathcal{B}=\mathbb{P}^1$ was introduced in~\cite{Hellerman:2002ax}
and it can be understood, for the heterotic theory,
from a duality with F-theory~\cite{McOrist:2010jw}. This is the most interesting situation,
since if one describes the auxiliary $T^2_{\rho}$ fibration as an
elliptic fibration, the corresponding line bundles on $\mathcal{B}$
that specify the fibration can be mapped explicitly to geometric compactifications
on the F-theory side. Reversing this map, one can determine the conditions
on the non-geometric $T_{\rho}^2$ fibration to obtain sensible string
vacua.
We will make precise the relation between these two approaches, and
we will provide a comprehensive classification of local geometries that
arise in this context. 

We will later generalize our results beyond
torus fibrations, motivated by the heterotic T-duality group with a
single Wilson line. We note that some of the results we obtain can
be applied in more general contexts as well.

\subsection{Monodromy and mapping tori}\label{subsec:mappingtori}

Let us review some elementary facts about monodromies of
torus fibrations. We leave some of the details to
appendix~\ref{app:mcg}. We identify the $SL(2,\mathbb{Z})$ group with the
mapping class group $M(T^2)$, the group of large diffeomorphisms
of a two torus $T^2$, by the standard bijective
map
\begin{equation}
\Phi: M(T^2) \rightarrow SL(2,\mathbb{Z}) \, .
\end{equation}
It will be useful to consider a fibration $T^2 \rightarrow
\mathcal{N}_{\phi} \rightarrow S^1$ of the $T^2$ over a
circle constructed as
\begin{equation}\label{defmt}
\mathcal{N}_{\phi} = \frac{T^2\times [0,2\pi]}{(x, 0)\sim (\phi(x),2\pi)} \, ,
\end{equation}
where $\phi \in
M(T^2)$. $\mathcal{N}_{\phi}$ is known as the mapping torus for $\phi$
and we call $\phi$ the monodromy of the fibration. We will often use
identification by $\Phi$ and write $\phi \in SL(2,\mathbb{Z})$.
It is a well known result that the geometry of the mapping torus is
completely determined by the trace of $\phi$ in $SL(2,\mathbb{Z})$, or
equivalently by the kind of induced diffeomorphism of $T^2$.

\begin{figure}
\begin{center}
\includegraphics[scale=0.45]{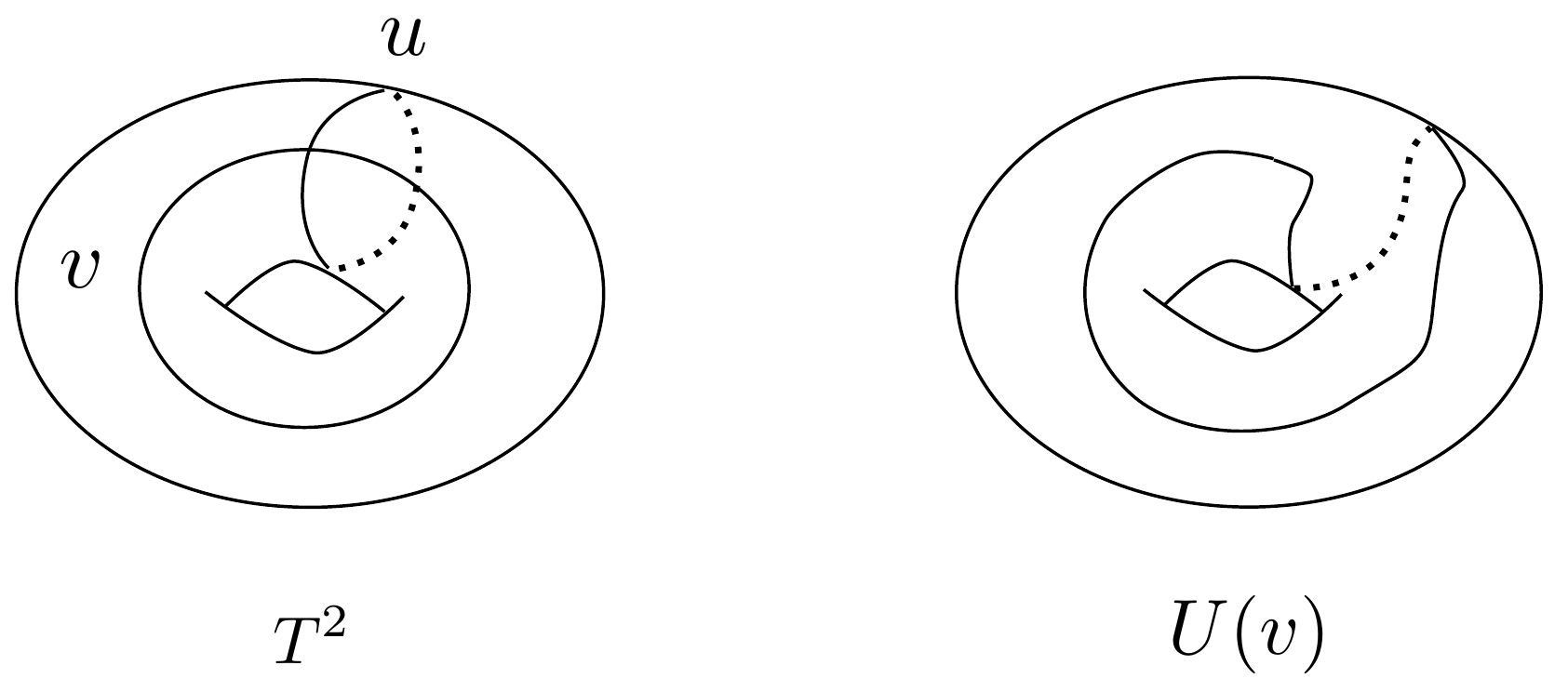}
\caption{Left: a torus with the cycles $u$, $v$. Right: the action on
  $v$ of a Dehn twist along the
  $u$-cycle, represented by the matrix $U$ in~\eqref{UVtwists}.}
\label{Fig:twistT2}
\end{center}
\end{figure}

Let us
briefly recall how the group $M(T^2)$ is generated. A presentation of
the mapping class group of the torus is given in terms of Dehn twists
along two closed curves $(u,v)$ with intersection number one, that we denote
by $U$, $V$. We then have:
\begin{equation}\label{mcgtorus}
M(T^2) \approx \langle U, V \, \vert \, UVU = VUV,  (UV)^6 =\mathbb{1} \rangle \, .
\end{equation}
Note that the first relation is a braid relation.
 A simple
choice for the curves is the standard basis for the homology, see
Figure~\ref{Fig:twistT2}.
This gives the following matrix representation of
the two Dehn twists:
\begin{equation}\label{UVtwists}
 U =  \begin{pmatrix} 1& 0 \\- 1 & 1 \end{pmatrix} \, ,\qquad 
V = \begin{pmatrix} 1 &  1 \\ 0 &  1 \end{pmatrix} \,  ,
\end{equation} 
which indeed satisfy~\eqref{mcgtorus}. We now study a useful classification
of elements in $M(T^2)$. This comes from the characterization of elements in
$SL(2,\mathbb{Z})$ inherited by the classification of
orientation-preserving isometries of
the hyperbolic space, from the identification $PSL(2,\mathbb{R})\approx
\text{Isom}^+(\mathbb{H}^2)\approx \text{Isom}^+(\text{Teich}(T^2))$ given by M\"obius transformation. For a
given isometry $\phi$ represented by a matrix $M$, the following
mutually exclusive possibilities are given:
\begin{enumerate}
\item \emph{Elliptic type}: $\phi$ has one fixed point in $\mathbb{H}^2$, or equivalently
  $|tr(M)| < 2$.
\item \emph{Parabolic type}: $\phi$ has no fixed points in
  $\mathbb{H}^2$ and exactly one fixed point on $\partial
  \mathbb{H}^2$, or equivalently $|tr(M)| = 2$.
\item \emph{Hyperbolic type}: $\phi$ has no fixed points in
  $\mathbb{H}^2$ and exactly two fixed points on
  $\partial\mathbb{H}^2$, or equivalently $|tr(M)|>2$.
\end{enumerate}
\begin{figure}[p]
\centering
\subfigure[]{
\includegraphics[width=0.35\textwidth]{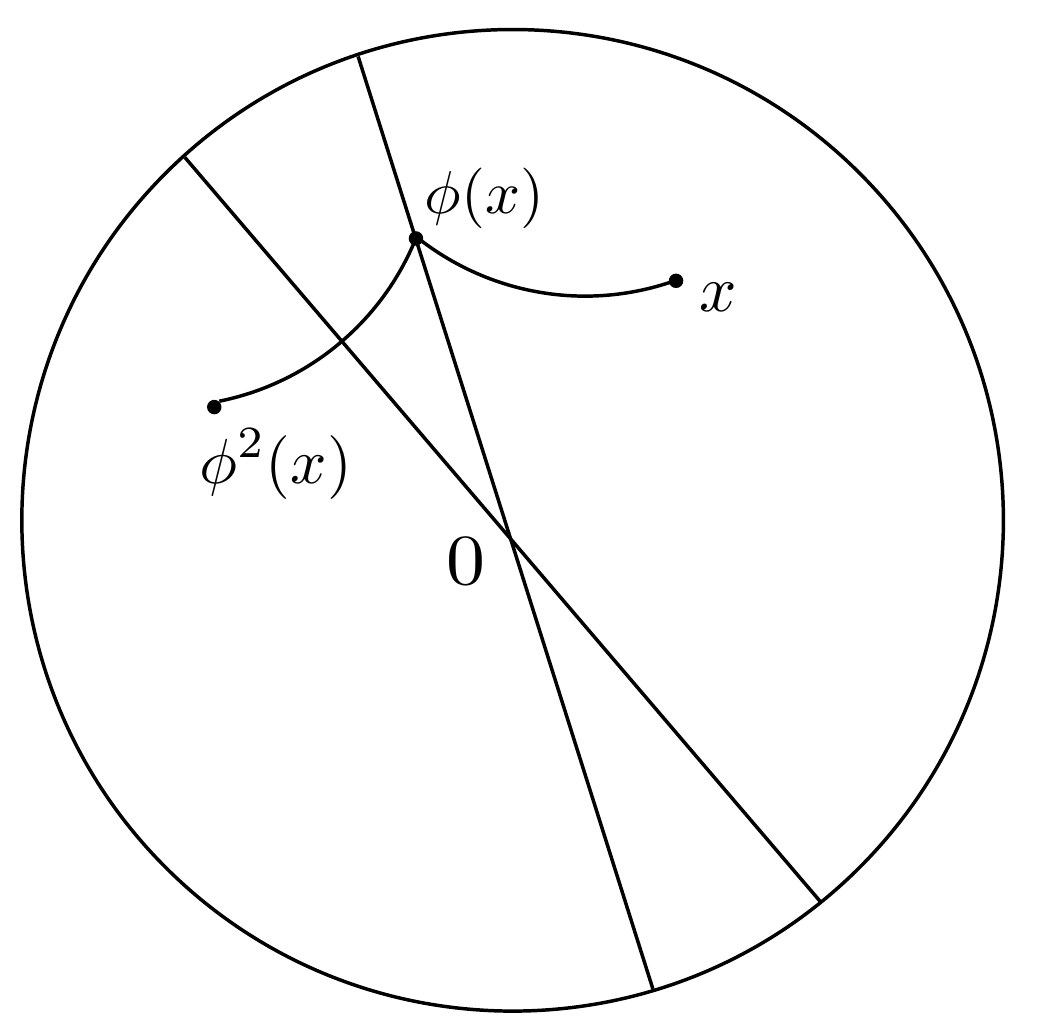}
 }
 \hspace{1.5cm}
 \subfigure[]{
\includegraphics[width=0.45\textwidth]{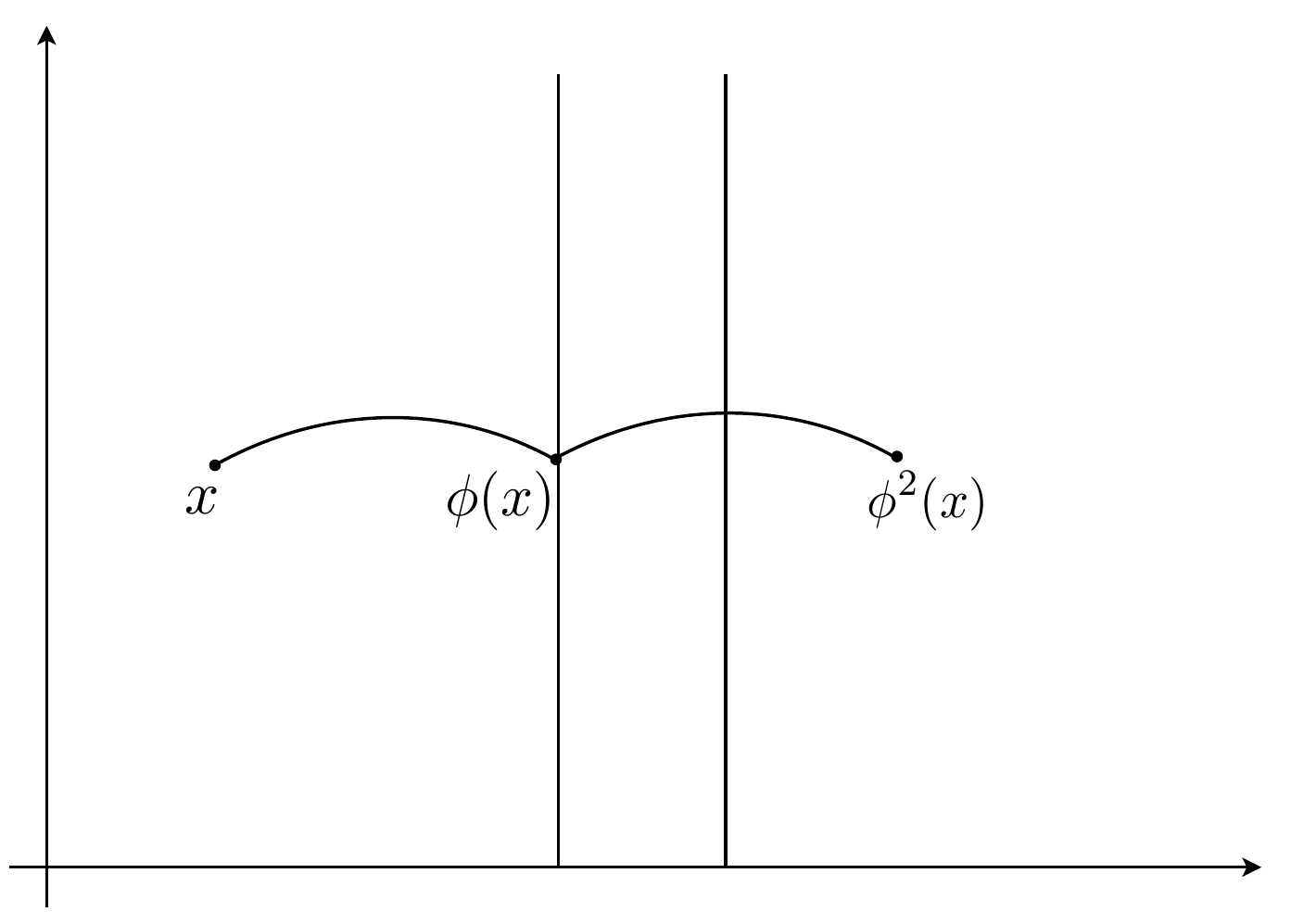}
 }\\\vspace{0.5cm}
\subfigure[]{
\includegraphics[width=0.45\textwidth]{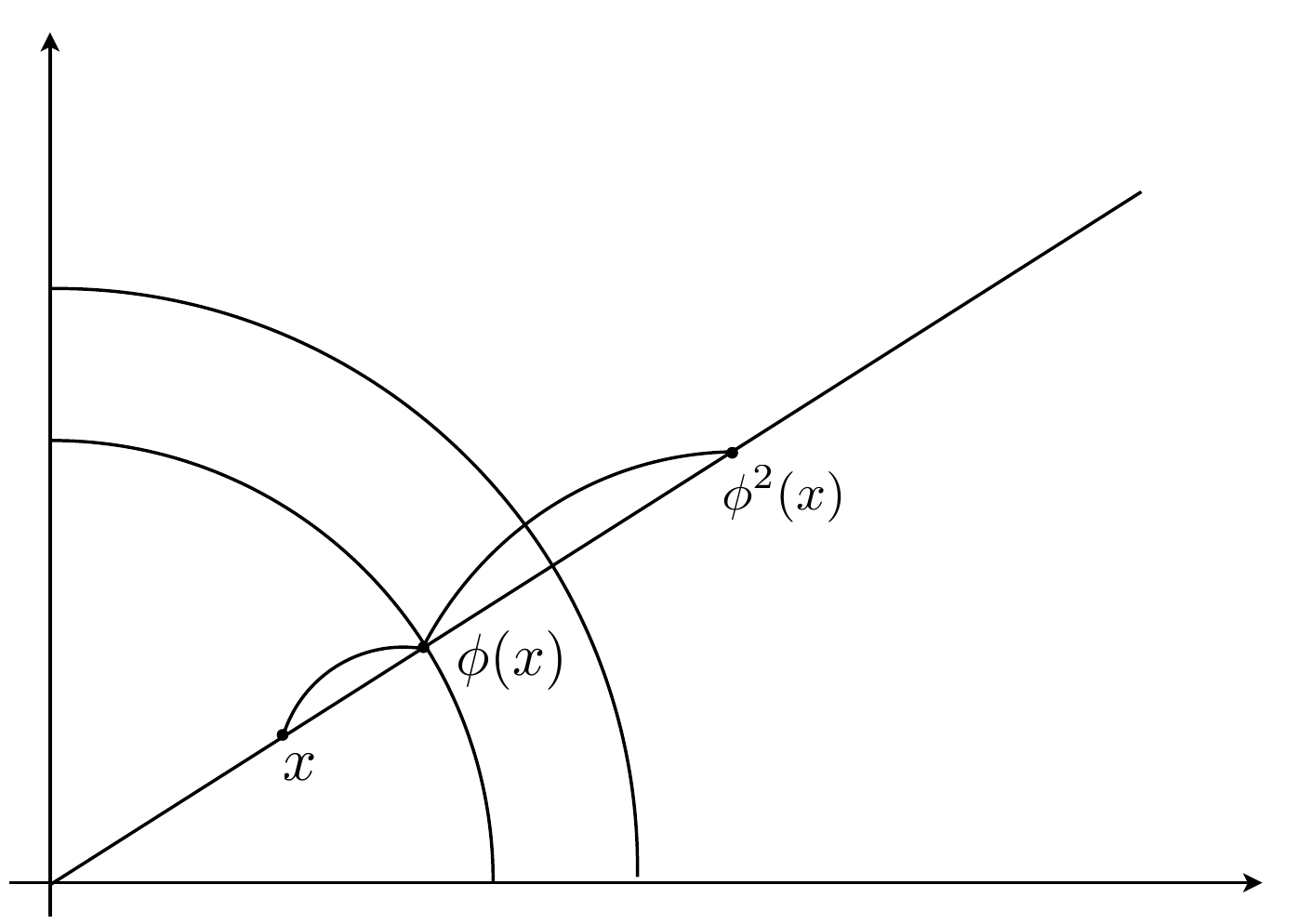}
 }
\caption{Illustration of the classification of isometries of a two
  dimensional hyperbolic space. The two lines are the axis of the
  segment $[\phi(x),\phi^2(x)]$, and the bisecting line of the angle
  between the two segments  $[x,\phi(x)]$ and $[\phi(x),\phi^2(x)]$. (a) The case of an elliptic map
  $\phi$ in the disk model $\mathbb{D}^2$, with fixed point $0\in
  \mathbb{D}^2$: the two lines are incident. (b) A
  parabolic map shown in the half-space: the lines are asymptotically parallel. (c) An
  hyperbolic map: the lines are ultra-parallel. The
  figures are adapted from~\cite{benedetti}.}
\label{fig:isometries}
\end{figure}
A geometric picture of this classification for $SL(2,\mathbb{R})$ is shown in Figure~\ref{fig:isometries}.
Note that elliptic elements are necessarily of
finite order.
We also remark that this trichotomy does not exhaust all conjugacy classes in
$SL(2,\mathbb{Z})$. Indeed, there are 6 conjugacy classes of elliptic
type, given by the matrices
\begin{equation}\label{ellipticsl2z}
(UV)^2 = \begin{pmatrix} 0& 1 \\-1 & -1 \end{pmatrix} \, , \qquad
UVU = \begin{pmatrix} 0& 1 \\-1 & 0 \end{pmatrix} \, ,  \qquad
UV = \begin{pmatrix} 1 & 1 \\-1 & 0 \end{pmatrix} \, , \qquad
\end{equation}
respectively of order 3, 4, 6, together with their inverses. Note
that from the relations~\eqref{mcgtorus}, the inverses are given by
\begin{align}
  (UV)^{-2}  &=(UV)^4  = \begin{pmatrix} -1& -1 \\1 & 0 \end{pmatrix} \, , \qquad
(UVU)^{-1} =  (UV)^4U= \begin{pmatrix} 0& -1 \\ 1 &
  0 \end{pmatrix}\, , \nonumber\\
(UV)^{-1} &= (UV)^5 = \begin{pmatrix} 0 & -1 \\ 1 &1 \end{pmatrix}
\, . \label{ellipticsl2zinv}
\end{align}
The only other finite order element is $(UV)^3 =
-\mathbb{1} $. For parabolic elements, which are of infinite order,
there are infinite conjugacy classes. Note that both the $U$ and $V$ Dehn twists are
in the same conjugacy class; a general
representative is 
\begin{equation}
\pm V^N =  \pm\begin{pmatrix} 1& N\\0 & 1 \end{pmatrix} \, .
\end{equation} 
For hyperbolic elements there is a conjugacy class
for each value of the trace, plus additional sporadic classes, whose
number can be computed by studying equivalence classes of binary
quadratic forms, as explained for example in~\cite{DeWolfe:1998pr}. Two
examples are given by
\begin{equation}\label{hypmapsgenus1}
V^{1-N} U V  = \begin{pmatrix} N & 1 \\-1 & 0 \end{pmatrix} \,
,\qquad U^{N+1}VU  = \begin{pmatrix} 0 & 1 \\ -1 & -N \end{pmatrix} \,
, \quad N \geq 3 \, .
\end{equation}
We also provide a Dehn twist decomposition for the sporadic
elements listed in~\cite{DeWolfe:1998pr} (some of the following matrices are
conjugated to their representative matrices):
\begin{align}
M_8 &= U^{-3} V^{2}  = \begin{pmatrix} 1 & 2 \\3 & 7 \end{pmatrix}\,
,\quad M_{10} = U^{-4} V^{2}=\begin{pmatrix} 1 & 2 \\ 4 &
  9 \end{pmatrix} \, , \quad M_{12} = U^{-5} V^{2} =\begin{pmatrix} 1 & 2 \\ 5 &
  11 \end{pmatrix} \, , \nonumber \\
M_{13} &= U^2V^2U^3V^3  = \begin{pmatrix} -5 &  -13 \\ 7 & 18 \end{pmatrix}\,
,\quad M_{14} =  U^{-6}  V^{2} = \begin{pmatrix} 1 & 2 \\6 &
  13 \end{pmatrix} 
\quad  + \text{ inverses} \, .
\end{align}
The subscript indicates the trace. We note that the decomposition
$M_{2N+2} = U^{-N}V^2$ seems not to be conjugate to the element~\eqref{hypmapsgenus1} with the same
trace, $V^{-2N-1}UV$, nor to its inverse for $N > 1$ so we conjecture that this gives a 
list of sporadic conjugacy classes for hyperbolic elements of even
trace $2N+2$, $N\geq 2$.\footnote{In Table 7 of~\cite{DeWolfe:1998pr} only two conjugacy
  classes are listed for the $Tr =6$ case. However, we could not find
  any element that conjugates the matrix $U^{-2}V^2$ to $V^{-5}UV$ or
  its inverse.}

The classification obtained from the trace of $SL(2,\mathbb{Z})$
elements is reflected in
a trichotomy of the corresponding torus diffeomorphisms. These are periodic, reducible
and Anosov maps. We refer to appendix~\ref{app:mcg} for a brief review
of mapping class groups.
A standard result in the theory of 3-manifolds is that the class of
torus diffeomorphism $\phi$ determines the geometry of the corresponding
mapping torus $\mathcal{N}_{\phi}$, which can be either Euclidean, Nil
or Sol. We will see an example of this in the next section.

\section{Manifolds and T-folds}\label{sec:Tfolds}

We can apply the discussion of the previous section to string theory
compactified on a two torus $T^2_{\tau}$. An additional fibration on a
circle generates a class of 3-manifolds which provides the simplest
examples of Nil- and Sol-manifolds.\footnote{See for
  example~\cite{scott} for a discussion of Nil and Sol geometries.} If we
also include monodromies in $SL(2,\mathbb{Z})_{\rho}$, classified by an
auxiliary $T^2_{\rho}$ fibration over the circle, the resulting
3-spaces will be in general non-geometric and are usually referred to
as T-folds~\cite{Hull:2006qs} (see for
example~\cite{Hull:2005hk,Hull:2006tp,Hull:2007jy,Hull:2009sg} for
a discussion of such spaces in various contexts).
 Usually, one arrives to such spaces by
performing a chain of T-duality transformations~\cite{Kachru:2002sk,Shelton:2005cf}. The simplest example
is to start with a three torus $ds^2 = dx^2 + dy^2 + d\theta^2$ and a
B-field $B = N \theta dx\wedge dy$. Busher rules on $x$ give a metric $ds^2 = (dx - N \theta dy)^2 +dy^2
+d\theta^2$ and $B=0$. To make sense of the
metric globally one identifies $(x,y,\theta)\sim (x+1, y, \theta) \sim (x, y+1,
\theta) \sim (x + N y, y, \theta+1)$. Note that later in this section
and in section~\ref{sec:exoticbranes} we will find more convenient to
use conventions for which the
period of $\theta$ is $2\pi$. The resulting 3-manifold is a Nil geometry
and it is identified with a mapping torus $\mathcal{N}_{\phi}$ for an element
$\phi \in SL(2,\mathbb{Z})_{\tau}$ that acts by $\tau \rightarrow \tau
+ N$. A further application of Busher rules on $y$ gives a metric
\begin{equation}
ds^2 =  f(\theta) (dx^2 + dy^2) + d\theta^2 \, ,
\end{equation} and a B-field $B = N \theta  f(\theta) 
dx\wedge dy$, with $f(\theta) =(1+ N^2 
\theta^2)^{-1}$. The monodromy is now an
element of $SL(2,\mathbb{Z})_{\rho}$ and acts by $\rho^{-1} \rightarrow
\rho^{-1} + N$. This has a non-trivial action on the volume, and the
corresponding 3-space is a T-fold. The three torus with B-field
obviously is not a string vacuum, so this simple example should be
considered as an illustrative toy model. An important caveat is the
fact that the Nilmanifold is not a principal torus fibration, and the
lack of a global $U(1)$ isometry makes the application of Busher rules
a priori not obvious. For a detailed discussion on such ``obstructed''
T-duality see for
example~\cite{Hull:2006qs,Belov:2007qj,Grana:2008yw,Hull:2009sg}, and~\cite{Plauschinn:2014nha} for
a discussion on how to possibly overcome this problem.

Our logic will be different and we do not rely on T-duality. In this
section, we will classify all possible geometric and non-geometric
3-spaces from the corresponding classification of mapping class groups
of the $T^2_{\tau}$ and $T^2_{\rho}$ fibrations. In the following
sections, we will construct explicit solutions for such T-folds and
we will use the results of this section to classify all possible local
geometries around defects arising from fibration of the T-duality group on $\mathbb{P}^1$.
The mapping tori we are considering will arise at a boundary of a
fibration on a
small disk encircling such T-fects. In particular, a subset of mapping
class group elements arises from degenerations of elliptic fibrations
dual to geometric F-theory compactifications.

We will first briefly review geometric torus bundles with twists in
$SL(2,\mathbb{Z})_{\tau}$ and we
will then discuss spaces with non-geometric twists.

\begin{figure}
\begin{center}
\includegraphics[scale=0.4]{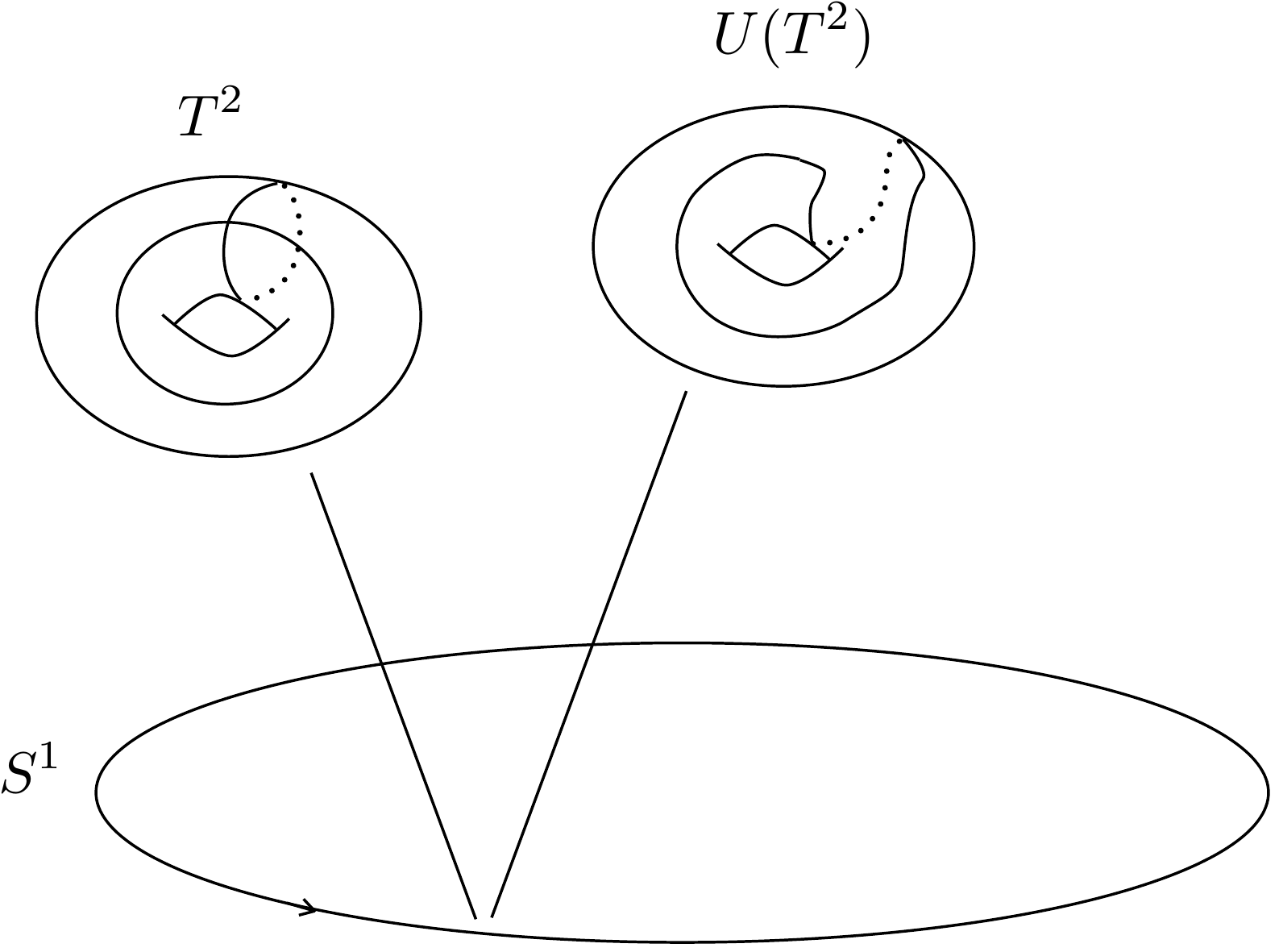}
\caption{The mapping torus of a $T^2$ compactification. The monodromy
  of the fibration, that determines the geometry of the total space,
  is a product of Dehn twists. The simplest example is a twist along
  one cycle of the homology basis: the total space is a Nilmanifold.}
\label{Fig:mappingtorus}
\end{center}
\end{figure}

\subsection{Geometric monodromies}\label{subsec:geometricmonodromies}

We now consider a torus fiber with $\rho= i$ fixed and with
a complex structure $\tau(\theta)$ varying on $\mathcal{B}=S^1$, as
shown in
Figure~\ref{Fig:mappingtorus}. This example is elementary and has
been discussed in some details in~\cite{Hull:2005hk} but it serves well
to illustrate
the role of monodromy. 
In order to construct the torus fibration, we need to determine the
function $\tau(\theta)$, given a 
monodromy $M \in SL(2,\mathbb{Z})_{\tau}$.  The metric of the
total space is then:
\begin{equation}\label{metrictaubundle}
ds_3^2 = d\theta^2 +G_{ab}(\tau)dx^adx^b \, , \quad a=1,2 \, ,
\end{equation}
where the torus metric is, from~\eqref{taurhodef},
\begin{equation}
G(\tau) = \frac{1}{\tau_2}\begin{pmatrix} |\tau|^2& \tau_{1} \\\tau_1 &
  1\end{pmatrix} \, ,
\end{equation}
we set $\tau=\tau_1+i\tau_2$ and we fix the radius of the base circle to
$R=1$. Through this section we will set $dx^1= dx$, $dx^2 = dy$.
The action of a given element $M$ on $\tau$ is by M\"obius transformation~\eqref{sl2action}.
We fix $\tau_0=\tau(0)$ and we demand that $\tau$ has the
monodromy $\tau(2\pi) = M[\tau_0]$. To find such a function we
consider the element $\mathbf{m}= \log(M)$ of the Lie algebra
$sl(2,\mathbb{R})$. Then, we construct the path $M(\theta)$ in $SL(2,\mathbb{R})$
given by the exponential map $M(\theta)= \text{exp}(\mathbf{m}\,
\theta/2\pi)$. By construction, $M(2\pi)=M$. We
define then
\begin{equation}\label{deftautheta}
\tau(\theta) = M(\theta)[\tau_0] \, ,
\end{equation}
which obviously has the desired monodromy properties: $\tau(0)=\tau_0$,
$\tau(2\pi)= M[\tau_0]$. We will discuss each of the three classes of
$SL(2,\mathbb{Z})_{\tau}$ monodromies, and the corresponding torus diffeomorphisms, separately.

\subsubsection*{Parabolic (reducible)}

We first consider a monodromy of parabolic type, namely an element
$M_{\tau} \in SL(2,\mathbb{Z})_{\tau}$ with $|tr(M_\tau)|=2$. The simplest
example is a shift $\tau \rightarrow \tau
+ N$:
\begin{equation}\label{tauV}
M_{\tau} = V^{N} = \begin{pmatrix} 1 & N\\ 0& 1 \end{pmatrix} \, ,
\qquad \mathbf{m}_{\tau}= \begin{pmatrix} 0& N \\0 & 0 \end{pmatrix} \, .
\end{equation}
This monodromy implements a $N$-th power of a Dehn
twist along the $v$ cycle of the torus once encircling the base and
the complex structure is given by $\tau(\theta) = \tau_0 +
N\theta/2\pi$, where $\tau_0$ is an arbitrary complex parameter. The corresponding total space is a
Nilmanifold. This can be seen from the metric~\eqref{metrictaubundle}
which (fixing $\tau_0=i$) becomes
\begin{equation}\label{Nil3}
ds^2_3 = d\theta^2 + dx^2 + \left( dy + \frac{N}{2\pi} \theta dx\right)^2 \, ,
\end{equation}
with the coordinate identification discussed above, that implements the desired monodromy.
The metric~\eqref{Nil3} can be written explicitly as a left-invariant
metric on a nilpotent Lie group $\mathcal{G}$ as follows. We introduce the Maurer-Cartan
forms $\eta^{\theta} = d\theta$ and $\eta^a = M(\theta)^a_{\phantom{a}
  b} dx^b$, which satisfy:
\begin{equation}\label{formnilman}
d\eta^{\theta} = 0 \, ,\quad d\eta^a = (\mathbf{m}_{\tau})_{\phantom{a}
  b}^{a} \,\eta^{\theta} \wedge \eta^b \, .
\end{equation}
The generators $(t_{\theta}, t_{1},t_2)$ of the corresponding Lie algebra $\mathfrak{g}$ then
satisfy 
\begin{equation}\label{nilpotalgebra}
[t_{\theta}, t_a] = -(\mathbf{m}_{\tau})_{a}^{\phantom{a} b}\,  t_b \,
,\quad [t_a,t_b] = 0 \, .
\end{equation}
We see from~\eqref{tauV} and~\eqref{formnilman} that $\mathfrak{g}$ is the Heisenberg
algebra and the torus fibration can also be seen as a principal circle bundle over
a torus. The global identification that makes the space compact is a
quotient of the Lie group $\mathcal{G}$ by a discrete subgroup. In the notation
of~\cite{Grana:2006kf} this space is referred to as  $(0,0,N\times \theta 1)$. The
existence of a compactification of the Lie group can be inferred from
the condition that the Lie algebra element $\mathbf{m}_{\tau}$ is traceless.
The previous example is the simplest example of a
Nil geometry and, as we discussed above, it can be obtained by T-duality from a three torus
with non-vanishing B-field. 
We can also consider a monodromy which implements a Dehn twist
around the $u$ cycle of the fiber $T^2$:
\begin{equation}
M_{\tau}=U^N =\begin{pmatrix} 1 &&0\\-N && 1 \end{pmatrix} \, , \qquad
\mathbf{m}_{\tau}= \begin{pmatrix} 0&0\\ -N & 0 \end{pmatrix} \, ,
\end{equation}
which corresponds to the following action on $\tau$:
\begin{equation}
\tau \rightarrow \frac{\tau}{1-N\tau} \, .
\end{equation}
Note that this corresponds to a shift of $\tau^{-1}$ instead of a shift
of $\tau$: $\tau^{-1} \rightarrow \tau^{-1} - N$.
More generally, for each $N$ the conjugacy class of a parabolic
monodromy is labeled by two coprime integers $(p,q)$:
\begin{equation}\label{nilpqtau}
M_{\tau} = L^{-1}V^{N} L=  \begin{pmatrix} 1+N p q & N p^2\\-N q^2  & 1-N p q \end{pmatrix}
\, ,\qquad
\mathbf{m}_{\tau}= \begin{pmatrix} N p q& N p^2\\ -N q^2 & -N pq \end{pmatrix} \, ,
\end{equation}
with $L \in SL(2,\mathbb{Z})$. The corresponding solution for
$\tau(\theta)$ is given by:
\begin{equation}\label{pqtau}
\tau(\theta) = \exp(\mathbf{m}_{\tau}\theta/2\pi)[\tau_0] = \frac{(2\pi+N pq \theta)\tau_0 +N p^2 \theta}{-N q^2
  \theta \, \tau_0 + (2\pi-N pq \theta)} \, .
\end{equation}
The corresponding metric on the total space is:
\begin{equation}
ds^2_3 = d\theta^2 + d\tilde x^2 +\left(d\tilde y + \frac{N}{2\pi
    }(p^2+q^2)\theta d\tilde x\right)^2 \, ,
\end{equation}
where 
\begin{equation}\label{rotatednil}
  \tilde x + i \tilde y = e^{i \varphi} (x + i y) \, , \quad \varphi =
  \text{arctan}\left(\frac{q}{p}\right) \, .
\end{equation}
This is of course the expected result, and the identification of the coordinates
now corresponds to gluing the $T^2$ after
performing a $N$-th power of a Dehn twist along a cycle represented by $p [v]
+ q [u]$. It is easy to see from~\eqref{nilpotalgebra} and~\eqref{nilpqtau} that 
the lower central series, namely the sequence of ideals
$\mathfrak{g}^0 = \mathfrak{g}, \, \mathfrak{g}^1 =
[\mathfrak{g},\mathfrak{g}^0], \, \mathfrak{g}^2 =
[\mathfrak{g},\mathfrak{g}^1], \dots$, terminates and the algebra is
nilpotent. One might think that different monodromies in the same
conjugacy class are uninteresting, however we will see by
repeating the previous analysis for twists in
$SL(2,\mathbb{Z})_{\rho}$, that parabolic elements in the same class give rise to very
different local metrics.

\subsubsection*{Elliptic (periodic)}

We now consider an elliptic conjugacy class represented by the
order 4 monodromy:
\begin{equation}
M_{\tau}=UVU= \begin{pmatrix} 0&1\\ -1 & 0 \end{pmatrix} \, , \,
\qquad \mathbf{m}_{\tau}= \begin{pmatrix} 0&\frac{\pi}{2}\\ -\frac{\pi}{2} & 0 \end{pmatrix} \, .
\end{equation}
From~\eqref{deftautheta} we then find the following solution for $\tau(\theta)$ (see also~\cite{Hull:2005hk}):
\begin{equation}\label{ellipticorder4}
\tau(\theta) = \frac{\cos \left(\theta/4\right) \tau_0 + \sin
\left( \theta/4\right) }{ -\sin \left(
    \theta/4\right)\tau_0 + \cos
\left( \theta/4\right) } \, .
\end{equation}
The corresponding total space is a compactification of the Lie group ISO(2), namely the symmetries of the Euclidean 2-plane. This can be seen from the local coordinates
representation of the left-invariant forms:
\begin{align}
\eta^{\theta} & = d\theta \, , \nonumber\\
\eta^2 & = \cos \left(\theta/4\right) dx + \sin
\left(\theta/4\right) dy \, ,\nonumber\\
\eta^3 & = -\sin \left(\theta/4\right) dx + \cos
\left(\theta/4\right) dy \, ,
\end{align}
which satisfy $d\eta^{\theta}=0, \, d\eta^1=
\frac{\pi}{2}\eta^{\theta}\wedge \eta^2, \,
d\eta^2=-\frac{\pi}{2}\eta^{\theta}\wedge \eta^1$. 
The left-invariant metric is then~\cite{Hull:2005hk} $ds_3^2 =(\eta^{\theta})^2 +  G_{a b}(\tau_0) \eta^a
\eta^b$.
Analogous results
apply to the remaining finite order conjugacy classes. For example, 
for the conjugacy classes of order 6 and 3, represented by $(UV)^k$, $k=1,2$, we have
\begin{equation}
\mathbf{m}_{\tau}^{UV} = \frac{\pi}{3\sqrt{3}}\begin{pmatrix} 1&2\\ -2
  &  -1 \end{pmatrix} \, ,\quad \mathbf{m}_{\tau}^{UVUV} = 2\, 
\mathbf{m}_{\tau}^{UV} \, ,
\end{equation}
and the complex structure is:
\begin{equation}\label{ellipticorder6}
\tau(\theta) =\frac{A^{+}\tau_0 +  \frac{2}{\sqrt{3}}\sin\left(k \,
    \theta/6\right)}{-\frac{2}{\sqrt{3}}\sin\left(k \,
    \theta/6\right)\tau_0 +A^{-}} \, , \quad A^{\pm} = \cos\left(k\,
    \theta/6\right)\pm\frac{1}{\sqrt{3}}\sin\left(k\,
    \theta/6\right) \, .
\end{equation}
All 3-manifolds obtained in this way are compactifications of Lie groups of a
unimodular solvable (but not nilpotent) algebra. Note that if the
modulus $\tau_0$ is chosen to be the fixed point of the monodromy
element, namely
\begin{equation}
UVU : \tau_0 = i \, , \quad UV : \tau_0 = e^{i \pi/3} \, , \quad UVUV
: \tau_0 = e^{2 i \pi/3} \, ,
\end{equation}
 the complex structure does not depend on the circle coordinate and we
 have $\tau(\theta)=\tau_0$. These
 fixed points are the minima of the potential for the moduli obtained
 by a reduction with duality twsits and at those points one has a CFT
 description in terms of symmetric orbifolds. More details can be
 found in~\cite{Dabholkar:2002sy,Hull:2006tp,Condeescu:2013yma,Grana:2013ila}.

\subsubsection*{Hyperbolic (Anosov)}

Lastly, we consider hyperbolic monodromies, corresponding to
Anosov diffeomorphisms of the fiber torus. Let us first give a simple example
with monodromy in $SL(2,\mathbb{R})$:
\begin{equation}\label{hyperbolicsl2r}
M_{\tau}= \begin{pmatrix} e^{\omega}& 0 \\ 0 & e^{-\omega} \end{pmatrix} \, , \,
\qquad \mathbf{m}_{\tau}= \begin{pmatrix} \omega&0\\ 0 & -\omega\end{pmatrix} \, .
\end{equation}
The corresponding modulus is given by $\tau(\theta)= e^{\omega
  \theta/\pi} \tau_0$ and the metric is that of a Sol-manifold. For
example, the metric 
on the total space assuming $\tau_0$ is an imaginary parameter is:
\begin{equation}\label{metricsolvsl2r}
ds_3^2 = e^{\omega \theta/\pi} dx^2 + e^{-\omega  \theta/\pi} dy^2 + d\theta^2 \, .
\end{equation}
From the relations~\eqref{nilpotalgebra} we see that the corresponding
algebra is solvable. However, the lower central series generated by
$[t_{\theta} , \mathfrak{g}]$ for the $sl(2,\mathbb{R})$
element~\eqref{hyperbolicsl2r} do not terminate and thus the algebra
is not nilpotent. The compact torus bundle is obtained by a quotient
of the Lie group Sol by a discrete subgroup (see for
example~\cite{scott} for more details).
Note that the background~\eqref{metricsolvsl2r} was discussed
in~\cite{Silverstein:2007ac} and since then Sol-manifolds have played
an important role in the search for de Sitter vacua in string theory.
 Some of
 the hyperbolic $SL(2,\mathbb{Z})$ conjugacy classes, labeled by an integer $N$ are
 given by:
\begin{equation}\label{hyperbolictaumonodromy}
M_{\tau}= V^{1-N}UV= \begin{pmatrix} N & 1 \\- 1 & 0 \end{pmatrix}
\, , \quad N\geq 3 \, .
\end{equation}
Note that the matrix
\begin{equation}
M=\begin{pmatrix} 2 &1 \\
  1&1\end{pmatrix} \, ,
\end{equation}
representing Arnold's cat map on the
torus, is conjugate to the matrix above for $N=3$. The monodromy
matrix can be diagonalized with two eigenvalues ($\lambda$, $1/\lambda$), with 
\begin{equation}\label{lambdahyp}
\lambda = \frac12\left( N+\sqrt{N^2-4}\right) \, , \quad N\geq 3 \, .
\end{equation}
Powers of the Anosov diffeomorphism stretch and contract
exponentially the two eigenspaces. For this reason, the repackaging of
the torus image under powers of an hyperbolic map into the fundamental domain results in a
complicate behavior that share many similarities with chaotic systems.
We refer to appendix~\ref{app:mcg} for a discussion on the dynamics of
hyperbolic
maps. In terms of $\lambda$ the algebra element is given
by
\begin{equation}
 \mathbf{m}_{\tau}= \frac{\log \lambda}{\lambda^2-1} \begin{pmatrix} \lambda^2
   +1 & 2\lambda\\ -2\lambda & - (\lambda^2+1) \end{pmatrix} \, ,
\end{equation}
and the corresponding
modulus is then given by:
\begin{equation}\label{tauhyp}
\tau(\theta) = \frac{\tau_0 +
  \lambda -\lambda^{1+\theta/\pi}(1+\lambda \tau_0)}{-\lambda (\tau_0+\lambda) + \lambda^{\theta/\pi}
  (1+\lambda \tau_0)} \, .
\end{equation}
This expression again determines a metric on the total
space of the torus bundle. Analogous solutions can be found for the remaining conjugacy
classes. Given the peculiar nature of Anosov diffeomorphisms, it
would be interesting to revisit phenomenological applications of string
models based on hyperbolic torus bundles.

\subsection{Non-geometric monodromies}\label{subsec:nongeometricmonodromy}

We now discuss the case of a torus bundle in which we fix
the complex structure $\tau=i$ of the fiber but we allow $\rho(\theta)$ to vary along the base
$\mathcal{B}=S^1$. A classification of such spaces can be obtained
formally by applying the results of the previous section to an
auxiliary $T^2_{\rho}$ fibration on the circle, that geometrizes the 
$SL(2,\mathbb{Z})_{\rho}$ factor of the T-duality group. A local
expression for the three dimensional metric and the B-field can be
then obtained from:
\begin{equation}\label{mappingtorusrhoansatz}
ds_3^2 = d\theta^2 + \rho_2(\theta) (dx^2+dy^2) \, , \quad B =\rho_1(\theta)
dx\wedge dy \, ,
\end{equation}
where $\rho = \rho_1 + i \rho_2$.
With the exception of gauge transformations for the
B-field, the monodromy mixes the volume and the B-field, and there is
an obstruction to glue the ends of a mapping cylinder with $\rho$ monodromy:
the resulting total space will not be a manifold. There exists
particular limits in moduli space at which we have a string
description of such spaces from asymmetric
orbifold CFTs, but otherwise there is no clear string description of such
T-folds. Our discussion for now will be formal, but as we will discuss in the next sections, the
non-geometric monodromies that we describe do arise in fibrations over
$\mathbb{P}^1$, and in some cases one could obtain evidence for the
existence of the corresponding string vacuum from string dualities.

\subsubsection*{Parabolic}

Let us consider the parabolic monodromy
\begin{equation}
M_{\rho}= V^N =\begin{pmatrix} 1 & N\\ 0 & 1 \end{pmatrix} \, .
\end{equation}
This corresponds to a shift of $\rho$, giving $\rho(\theta) = \rho_0 +
N \theta/2\pi$ and can be understood geometrically as a power of a Dehn twist along
the $v$ cycle of the auxiliary $T^2_{\rho}$. The corresponding total
space is just a three torus $T^3 = T^2_{\tau} \times
S^1$ with $N$ units of $H$-flux, and can be obtained by T-duality from
the Nilmanifold~\eqref{Nil3} along the global $U(1)$ isometry as we discussed at the beginning of this section.
If we consider the conjugate monodromy 
\begin{equation}
M_{\rho}= U^N=\begin{pmatrix} 1 & 0\\ -N & 1 \end{pmatrix} \, ,
\end{equation}
we obtain the following solution for $\rho$:
\begin{equation}\label{rhofibrations1}
 \rho(\theta) = \frac{2\pi\rho_0}{- N \theta \rho_0 +2\pi} \, .
\end{equation}
The gluing condition now mixes the volume of the fiber torus with the
B-field, as follows from the action on the volume: 
\begin{equation}
 \sqrt{G} \rightarrow
\frac{\sqrt{G}}{N^2 G +(N B-1)^2} \, .
\end{equation}
We could consider a fibration of the torus on a interval, but we
encounter an obstruction to glue the endpoints to obtain a torus bundle. Note that for $\rho_0 = i$,
from~\eqref{rhofibrations1} we obtain the following metric:
\begin{equation}
ds_3^2 = d\theta^2 + \frac{4\pi}{4\pi+N^2\theta^2}\left(dx^2+dy^2\right) \, , \quad B =
-\frac{2\pi N\theta}{4\pi^2+N^2\theta^2}dx\wedge dy \, ,
\end{equation}
which coincides with the T-fold metric we obtain from an obstructed T-duality on
the Nilmanifold.
 We can also consider the general $(p,q)$ parabolic
conjugacy class as in~\eqref{pqtau}, thus obtaining a class of T-fold
metrics that interpolate between the 3-torus and the T-fold.
 
\subsubsection*{Elliptic}

We can now repeat the analysis of finite order elements done for the
geometric $\tau$ monodromies. The corresponding solutions for
$\rho(\theta)$, for monodromies of order 3, 4 and 6, are obtained
by~\eqref{ellipticorder4} and~\eqref{ellipticorder6} by a fiberwise
mirror symmetry $\tau\rightarrow \rho$. The corresponding non-geometric T-folds have been
discussed in the context of generalized Scherk-Schwarz reduction
within double field theory in~\cite{Hassler:2014sba,Hassler:2014mla}. The
corresponding potential $V(\rho_0)$ admits a minimum at the fixed point
of the monodromy matrix. At such minimum, there exists a 
description in terms of an asymmetric
orbifold CFT~\cite{Condeescu:2012sp,Condeescu:2013yma}. We note that
in the CFT description (as well as in double field theory) we see the presence of both $H$-flux and
$Q$-flux. These fluxes should be identified by the corresponding
monodromy, the perturbative shift $V_{\rho}\sim H$ and the
non-geometric twist $U_{\rho}\sim Q$. The simultaneous presence of
both fluxes fits well with the monodromy decomposition of the elliptic
monodromy of order 4, $M_{\rho}^{ell} = U_{\rho} V_{\rho}
U_{\rho}$. As we will discuss in the next section, such decompositions
are defined up to some redundancy, which basically follows from a
braid relation of the Dehn twist decomposition. In this case we have for
example, from~\eqref{mcgtorus}, $U_{\rho} V_{\rho}U_{\rho} = V_{\rho}
U_{\rho} V_{\rho}$, so it is not simple to match the monodromy
decomposition with the corresponding flux parameters. We will come
back to this point when we will study fibrations on a $\mathbb{P}^1$
base and we will identify the sources for the fluxes.

\subsubsection*{Hyperbolic}

The last case is that of a hyperbolic monodromy in $\rho$. A simple
example is again obtained by taking the $SL(2,\mathbb{R})$
monodromy~\eqref{hyperbolicsl2r}, which for a choice of parameters gives a metric $ds_3^2 =
d\theta^2 + e^{2\theta \omega}(dx^2+dy^2)$. Let us
discuss the action on the volume for an $SL(2,\mathbb{Z})$ representative element
\begin{equation}\label{rhohypmonodromy}
M_{\rho}= \begin{pmatrix} N & 1 \\- 1 & 0 \end{pmatrix}
\, , \quad N\geq 3 \, .
\end{equation}
The action on $\rho$ is given by
\begin{equation}\label{rhohyp}
\rho \rightarrow -\frac{1}{\rho} - N \, .
\end{equation}
We note that the
action on the volume given by~\eqref{rhohyp} is the same as for the
elliptic case:
\begin{equation}
\sqrt{G}\rightarrow \frac{\sqrt{G}}{B^2+G} \, .
\end{equation}
However, iterations of
the map are not periodic and act in a complicated way. For example,
successive iterations act as:
\begin{equation}
\rho \rightarrow  - N  - \cfrac{1}{-N-\cfrac{1}{\ddots \, -N
    -\cfrac{1}{\rho}}} \,  ,
\end{equation} 
and for $N\ge 3$ the continued fraction does not terminate since
the fixed points of the transformation are quadratic irrational numbers.
The exact expression can be found
by setting $\theta=2\pi\, n$ in~\eqref{tauhyp} with $\tau \rightarrow
\rho$. In terms of the largest eigenvalue $\lambda$~\eqref{lambdahyp}
we find the following action on the volume:
\begin{equation}
\sqrt{G} \rightarrow
\frac{\sqrt{G}\, \lambda^{2n}\,\lambda_1^2}{(\lambda^{2n}-\lambda^2)^2+2B(\lambda^{2n}-\lambda^2)\lambda\lambda_{n}+(B^2+G)\lambda^2\lambda_n^2}
\, ,\quad \lambda_k = \lambda^{2k}-1 \, .
\end{equation}
Again, we refer to appendix~\ref{app:mcg} for more details on the
dynamics of hyperbolic diffeomorphisms of surfaces.

\subsection{General case}

Finally, we briefly discuss the situation in which the full T-duality
group is fibered over the circle, and both $\tau(\theta)$ and
$\rho(\theta)$ vary. If such configurations are allowed, we can get
non-geometric backgrounds which are not fiberwise mirror symmetric to
any geometric background. We will discuss the consistency of such
configurations in the next sections in relation to heterotic/F-theory
duality. We point out that a torus fibration with both $\tau$ and $\rho$
elliptic monodromies has been studied recently
in~\cite{Condeescu:2013yma,Hassler:2014sba}. An example is the order 4 solution:
\begin{align}
\tau(\theta) &= \frac{\cos \left(f \theta\right) \tau_0 + \sin
\left(f \theta\right) }{ -\sin \left(
    f \theta\right)\tau_0 + \cos
\left( f\theta\right) } \, , \quad f \in \frac14 + \mathbb{Z} \, , \nonumber\\
\rho(\theta) &= \frac{\cos \left(g\theta\right) \rho_0 + \sin
\left( g\theta\right) }{ -\sin \left(
    g\theta\right)\rho_0 + \cos
\left( g\theta\right) } \, ,\quad  g \in \frac14 + \mathbb{Z} \, .\label{doubleellipticorder4}
\end{align}
Here the parameters $f$ and $g$ should be identified with geometric
and non-geometric 
fluxes.
Such configuration, being not
T-dual to a geometric space, can be described in double field theory
at the price of relaxing the strong constraint for the generalized
Killing vectors. The potential for the moduli $(\tau_0, \rho_0)$ has a
minimum at the fixed points of the monodromy $(\tau_0, \rho_0) =(i,i)$. At this minimum, an asymmetric
orbifold description was studied in~\cite{Condeescu:2013yma}. We note that
the presence of both kind of geometric and non-geometric fluxes
should be related in the present context to the factorization of the
monodromy in terms of Dehn twists. As we will discuss in more details
in the following sections, we should really think in terms of mapping
class group of a genus 2 surface obtained by gluing the two tori
$T^2_{\tau}$, $T^2_{\rho}$ with a trivial map. If we then embed the
representative matrices $M_{\tau}$, $M_{\rho}$ in $4\times 4$ matrices
acting on the homology of the genus-2 surface, we can think about the
double elliptic monodromy as being decomposed into
$M_{\tau,\rho}^{ell}
=U_{\tau}V_{\tau}U_{\tau}U_{\rho}V_{\rho}U_{\rho}$. This would
corresponds to having two kind of geometric fluxes, as well as $H$ and
$Q$-fluxes, in agreement with the analysis
of~\cite{Condeescu:2013yma}. We will discuss more about genus 2 fibrations in the last section.

We will now turn to the construction of explicit solutions that
provide a concrete realization of the spaces described in this section.

\section{T-folds across the wall}\label{sec:T-walls}

\begin{figure}
\begin{center}
\includegraphics[scale=0.4]{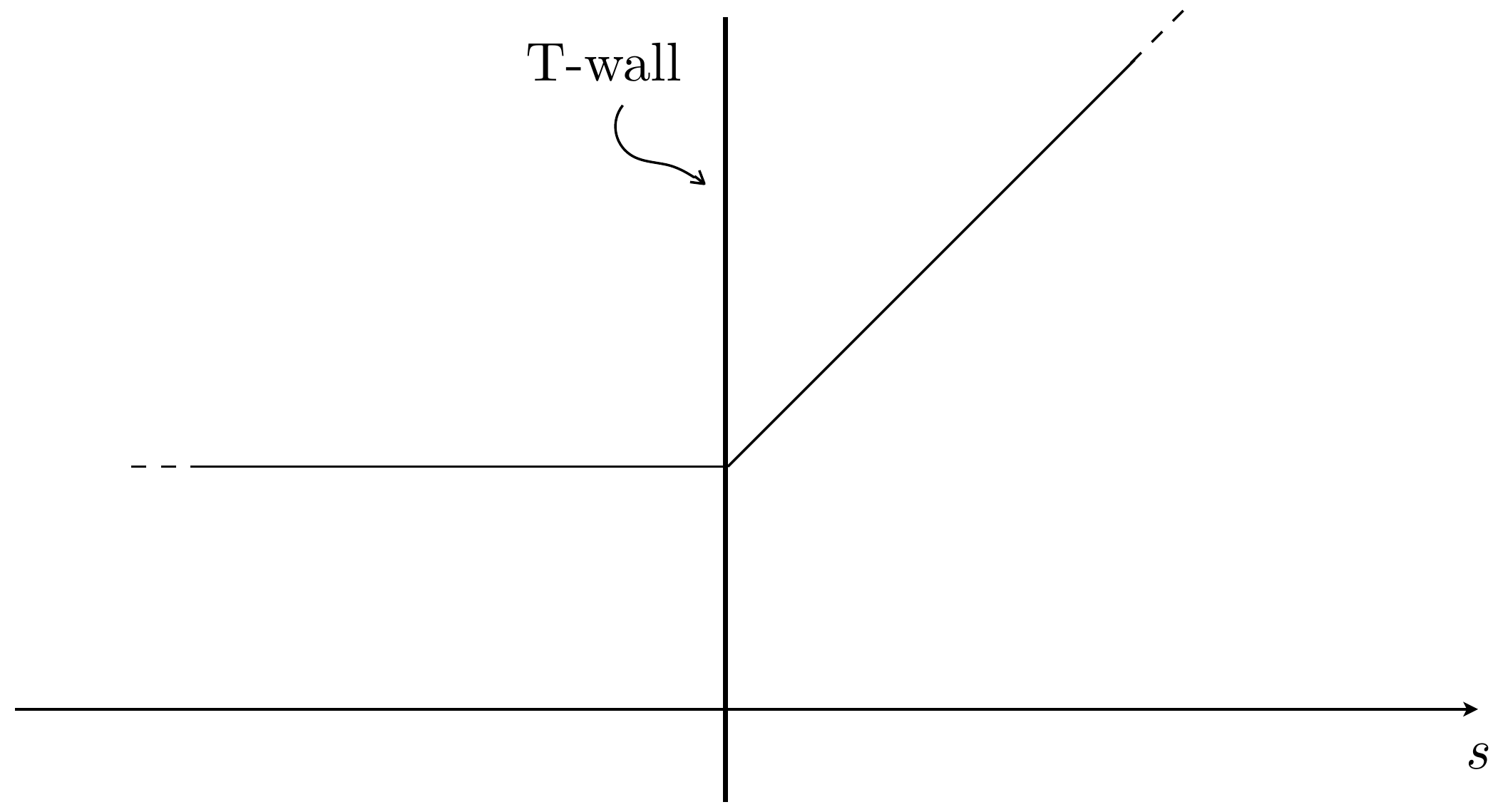}
\caption{The profile of the harmonic function $h(s)$ for the domain
  wall considered in the text.}\label{Fig:Twall}
\end{center}
\end{figure}

Before studying in details fibrations over two-dimensional bases, we
mention here a way to embed the 3-spaces discussed in the previous
section into backgrounds that solve the supergravity equations of
motion. This construction was described in a different context by Hull
in~\cite{Hull:1998vy} and further discussed
in~\cite{Ellwood:2006my,Schulgin:2008fv}. Let us consider the mapping
torus with a parabolic monodromy $\tau \rightarrow \tau + N$, namely a
Nil-manifold, whose metric is given in~\eqref{Nil3}. The latter is
determined by the complex modulus $\tau(\theta) = N \theta +
\tau_0$ (for simplicity in this section we take a periodicity $\theta
\sim \theta+1$). We can now make an ansatz for a ten-dimensional metric of the
form $\mathbb{R}^{5,1}\times \mathbb{R}\times \mathcal{N}_3$, where
the mapping torus is fibered over a line parametrized by a coordinate
$s$. We then assume that $\tau_0$ is a function of $s$:
\begin{equation}
ds^2 = \eta_{\mu \nu} dx^{\mu}dx^{\nu} + h(s) ds^2 +  R^2(s) d\theta^2
+ G_{ab}(\tau(\theta,s))dx^adx^b \, ,
\end{equation}
and we further assume that $R^2(s) =h(s)$ and $\tau(s) =N \theta + i
h(s)$~\cite{Hull:1998vy}. The Ricci curvature is then given by
\begin{equation}
R = -\frac{N^2-(\partial_s h)^2+ 2 h \partial_s^2 h}{2h^3} \, .
\end{equation}
It is easy to see that a flat solution which also solves the
Ricci-flatness condition
is given by $h(s) = N s + C$, with $C$ an arbitrary constant (see Figure~\ref{Fig:Twall}). To
obtain a sensible solution we can then take the following functions:
\begin{equation}
h(s) = C +  s  \left[N\Theta(s) - M\Theta(-s)\right] \, ,\quad \tau =
\left[N\Theta(s) - M\Theta(-s)\right] (\theta + i s) + i C\, ,
\end{equation}
where $\Theta$ is the step function. For $M=0$ the background becomes:
\begin{equation}\label{Nilwall}
ds^2 = \eta_{\mu \nu} dx^{\mu}dx^{\nu} +
h(s)\left[ds^2+d\theta^2 + dx^2\right] +\frac{1}{h(s)}\left[ N
  \Theta(s) \theta dx+ dy\right]^2 \, .
\end{equation}
Note that at $s=0$ the solution breaks down and there is a curvature
singularity $R \sim - N\delta(s) $. As we will discuss in more details
in the next section, the background~\eqref{Nilwall} corresponds to a
KK monopole smeared on two of the $\mathbb{R}^3$ directions, and so we
should accept the singularity at $s=0$. Alternatively, one can obtain
the same solution by first smearing a NS5 brane on a transverse $T^3$
and then T-dualize. The solutions clearly make no sense at infinity,
much like a D8 brane, and should be regarded as local approximations. 

For general $N,M$, the $s=0$ degeneration acts much like the
domain-wall D8 brane: it separates two regions with mapping tori of
different monodromy.

It is easy to construct the analogous solution with parabolic
monodromies in $\rho$. The geometric monodromy $\rho \rightarrow \rho
+ N$ gives indeed the NS5 brane smeared on a $T^3$:
\begin{align}
ds^2 &= \eta_{\mu \nu} dx^{\mu \nu} + h(s) \left[ ds^2+d\theta^2+
  dx^2+dy^2\right] \, ,\\
B & = N \theta \Theta(s) dx \wedge dy \, , \quad e^{2\Phi} = h(s) \, ,
\end{align}
where $h(s) = C+sN\Theta(s)$ and we made a gauge choice for the
$B$-field. A T-fold solution can be obtained starting from the
solution~\eqref{rhofibrations1}. Since the monodromy is now a
perturbative shift in $\rho^{-1}$, we take the following ansatz: $\rho_0(s) = h(s)^{-1}$. We then
find the following solution:
\begin{align}
ds^2 &= \eta_{\mu \nu} + h(s)\left[ ds^2 + d\theta^2\right] +
  \frac{h(s)}{1 + N^2 h(s)^2}\left[dx^2+dy^2\right] \, ,\\
B & = \frac{N \theta}{h(s)^2+ N^2\theta^2} dx\wedge dy \, ,\quad
    e^{2\Phi} = \frac{h(s)}{1 + N^2 h(s)^2} \, .
\end{align}
This solution coincides with the T-dual along $y$ of the smeared KK monopole~\eqref{Nilwall}.
Again one could glue T-folds with different monodromies on the two
sides of the domain wall. For this reason such defects could be called
\emph{T-walls}.

In principle it should be possible to generalize this construction to
obtain T-walls with arbitrary T-folds (see~\cite{Schulgin:2008fv} for
an example with hyperbolic monodromy). However, embedding such local
solutions into a global consistent background is not simple, and unless one can
understand the degenerations as T-dual of known ones, it is not clear
how to discern good and bad singularities. In the next section we will
study codimension-2 defects, where a systematic construction of local
supergravity solutions is possible for arbitrary T-duality
monodromies. Moreover, in some cases the non-geometric fibrations can
be understood from string dualities, and this makes possible to
eventually put the existence of such backgrounds on more firm ground.

\section{Torus fibrations and T-fects}\label{sec:exoticbranes}

\begin{figure}
\begin{center}
\includegraphics[scale=0.45]{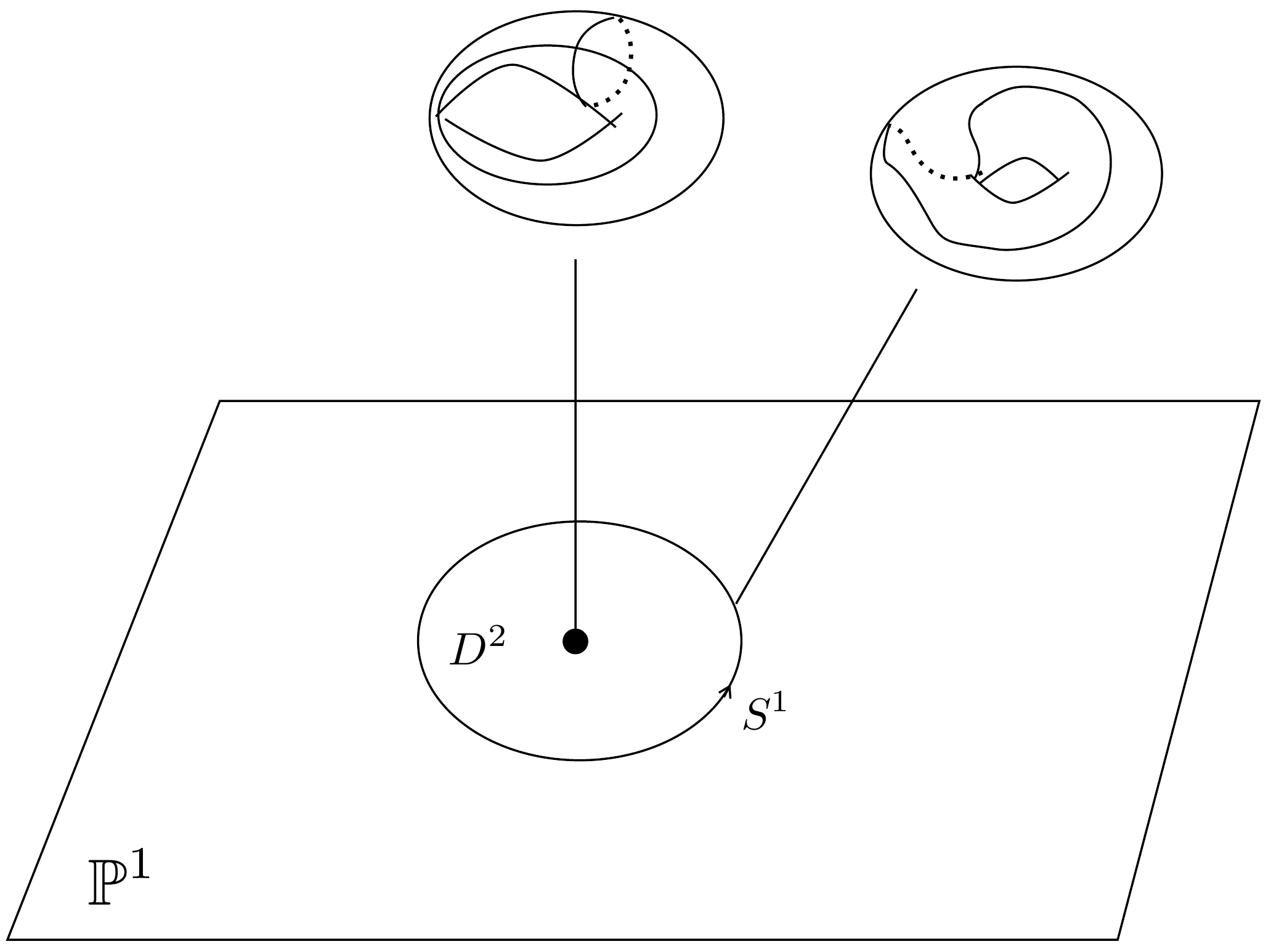}
\caption{Torus fibration over $\mathbb{P}^1$. The mapping torus at the
boundary of a small disk $D^2$ that encircle a degeneration point determines the
local geometry.}\label{Fig:torusfibration2d}
\end{center}
\end{figure}

In this section we generalize the previous discussion to the case of a
fibration of the T-duality group $O(2,2,\mathbb{Z})$ over
 a two dimensional base $\mathcal{B}$. This configuration
is well known from stringy cosmic strings~\cite{Greene:1989ya} and was
discussed in the context of non-geometric backgrounds
in~\cite{Hellerman:2002ax}. A complementary point of view, recently
described in~\cite{McOrist:2010jw,Malmendier:2014uka,Gu:2014ova}, is
to describe the fibrations of the complex structure $\tau$ and the
complexified K\"ahler form
$\rho$ in heterotic string theory as elliptic fibrations, and map the
corresponding line bundles 
to an F-theory compactification via heterotic/F-theory duality. A
surprising result of this analysis is that even spaces with non-geometric $\rho$
monodromies are mapped to geometric compactifications of F-theory.
In such configurations $\tau$ and $\rho$ are meromorphic functions on
the base. There is a set of points, that we denote $\Delta=(x_0, \dots,
x_n)$, at which the fibration degenerates and a defect
sits (see Figure~\ref{Fig:torusfibration2d}). These points are branch points for $\tau$ and $\rho$ and there is
a nontrivial monodromy around them.
Possible degenerations of the elliptic fibrations are described
by the Kodaira list as it is familiar in F-theory. Since such defects
have T-duality monodromy around them, we will call them generically
\emph{T-fects}.

Here we take a slightly different point of view, with the goal of
understanding the relation between the T-folds described in the
previous sections and the defects that arise in the case of fibrations
over $\mathbb{P}^1$. We will consider all possible monodromies in the
mapping class group of the compactification torus $T_{\tau}^2$ and the
auxiliary torus $T^2_{\rho}$ and classify the corresponding local solutions. In principle, this classification is more general
then the Kodaira classification, since not all of the
$SL(2,\mathbb{Z})$ conjugacy classes arise from degenerations of
elliptic curves. An illustration of this in F-theory is the case
of non-collapsible
singularities studied in~\cite{DeWolfe:1998pr,DeWolfe:1998eu}. We will clarify the relation between these two approaches
in a way that can be easily generalized to higher genus
fibrations. The latter situation, for genus $g=2$, is of interest in the case of
heterotic compactification on a $T^2$ with a Wilson
line~\cite{Malmendier:2014uka} and we outline an interesting geometric
perspective on the algebraic classification of genus 2
degenerations~\cite{Namikawa:1973yq}, as well as a generalization to
include non-collapsible defects in this case.

Another outcome of our analysis is the list of all possible geometries
arising in a neighborhood of a given $\tau$ and $\rho$
degeneration. In the case of a geometric fibration, some of such geometries
can be interpreted as a semi-flat
approximation~\cite{Strominger:1996it,Vegh:2008jn} of a class of known ten-dimensional
brane solutions, such as the NS5 brane and the KK monopole. In the case of non-geometric
fibrations, we recover the exotic brane solution recently discussed
in~\cite{deBoer:2010ud,deBoer:2012ma}. Our list
contains new exotic solutions corresponding to T-fects described
by different
$SL(2,\mathbb{Z})$ conjugacy classes. As in the case of torus
fibrations over a circle, when both $\tau$ and $\rho$ degenerations
collide, we found solutions which are not T-dual to a geometric
one. 

A crucial point that need to be emphasized is about the viability of such
local solutions. The solutions alone cannot be taken as automatic
evidence that the corresponding degeneration exists in string
theory. Put in another
way~\cite{Vafa:1996xn,Ooguri:1995wj,Hellerman:2002ax}, one has to supplement the semi-flat
solutions with the knowledge of the microscopic description
of a given degeneration, at which the semi-flat approximation breaks down. Once this is understood, the semi-flat torus
fibration is a powerful tool to understand physical properties of the
corresponding T-fects. One possible point of view is that heterotic/F-theory
duality can provide the evidence, even if quite indirect, that
non-geometric degenerations should be allowed and in the situations in which
enough supersymmetry is preserved, the supergravity solutions should
provide a qualitatively correct description. 

\paragraph*{Relation to previous works}

We note that some of the exotic brane solutions, which correspond to a subset of
the solutions for our T-fects, have been described in various
places, from different point of views and with a plehtora of different
names. We give a short guide to the literature in order to facilitate the
comparison with our discussion. As we already mentioned, a parabolic exotic brane
usually referred to as $5^2_2$-brane has been studied in details
in~\cite{deBoer:2010ud,deBoer:2012ma} and in~\cite{Hassler:2013wsa},
where it was named Q-brane. Early works include~\cite{Obers:1998fb,Blau:1997du,LozanoTellechea:2000mc}. In~\cite{deBoer:2014iba}, parabolic,
elliptic and hyperbolic branes were called N, K and A-branes
respectively (following an Iwasawa decomposition). Solutions for
elliptic and hyperbolic branes for the S-duality $SL(2,\mathbb{Z})$ group have been discussed
in~\cite{Bergshoeff:2006jj} for the 7-branes of F-theory,
where they are referred to as Q7-branes,
and in~\cite{Raeymaekers:2014bqa} as solutions of five dimensional
supergravity. Elliptic branes were used in~\cite{Dong:2010pm} as
an ingredient to uplift holography and they were referred to as
SC5-branes (stringy cosmic five-branes). An elliptic brane was studied in~\cite{Kikuchi:2012za}
and called $T_{34}$-brane. There have been also more general
investigations of codimension-2 solutions, sometimes referred to as
``defect branes'' (see for
example~\cite{Bergshoeff:2010xc,Bergshoeff:2011zk,Bergshoeff:2011se}). Here
we will simply refer to all the codimension-2 defects with T-duality monodromies
as T-fects, emphasizing the relation with the T-folds
described in section~\ref{sec:Tfolds}.

%%%%%%%%%%%%%%%%%%%%%
%%%%%%%%%%%%%%%%%%%%%% 

\subsection{Degenerations and monodromy factorization}\label{subsec:monodromy}

Let us briefly describe torus fibrations from the point of view of
the mapping class group. The following discussion is in part inspired
by the theory of Lefschetz fibrations~\cite{auroux1,2004math5300C}, and we stress that it can
be applied to fibrations of surfaces of arbitrary genus.
We consider a torus fibration on the punctured sphere
$\mathbb{P}^1\setminus \Delta$, with $\Delta = (x_1, \dots, x_n)$. Let
us suppose that at a point $x_i$ the fiber degeneration is such that a
cycle $p [u] + q [v]$, where $u$ and $v$ is the basis used in section~\ref{Sec:torusbundles}, shrinks to zero. The monodromy of the torus
fiber around $x_i$ is a positive Dehn twist $T_{i}$ along the vanishing
cycle and the corresponding action on the homology is given by the Picard-Lefschetz
formula (we refer to appendix~\ref{app:mcg} for details).

\begin{figure}[h]
\begin{center}
\includegraphics[scale=0.45]{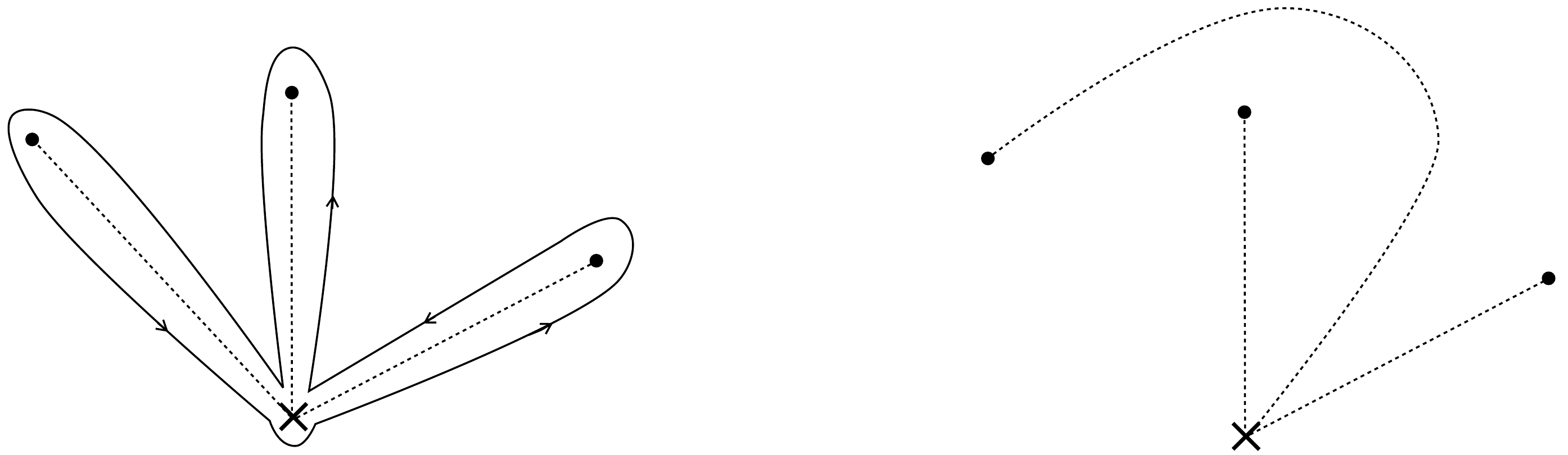}
\caption{Left: a path that encircle degeneration points (black points)
  at which a cycle shrinks. A reference smooth point (the cross) is
  chosen to define the local monodromies. Right: switching the arcs
  that connect degeneration points and the
  reference point induce an action by Hurwitz moves on the
  corresponding monodromy factorization.}
\label{Fig:monodromy}
\end{center}
\end{figure}

Consider now a disk that encircles some of the degeneration points. If
the disk contains the whole $\Delta$, the monodromy around the disk
should be homotopic to the identity in order to have a globally
defined fibration; otherwise we will have a monodromy $M$, that we can
factorize as the product of Dehn twists around each degeneration
point: $M = T_1\dots T_n$, as shown in Figure~\ref{Fig:monodromy}. 
This factorization has some gauge
redundancy. Locally, we make a choice for defining the monodromy
around each points, for example by choosing an arc that connects $x_i$ to
a given smooth reference point $x_0$. The choice is arbitrary, and this results in a equivalence class of
monodromy factorizations obtained from the list of vanishing cycles by
the action of a braid group $B_n$:
\begin{equation}
T_1 \dots T_i T_{i+1} \dots T_ n \sim T_1\dots T_i T_{i+1}
T_i^{-1} T_i \dots T_n \, , \quad i < n \, .
\end{equation}
This action is usually called an \emph{Hurwitz} move and it clearly
preserves the global monodromy. 
 We illustrate this in more details in
appendix~\ref{app:braid}, where we use it to derive the relation between our
Dehn twist factorization for the $SL(2,\mathbb{Z})$
monodromies~\eqref{ellipticsl2z},~\eqref{ellipticsl2zinv} and the ABC
factorization commonly used in F-theory (in this context the Hurwitz moves arise by moving branch cuts across branes, as in~\cite{Gaberdiel:1998mv}). There is an additional freedom in
the choice of the smooth reference fiber over $x_0$, which
is just a global conjugation on $M$. We arrive at the conclusion that the
torus fibration with a fixed number of elementary degenerations is defined by the factorization of the global
monodromy up to Hurwitz moves and global conjugation.

In general, the degeneration of a torus fiber can be worse then a
simple shrinking cycle. The classification of such degenerations
are well known from the Kodaira classification of singular fibers for
elliptic fibrations that we briefly review in
appendix~\ref{app:braid}. However, the generalization to fibrations of higher genus is much more involved,
see~\cite{Namikawa:1973yq,Malmendier:2014uka,Gu:2014ova} for a
discussion of the genus 2 case. 

Given the
geometric group approach we described above, it is natural to ask whether
one can give a more geometric description of such classification
from the point of view of fiber diffeomorphisms. Note that
for the torus fibration a classification of periodic diffeomorphisms,
i.e. elliptic $SL(2,\mathbb{Z})$ conjugacy classes, is in
correspondence with the Kodaira list with the exception of $I,
I^{\ast}$ fibers, which instead corresponds to the parabolic class,
but no hyperbolic (Anosov) monodromy arises from
degenerating elliptic fibers. It turns out that such a geometric
approach exists as a result of an elaborated
theorem~\cite{MatMont}. For genus $g\geq 2$, there is a bijection
between degenerations of a family of Riemann surfaces (up to an
appropriately defined topological equivalence relation)
and a subset of conjugacy classes of the mapping class group of the
fiber, given by pseudo-periodic elements. For the torus, the map
is surjective.\footnote{The kernel consists of multiple fibers, namely
the Kodaira type $_mI_b$, $m\ge 2$.} We refer to appendix~\ref{app:mcg} for a definition of
pseudo-periodic mapping classes, but this subset basically contains parabolic and
elliptic elements. We anticipate that this geometric
point of view will be useful to generalize our presentation to the
genus 2 case, relevant for non-geometric backgrounds of the heterotic
theory with a Wilson line. 

Note that for lower genus fibrations, known facts about mapping
class groups can be also helpful to determine possible global models. For
example, one can show that for torus fibrations it is not possible to
factorize the identity with less then 12 Dehn twists. This corresponds
to the defining relation~\eqref{mcgtorus} for the
presentation of the
mapping class group $M(T^2)$. We can think of a word in the $U,V$
alphabet that factorizes the identity as defining a global fibration in
which the total charge is canceled. Some of the elementary constituents
can then be collapsed into one of the Kodaira degenerations if their total
monodromy is elliptic or parabolic. However, one can still imagine
situations in which a group of non-collapsed branes gives rise to a hyperbolic
monodromy.

We now turn to a classification of all possible local geometries
that arise in our model, both for the geometric $T^2_{\tau}$ and the
auxiliary $T^2_{\rho}$ fibrations. The complex and K\"ahler moduli
$\tau$ and $\rho$ should really be thought as functions on an
appropriate multi-sheeted Riemann surface, where the transition
between sheets is determined by a given $SL(2,\mathbb{Z})$ element,
that we identify with an element in $M(T^2)$.\footnote{Note that the solutions
will be actually classified by $PSL(2,\mathbb{Z})$ elements, but our
discussion regarding mapping class groups is general.} 
We construct the local solutions by
taking the mapping torus $\mathcal{N}_{\phi}$~\eqref{defmt} for a
given $\phi \in M(T^2)$ in each conjugacy class, and we then solve the Cauchy-Riemann
equations to obtain a meromorphic function with the desired
monodromy. This determines a semi-flat approximation of the
corresponding T-fect. For parabolic and elliptic conjugacy class, we
will recover in this way the known local solutions that already appeared in the work
of Kodaira~\cite{kodaira123}, see Table~\ref{tab:localsolution}. These can be obtained by inverting
Klein's $j$ modular function that specify a given global elliptic
fibration. Inserted in the appropriate semi-flat
ansatz, we obtain the corresponding geometric and non-geometric local
solutions. Note however that interesting solutions can arise from
different monodromies in the same conjugacy class, so we also need to
determine the general form of the meromorphic functions. Solutions for
hyperbolic monodromies are substantially more complicated and cannot
be found by other simple methods.

\begin{table}[t]\begin{center}
\begin{tabular}{|c|c|c|c|}
\hline
Class &      Type                       &  Monodromy   &    Local model    \\
\hline
  Parabolic & $I_n$ &    $V^n = \begin{pmatrix} 1 & n\\0&
    1   \end{pmatrix}  $ &    $\displaystyle \frac{n}{2 \pi i}\log z
                           \, , \quad n> 0$
  \\[20pt]
   Elliptic order 6    &     $II$  &
                                                    $UV=\begin{pmatrix}
                                                      1 & 1\\ -1&
                                                      0   \end{pmatrix}
                                                                  $
                                        &$\displaystyle \frac{\eta-\eta^2z^{1/3}}{1-z^{1/3}} \, ,
                                          \quad \eta = e^{2\pi i/3} $                 \\[20pt]
     Elliptic order 4    &     $III$  &
                                                    $UVU = \begin{pmatrix}
                                                      0 & 1\\ -1&
                                                      0   \end{pmatrix}
                                                                  $
                                        &  $\displaystyle \frac{i + i
                                          \sqrt{z}}{1-\sqrt{z}} $
  \\[20pt]
 Elliptic order 3   &    $IV$   &
                                                    $UVUV = \begin{pmatrix}
                                                      0 & 1\\ -1&
                                                      -1  \end{pmatrix}
                                                                  $
                                        &  $\displaystyle
                                          \frac{\eta-\eta^2z^{2/3}}{1-z^{2/3}}$                 \\

\hline 
\end{tabular}\caption{Local solutions around degenerations of Kodaira type $I_n$, $II$, $III$, $IV$.}\label{tab:localsolution}\end{center}\end{table}

\subsection{Geometric $\tau$-fects}\label{sec:taubranes}

We now discuss local solutions in neighborhoods of degenerations for the
$\tau$ fibration. We consider the complexified K\"ahler parameter
fixed to $\rho=i$, while $\tau(z)$ will now be a function on a two
dimensional base $\mathcal{B}$ with complex coordinate $z = r e^{i
  \theta}$. We assume a semi-flat metric ansatz
$\mathbb{R}^{1,5}\times \mathcal{B}\times_{\varphi} T^2$, where $T^2$ is fibered
over $\mathcal{B}$ and we preserves the $U(1)\times U(1)$
isometries of the fiber torus:
\begin{align}
ds^2_{10} &= \eta_{\mu \nu} dx^{\mu}dx^{\nu}+ e^{2\varphi_1} \tau_2 dzd\bar z
+G_{ab}(z) dx^adx^b \, , \nonumber\\
\qquad e^{2\Phi} &= \text{const} \, ,\qquad B=0 \, \qquad \mu, \nu =
                  0,\dots 5 \, , \quad a,b=8,9\, ,\label{geometrictaufibrationansatz}
\end{align}
where 
\begin{equation}\label{Habtaubrane}
G(z) = \frac{1}{\tau_2} \begin{pmatrix}  |\tau |^2 & \tau_1 \\ \tau_1 &
 1\end{pmatrix} \, .
\end{equation}
Here $\varphi= \varphi_1 + i \varphi_2$ and $\tau= \tau_1 +
i\tau_2$ are meromorphic functions on $\mathcal{B}$. We will mainly
focus on a local description with $\mathcal{B}$ a disk. Such local
solutions can then be glued to obtain a global fibration on $\mathbb{P}^1$.
As we will
discuss later for the general case of arbitrary $\rho$, it is easy to show that
the second-order equations of motion are implied by holomorphicity of
$\tau$ and $\varphi$. One can also show that such condition is enough to
preserve half of the supersymmetries~\cite{Hellerman:2002ax,Bergshoeff:2006jj,deBoer:2012ma}.
We assume a branch point at the origin where the description will
break down and we have to resort to a string description of the
degeneration. Given a small disk $D^2: 0<\lvert z \rvert < R_0$ that
contains the origin, we consider the smooth torus fibration on the
boundary $S^1=\partial D^2$. These are the fibrations that we classified in
section~\ref{sec:Tfolds}, namely mapping tori for a given
mapping class group element, that we identify with the monodromy of
the multi-valued function $\tau(z)$, up to the kernel $\{\pm\mathbb{1}
\}$. Given such monodromy, a local solution for $\tau(z)$ can be found
by solving the Cauchy-Riemann equations by separation of
variables. Namely, we start with the solutions for mapping tori
derived in section~\ref{subsec:geometricmonodromies} and we promote the free modulus
$\tau_0$ of such solutions to a function of $r=\lvert z \rvert$. The
equations then reduce to an ordinary differential equation for
$\tau_0(r)$.
We will see that a solution always exists for
all the three different classes of the fiber diffeomorphism on
$\partial D^2$. Thus, given a monodromy
\begin{equation}
M_{\tau} = \begin{pmatrix} a & b \\ c& d \end{pmatrix} \, , \quad ad-bc=1 \, ,
\end{equation}
we obtain a function $\tau(z)$ such that the analytic continuation
along an arc that encircles the origin is 
\begin{equation}
\tau(z) \rightarrow \frac{a \tau(z) + b}{c\tau(z) + d} \, .
\end{equation}

We still have to solve for the warping factor $\varphi(z)$. By various
arguments~\cite{Hellerman:2002ax,Bergshoeff:2006jj,deBoer:2012ma}, one
can show that in order for the metric on $\mathcal{B}$ to be
single-valued the analytic continuation for $\varphi$ should be
\begin{equation}
e^{\varphi(z)} \rightarrow e^{\varphi(z)} (c\,\tau(z) + d) \, .
\end{equation}
Again, once $\tau(z)$ is known, one can solve the equation for
$\varphi$ by separation of variables. We note that in many cases one
can easily guess the solution with the given monodromy, but there are
few exceptions which are more complicated.

%----------------------------------- HERE ------------------------------------

\subsubsection*{Parabolic $\tau$-fects and KK monopoles}

\begin{figure}[h]
\begin{center}
\vspace{0.3cm}
\includegraphics[scale=0.5]{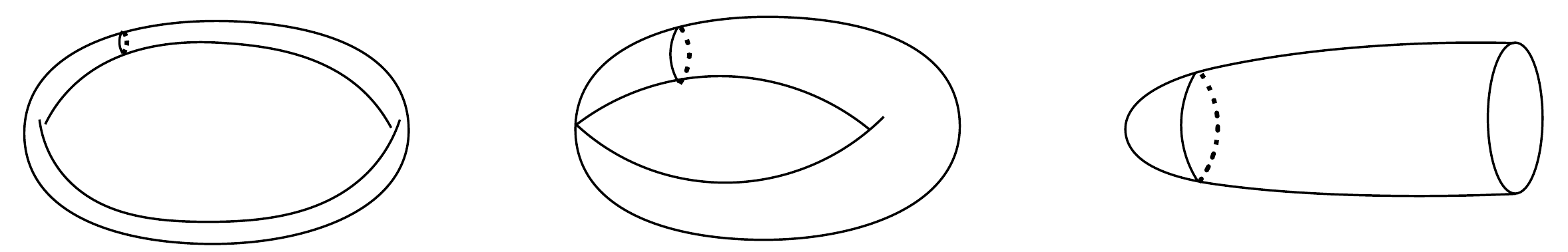}
\caption{From left to right: the semi-flat degeneration of a KK
  monopole, the
 localized solution and the decompactified half-cigar of the Taub-NUT space.}
\label{Fig:kkm}
\end{center}
\end{figure}

We consider the situation in which the tours fiber degenerates at $z_0=0$
by shrinking a cycle $p [v]+q [u]$. If $q=0$ and $p=1$ we have the
monodromy~\eqref{tauV} for $N=1$, $M_{\tau} = V$. The corresponding action
on $\tau$ is a shift:
\begin{equation} 
\tau \rightarrow \tau +1 \, .
\end{equation}
The corresponding solution is of course well known~\cite{Greene:1989ya}:
\begin{equation}\label{smearedKKM}
\tau(z) =  \frac{i}{2\pi}\log\left(\frac{\mu}{z} \right) \, , \qquad e^{\varphi} = 1 \, ,
\end{equation}
where $\mu$ is an integration constant. The corresponding metric in
polar coordinates on $\mathcal{B}$ is thus:
\begin{align}
ds^2 &= \eta_{\mu \nu}dx^{\mu}dx^{\nu}  + \frac{1}{2\pi}
\log\left(\frac{\mu}{r}\right) \left[d\theta^2 + r^2 dr^2 +
  (dx^8)^2\right] \nonumber \\
&\quad +\frac{2\pi}{\log\left(\frac{\mu}{r}\right)}\left(
  dx^9+\frac{\theta}{2\pi}dx^8 \right)^2 \, . \label{KKMsemiflat}
\end{align}
It is easy to check that this is a semi-flat
approximation of a KK monopole with compact circle $x^9\approx
u$. Indeed, this is precisely the analysis done
in~\cite{Ooguri:1996me}.\footnote{See also~\cite{Grimm:2012rg} for an extensive
discussion, although in a different contex.} To
show this, let us start from the Taub-NUT metric:
\begin{equation}\label{Taub-NUTmetric}
ds^2_{KKM} = \eta_{\mu \nu}dx^{\mu}dx^{\nu} + V(\vec{x}) \,d\vec x^2 + \frac{1}{V(\vec{x})}
\left (dx^9 + A\right)^2 \, ,\qquad V(\vec{x}) = 1 + \frac{R_9}{2 |\vec{x}|} \, ,
\end{equation}
where $A= \vec{A} \cdot d\vec{x}$, $\vec{x} = (z, \bar{z}, x^8)$, and $dA =*_3dV$. We will set
the radius $R_9=1$. We now
compactify on $x^8$. This corresponds to have an infinite array of
sources on the covering space, resulting in the potential:
\begin{equation}
V = \frac{1}{2}\sum_{n=-\infty}^{\infty} \left[ \frac{1}{\sqrt{(x_8 -2\pi n)^2+ r^2}}
  - \frac{1}{\vert 2\pi n \rvert}\right] \, ,
\end{equation}
where we added a regulator.
This sum can be performed exactly by Poisson resummation~\cite{Ooguri:1996me}, resulting in:
\begin{equation}
V = \frac{1}{2\pi}\left[ \log(\mu/r) +\sum_{n\neq 0} e^{ i n
    x^8}K_0\left( \lvert n \rvert r\right)\right]\, ,
\end{equation}
where $K_0$ is the modified Bessel function. Away from the origin, the leading order is a very good
approximation (up to the exponentially suppressed terms), and the metric
with this leading order approximation to $V$ reduces precisely to the
semi-flat metric~\eqref{KKMsemiflat}. The half-cigar of the Taub-NUT
space became a pinched torus obtained from the shrinking of the $u$
cycle (see Figure~\ref{Fig:kkm}).

This analysis can be easily
extended to the monodromy $V^{N}$, which corresponds to having a stack
of $N$ KK monopoles at the degeneration point.
We now consider the generic case corresponding to the $(p,q)$
parabolic conjugacy class, with $q\neq 0$:
\begin{equation}
M_{\tau} = \begin{pmatrix} 1+ p q &  p^2\\- q^2  & 1- p q \end{pmatrix}
\, .
\end{equation}
We also list the embedding into the $O(2,2,\mathbb{Z})$ group (see
appendix~\ref{app:T-duality} for a short review):
\begin{equation}
\Omega_{O(2,2)} =\begin{pmatrix}
  1+p q & p^2 & 0& 0\\-q^2&1-pq&0&0\\0&0&1-pq&q^2\\0&0&-p^2&1+pq\end{pmatrix} \, .
\end{equation}
 To find the appropriate meromorphic function we consider the following ansatz for $\tau(r,\theta)$:
\begin{equation}\label{taubraneansatz}
\tau(r,\theta)= \exp(\mathbf{m}_{\tau} \theta/2\pi)[\tau_0(r)] \, ,
\end{equation}
where $\mathbf{m}_{\tau}$ is the element of the $sl(2,\mathbb{Z})$ Lie
algebra defined in~\eqref{nilpqtau}.
By plugging this ansatz into the Cauchy-Riemann equations
\begin{equation}
\frac{\partial \tau(r,\theta)}{\partial r} - \frac{1}{i r}\frac{\partial
  \tau(r,\theta)}{\partial \theta}=0 \, ,
\end{equation}
we see that they reduce to a single ordinary differential equation for
$\tau_0(r)$:
\begin{equation}
2\pi i r \frac{d\tau_0(r)}{dr} = \big[p + q \tau_0(r)\big]^2 \, .
\end{equation}
 We then find a solution for $\tau(z)$:
\begin{equation}\label{taupqbrane}
\tau(z) = \frac{2\pi i}{q^{2}\log\left(\frac{\mu}{z}\right)}  -\frac{p}{q} \, .
\end{equation}
By making an analogous ansatz for $e^{\varphi(z)}$ we obtain:
\begin{equation}\label{phipqbrane}
e^{\varphi(z)} = i\kappa \log\left(\frac{\mu}{z}\right) \, ,
\end{equation}
where $\kappa$ is an arbitrary integration constant. Since the
monodromy corresponds now to a Dehn twist along the vanishing cycle $p
[u] + q [v]$, we obtain a semi-flat approximation of a KK
monopole with special circle an oblique direction in the $(x^8, \, x^9)$
plane whose slope is determined by the ratio $q/p$, as in~\eqref{rotatednil}.  The semi-flat approximation arises after compactifying the
orthogonal direction and taking the leading approximation for V as
discussed before.

\subsubsection*{Elliptic $\tau$-fects}

We now consider the case in which the monodromy around the
degeneration is of finite order, corresponding to a periodic
diffeomorphism of the torus. Different degenerations are classified by
the conjugacy classes of elliptic $SL(2,\mathbb{Z})$ elements and are
listed in~\eqref{ellipticsl2z},~\eqref{ellipticsl2zinv}. We consider
for illustration the monodromy of order 4: $M_{\tau}= UVU$ ($\tau
\rightarrow -1/\tau$), that corresponds to the following
$O(2,2,\mathbb{Z})$ element:
\begin{equation}
\Omega_{O(2,2)} =\begin{pmatrix}
  0 &1 &0&0\\-1&0&0&0\\0&0&0&1\\0&0&-1&0\end{pmatrix} \, .
\end{equation}
 The solution for $\tau(\theta)$ for a fibration
on a circle is~\eqref{ellipticorder4}. The Cauchy-Riemann
equations coming from the ansatz~\eqref{taubraneansatz} and the
analogous for $e^{\varphi}$ reduce to:
\begin{equation}
4 r\frac{d\tau_0}{dr} + i (\tau_0^2+1)=0 \, ,\qquad 4 r\frac{d
  \varphi_0}{dr}= i \tau_0 \phi_0\, .
\end{equation}
From this we obtain the following solutions:
\begin{equation}\label{tauelliptic4}
\tau(z) =  \tan \left[C - \frac{i}{4} \log (z)\right] \, , \qquad
e^{\varphi(z)} = \kappa \cos \left[C -\frac{i}{4} \log (z)\right] \, ,
\end{equation}
where $C$ and $\kappa$ are integration constants. Note that the
solution for $\tau$ can be rewritten as
 \begin{equation}
 \tau(z) = \frac{i - ie^{2 i C} \sqrt{z}}{1- ie^{2 i C} \sqrt{z}} = i - 2i e^{2 i C}\sqrt{\frac{z}{\mu}}
 + 2 i  e^{4 i C} z+\mathcal{O}(z^{3/2})  \, ,
 \end{equation}
% \begin{align}
% \tau(z) &= \frac{i - ie^{2 i C} \sqrt{z}}{1- ie^{2 i C} \sqrt{z}} = i - 2i e^{2 i C}\sqrt{\frac{z}{\mu}}
% + 2 i  e^{4 i C} z+\mathcal{O}(z^{3/2})  \, ,\\
% e^{\varphi(z)} &=   e^{-i C}\kappa z^{-1/4} +\frac12 e^{i C}\kappa z^{1/4} \, .
% \end{align}
For $C = \pi/2$ the solution for $\tau$ reproduces the local model
shown in Table~\ref{tab:localsolution}.
At the degeneration point the complex modulus is given, as expected,
by the fixed point of the elliptic monodromy, $\tau_0 = i$. The metric
for the total space can then be easily found from~\eqref{geometrictaufibrationansatz}.
%The corresponding metric is then:
% \begin{align}
% ds^2  &= \eta_{\mu \nu} dx^{\mu}dx^{\nu} +
% \frac{\kappa^2}{4}\frac{r-\mu}{\sqrt{\mu r}} (dr^2+ r^2d\theta^2)  \\ & \qquad -
%         \frac{r+\mu}{r-\mu}\left[(dx^8)^2 +(dx^9)^2\right]+ \frac{2\sqrt{\mu r}}{r-\mu}M_{a b}(\theta)dx^a
% dx^b \, ,
% \end{align}
% where
% \begin{equation}
% M(\theta) = \begin{pmatrix} \cos(\theta/2)&
%   -\sin(\theta/2)\\-\sin(\theta/2)& - \cos(\theta/2)\end{pmatrix} \, .
% \end{equation}
 
As for the parabolic case, we could also consider solutions associated
to the general conjugacy class of the particular elliptic
monodromy. This is specified by three parameters $(p,q,w)$:
\begin{equation}
M_{p,q,w} = L^{-1} UVU L=\begin{pmatrix} \displaystyle pq + \frac{w (1+qw)}{p} &\displaystyle p^2+w^2\\
 \displaystyle -q^2 - \frac{ (1+qw)^2}{p^2} & \displaystyle-\frac{w+q(p^2+w^2)}{p}\end{pmatrix} \, ,
\end{equation}
for $L \in SL(2,\mathbb{Z})$.
By constructing the fibration on a $S^1$ from~\eqref{deftautheta}, we
find a general solution of the Cauchy-Riemann equations for $\tau(z)$ and
$e^{\varphi(z)}$:
\begin{align}
\tau(z) &= w \left[\frac{1}{p+ w \tan\left[
         p^2\,  C- \frac{i}{4}\log(z)\right]}-\frac{1+w q}{p} \right]^{-1} \,
          ,\\[10pt] \label{solutionellipticconjugacyclass}
e^{\varphi(z)} &= \kappa \,p\, q
                 \cos\left[p^2 \, C -
                 \frac{i}{4}\log (z)\right]+\kappa \, 
                 i(1+p\, w)
                 \sin\left[p^2 \, C -\frac{i}{4}\log(z)\right]
                 \, .
\end{align}
A similar analysis can be done for all the remaining elliptic
conjugacy classes. For example, the solutions corresponding to the
monodromy $UV$ and $(UV)^2$ can be found by starting from the
solution~\eqref{ellipticorder6}. The results are:
\begin{align}\label{tauelliptic6}
UV & : \quad \tau(z) = \frac{\eta + \eta^2 e^{i C}z^{1/3}}{1+ e^{i C}z^{1/3}}\, , \qquad
e^{\varphi(z)} = \kappa \left( e^{-iC/2} z^{-1/6} + e^{i C/2}z^{1/6}\right)\, , \\
UVUV&:\quad \tau(z) =  \frac{\eta + \eta^2 e^{i C}z^{2/3}}{1+ e^{i C}z^{2/3}}\, , \qquad
e^{\varphi(z)} = \kappa \left( e^{-iC/2} z^{-1/3} + e^{i C/2}z^{1/3}\right)\, , 
\end{align}
from which the metric can be easily reconstructed. For $C= \pi$ these
solutions are listed in Table~\ref{tab:localsolution}.

We note that in a given
global model for an elliptic fibration specified by a Weierstrass form
\begin{equation}
y^2 = x^3 + f(z) x +g(z) \, ,
\end{equation} 
the function $\tau(z)$ is specified implicitly by Klein's $j$-invariant:
\begin{equation}
j(\tau) = \frac{(12 f)^3}{4 f^3+27 g^2} \, .
\end{equation}
In general, there will be degeneration points located at the
discriminant locus $\Delta: 4 f^3 + 27 g^2 = 0$. The solution for
$\tau$ around poles of the $j$-function are the parabolic solutions
(namely KK monopoles in our case), while around removable
singularities for $j(\tau)$ the solutions are approximated by the elliptic solutions
described above. In the next paragraph, we will study hyperbolic
solutions that cannot arise in this simple situation.

\subsubsection*{Hyperbolic $\tau$-fects}
%\label{subsubsec:hyperbolictau}

The last example is that of a hyperbolic monodromy, corresponding to an
Anosov map of the fiber torus. A simple example with monodromy in
$SL(2,\mathbb{R})$ is given by the
matrix~\eqref{hyperbolicsl2r}. In this case we trivially obtain:
\begin{equation}
\tau(z) = z^{-i w/\pi} \, ,\qquad  e^{\varphi(z)} =
z^{i w/2\pi}\, .
\end{equation}
Note that the imaginary part of $\tau$ is a periodic function of
$\log(r)$, so it has a highly oscillatory behavior near the origin. 
The solutions for conjugacy classes of hyperbolic $SL(2,\mathbb{Z})$
monodromies are more involved but share the same problem. As an example, for the 
monodromy~\eqref{hyperbolictaumonodromy}, which corresponds to the
$O(2,2,\mathbb{Z})$ element
\begin{equation}
\Omega_{O(2,2)} =\begin{pmatrix}
 N& 1 & 0& 0\\-1&0&0&0\\0&0&0&1\\0&0&-1&N\end{pmatrix} \, ,
\end{equation}
we obtain the following differential equations:
\begin{align}
\pi r (\lambda^2-1)\frac{d\tau_0(r)}{d r}+ i
  \log(\lambda)\left[\lambda
  +\tau_0(r)\right]\left[\lambda\tau_0(r)+1\right] & =0 \, ,\\
2\pi r (\lambda^2-1)\frac{d\phi_0(r)}{d r} - i
  \log(\lambda)\left[1+\lambda^2+2\lambda \tau_0(r)\right]\phi_0(r)
                                                   &=0 \, .
\end{align}
 The solutions can be written explicitly:
\begin{align}\label{tauhyperbolic}
\tau(z) & =\frac{\sigma^{\lambda^2}(\lambda^2-1)}{\lambda(\sigma e^{i
          \tilde z}-
          \sigma^{\lambda^2})}-\frac{1}{\lambda}\,
  , \\[10pt]
e^{\phi(z)} &= \kappa \sqrt{\lambda}  e^{\frac{i}{2}(\mu-\tilde
              z\lambda^2)}\left[\sigma^{\lambda^2}-\sigma e^{i \tilde
              z}\right] \, ,
\end{align}
where $\sigma,$ $\mu$, $\kappa$ are integration constants and we defined
\begin{equation}\label{hyperbolicztilde}
\tilde z = \mu(1-\lambda^2)+\frac{1}{\pi}\log\lambda\log\left[\pi
  z(\lambda^2-1)\right] \, .
\end{equation}
It is not obvious how to make sense of such solutions near the
origin, since there are infinite intervals at which $\tau_2<0$. This
fits well with
the fact that Anosov diffeomorphisms cannot be obtained as monodromies
of a
degenerating family of curves, and thus cannot be associated with a
degeneration point. The F-theory analogous of such hyperbolic branes
appeared in~\cite{DeWolfe:1998eu,DeWolfe:1998pr}. In the F-theory
context, there are separated 7-branes, and the associated massive
states gives rise to infinite dimensional algebras. We note that the previous
solution might approximates a non-collapsed group of branes, that can be
identified by the Dehn twist decomposition given in
section~\ref{Sec:torusbundles}. In this way there is a scale
representing the size of the brane distribution which serves as a
cutoff around the origin for our solution. This is reminiscent of an
enhan\c on mechanism~\cite{Johnson:1999qt}. To cure the problem at
infinity, one should put additional defects in order to cancel the
total charge, as for all the other co-dimension 2 defects. To show
that this is possible in the presence of a hyperbolic monodromy, let
us for example consider a model with the following $SL(2,\mathbb{Z})$ elements:
\begin{equation}\nonumber
M_H = U^4 V U = \begin{pmatrix} 0 & 1\\ -1 & -3 \end{pmatrix} \, ,
\quad M_1 = U V^2 = \begin{pmatrix} 1 & 2\\ -1 & -1 \end{pmatrix} \,
, \quad M_2 =  U^2 V = \begin{pmatrix} 1 & 1\\ -2 & -1 \end{pmatrix} \, ,
\end{equation}
where $M_{1}$ and $M_2$, are elliptic monodromies of order 4, both conjugate
to the Kodaira degeneration $III = UVU$.
We then check that they provide a factorization of the identity:
\begin{equation}
M_H M_1 M_2 = \mathbb{1} \, .
\end{equation}
Note that as expected, the total number of elementary branes is
12. One can add an additional factorization $(UV)^6 = \mathbb{1}$ to
obtain a model with 24 branes. It would be
interesting to understand better the nature of the hyperbolic
solutions in the present context.

\subsection{Exotic $\rho$-fects}

In the previous section we discussed a classification of local
geometries associated to monodromies filling the
$SL(2,\mathbb{Z})_{\tau}$ mapping class group of the compactification
torus. Global models of such $\tau$ fibrations give rise to
geometric six dimensional compactifications. We now consider the case of a $\rho$
fibration, that can be geometrized as a mapping class group of an
auxiliary torus. The $\rho$ fibrations, at constant
$\tau=i$, are the fiberwise mirror-symmetric of the geometric
solutions discussed above. The semi-flat metric ansatz in this case reads:
\begin{align}
ds^2_{10} &= \eta_{\mu \nu} dx^{\mu}dx^{\nu} + e^{2\varphi_1}\rho_2
dz d\bar z + \rho_2 (dx_8^2 + dx_9^2) \, , \label{ansatzrho}\\
B & = \rho_1 dx_8\wedge dx_9 \, ,\nonumber
\qquad e^{2\Phi} = \rho_2 \, .
\end{align}
Note that this is a concrete realization of the
fibration~\eqref{mappingtorusrhoansatz}. The fact that the base circle
now becomes contractible in the two dimensional base $\mathcal{B}$
provides the necessary gradient terms to solve the equations of
motion. This is similar to the T-walls discussed in
section~\ref{sec:T-walls} but we now have codimension-2 defects at
points where the $\rho$ fibration degenerates. Again, the second-order equations are implied by the
Cauchy-Riemann equations for $\varphi(z)$ and $\rho(z)$, as can be
shown from the supersymmetry analysis performed in details
in~\cite{Hellerman:2002ax,deBoer:2012ma,Becker:2009df}. In
appendix~\ref{app:eom} we independently check that the previous ansatz
solves the equations of motion. One should keep in mind that at the
degeneration points of the auxiliary $T^2_{\rho}$ fibration, namely
branch points for $\rho(z)$, this description breaks down and we have
to resort to a string description. Note that since the dilaton is
running, one should also check that it transforms in the correct way
under T-duality once encircling the defect. This is indeed the case
for the ansatz~\eqref{ansatzrho}.

\subsubsection*{Parabolic $\rho$-fects: NS5 and $5^2_2$ branes}

The meromorphic solutions derived in the previous section
automatically define, by just exchanging $\tau \rightarrow \rho$, a
local supersymmetric solution of the supergravity equations, of the
form~\eqref{ansatzrho}. The simplest situation is that of a parabolic
monodromy of type $V^N$. This corresponds to gluing the torus with a
gauge transformation for the B-field, so
the corresponding solution is geometric. We have
\begin{equation}
\rho(z) =  \frac{i}{2\pi}\log\left(\frac{\mu}{z} \right) \, , \qquad e^{\varphi}
= 1 \, ,
\end{equation}
and the background metric is then:
\begin{align}
ds^2 & = \eta_{\mu \nu}dx^{\mu}dx^{\nu}  + \frac{1}{2\pi}
\log\left(\frac{\mu}{r}\right) \left[d\theta^2 + r^2 dr^2 + (dx^8)^2 +
 (dx^9)^2\right] \, .  \label{NS5semiflat}\\
 B & = \frac{\theta}{2\pi}dx^8 \wedge dx^9 \, , \qquad e^{2\Phi} =\frac{1}{2\pi}
\log\left(\frac{\mu}{r}\right) \, .
\end{align}
 As it is well known, this solution can
be identified with the semi-flat approximation of a stack of NS5
branes~\cite{Ooguri:1995wj} (see also
~\cite{Vegh:2008jn,deBoer:2012ma}). To show this we can follow the
same steps as for the case of the KK monopole in the previous
section. The solution for a stack of NS5 branes
localized on $\mathbb{R}^2 \times T^2$ can be found by starting from
the NS5 harmonic function $h$ and compactify two directions, which is
equivalent to consider an array of sources on the $(x^8,x^9)$ plane. This gives
\begin{equation}
h(r) = \sum_{n,m} \frac{1}{(x^8-2\pi n)^2 + (x^9-2\pi m)^2+ r^2} + \text{regulator}\, .
\end{equation}
At distances large compared to the distance between the sources, the
result for the harmonic function is clearly the same as the smeared
KK monopole, and we obtain the metric~\eqref{NS5semiflat}. However,
note that the corrections to the semi-flat approximation involve the
breaking of both the $U(1)$ isometries of the torus
(see~\cite{Becker:2009df} for a field theory computation of such corrections) . It would be
interesting to understand if this localization is encoded in the
auxiliary $T^2_{\rho}$ fibration. Since the degeneration is the same
as for the KKM, it seems that information about the breaking of the
second $U(1)$ isometry is missing. We will come back shortly on this
point. First, we consider the solutions for the general 
conjugacy class of parabolic $\rho$ monodromies, namely the ones
associated to a $(p,q)$ monodromy
\begin{equation}
M_{\rho} = \begin{pmatrix} 1+ p q &  p^2\\- q^2  & 1- p q \end{pmatrix}
\, .
\end{equation}
We also show the corresponding $O(2,2,\mathbb{Z})$ element:
\begin{equation}
\Omega_{O(2,2)} =\begin{pmatrix}
  1+p q & 0 &0&p^2\\0&1+pq&-p^2&0\\0&q^2&1-pq&0\\-q^2&0&0&1-pq\end{pmatrix} \, .
\end{equation}
Note that for $p=0$ the monodromy is a $\beta$-transformation (see
Appendix~\ref{app:T-duality}), while for $q=0$ this is just a
B-transformation. For general $p,q$ both transformations are present
at the same time.
The NS5 brane corresponds to a $(1,0)$ brane, while the general
solution for $q\neq 0$ can be obtained
from~\eqref{taupqbrane},~\eqref{phipqbrane}. 
The $(0,1)$ solution, with monodromy $U^N$ ($1/\rho \rightarrow
1/\rho +1$) is
\begin{equation}
\rho(z) = \frac{2\pi i}{\log\left(\frac{\mu}{z}\right)} \, , \qquad
e^{\varphi} = i  \sigma \log{\left(\frac{\mu}{z}\right)} \, .
\end{equation}
We then get the following solution:
\begin{align}
ds^2_{10} &= \eta_{\mu \nu} dx^{\mu}dx^{\nu} +2\pi \sigma^2 h(r)
(dr^2 + r^2 d\theta^2) + \frac{2\pi h(r)}{h(r)^2 + \theta^2} \left[ (dx^8)^2 +
 (dx^9)^2\right]  \, , \label{Qbranesolution}\\
B & = -\frac{2\pi\theta}{h(r)^2 + \theta^2}dx^8\wedge dx^9  \, ,\nonumber
\qquad e^{2\Phi} = \frac{2\pi h(r)}{h(r)^2 + \theta^2}  \, ,
\end{align}
where
\begin{equation}
h(r) = \log\left(\frac{\mu}{r}\right) \, .
\end{equation}
We see that the monodromy acts non-trivially on the volume, so this
solution is non-geometric. This is the exotic brane solution recently
discussed in length in~\cite{deBoer:2012ma} (see
also~\cite{Hassler:2013wsa}) and usually called Q-brane or
$5^2_2$-brane in the notation of~\cite{Obers:1998fb}. We have
explicitly shown that at the boundary of a small neighborhood of such
exotic brane, the torus fibers on the boundary of the disk as in the
parabolic T-fold fibrations. Since the latter background con only be
obtained from an obstructed T-duality from a Nilmanifold, it is
interesting to obtain a solution with the same monodromy from a
different perspective.

Note that also the metric~\eqref{Qbranesolution} can be obtained by
applying Busher rules to the semi-flat KKM
metric~\eqref{smearedKKM}, in the same way the NS5 smeared on one
circle is T-dual to the KKM solution~\eqref{Taub-NUTmetric}.
In the latter case, correction to the Busher rules coming
from the breaking of the $U(1)$ smearing isometry can be computed
as worldsheet instantons
corrections~\cite{Gregory:1997te,Tong:2002rq}. It is clearly very
important to understand the breaking of the $U(1)$ isometry in the
case of the second T-duality that relate the KKM to the
metric~\eqref{Qbranesolution}. This is analogous to understand the
corrections to the semi-flat approximation for the auxiliary $\rho$ fibrations.

\subsubsection*{Elliptic $\rho$-fects} 

If we now consider elliptic monodromies of finite orders, we can again
obtain local solutions from the
functions~\eqref{tauelliptic4},~\eqref{solutionellipticconjugacyclass},~\eqref{tauelliptic6} and analogous
solutions for other conjugacy classes. For example, the
order 4 monodromy $M_{\rho} = UVU$ corresponds to the following background:
\begin{align}
ds^2 &= \eta_{\mu \nu} dx^{\mu}dx^{\nu} -\frac12 \sinh\left[\frac12 \log\left(\frac{r}{\mu}\right)\right]
(dr^2 + r^2 d\theta^2)\nonumber\\
& \quad - \frac{\sinh\left[\frac12 \log\left(\frac{r}{\mu}\right)\right]}{\cos\left(\frac{
  \theta}{2} +\sigma\right)+ \cosh\left[\frac12\log\left(\frac{r}{\mu}\right)\right]} \left[ (dx^8)^2 +
 (dx^9)^2\right] \, , \label{ellipticrhosolution}\\
B_2 & = -\frac{\sin(\theta/2+\sigma)}{\cos(\theta/2+\sigma)+\cosh \left[\frac12 \log\left(\frac{\mu}{r}\right)\right]}dx^8\wedge dx^9  \, ,\nonumber\\
 e^{2\Phi} &=  - \frac{\sinh\left[\frac12 \log\left(\frac{r}{\mu}\right)\right]}{\cos\left(\frac{
  \theta}{2}+\sigma\right)+ \cosh\left[\frac12\log\left(\frac{r}{\mu}\right)\right]}  \, .
\end{align}
Compared to the solution~\eqref{tauelliptic4} we redefined $C_1=
\sigma $, $C_2 = 1/4 \log(\mu)$.
Let us check explicitly that the solution has the desired
monodromy. The action of the elliptic transformation is $\rho
\rightarrow -1/\rho$. Recalling that $\rho=B+ i \sqrt{G}$, this corresponds to:
\begin{equation}\label{BVelliptictransf}
B(2\pi) = -\frac{B(0)}{B(0)^2+ G(0)} \, , \qquad \sqrt{G}(2\pi) =\frac{\sqrt{G}(0)}{B(0)^2+G(0)} \, ,\end{equation}
where we indicated in brackets the value of the angle $\theta$ at which
the fields are evaluated. It is not difficult to check that indeed the
B-field and the fiber torus metric in the
solution~\eqref{ellipticrhosolution} satisfy the
relations~\eqref{BVelliptictransf}. Since the metric and B-field are
mixed by the monodromy the solution is non-geometric, as in the case of
the $5^2_2$ parabolic brane. The corresponding $O(2,2)$ monodromy is:
\begin{equation}
\Omega_{O(2,2)} =\begin{pmatrix}
  0 &0 &0&1\\0&0&-1&0\\0&1&0&0\\-1&0&0&0\end{pmatrix} \, .
\end{equation}
Let us discuss the regime of validity of the
solution~\eqref{ellipticrhosolution}. As for the other codimension two
solutions, there is a scale at which the Ricci scalar blows up. This
scale is determined by the parameter $\mu$ and can be taken to be
large. In a global model this would be related to the scale at which
the local solution breaks down and is glued to the global
solution. Additionally, it would be very important to understand
corrections to the semi-flat approximations near the degeneration.

We conclude by discussing the relation between the elliptic defect and
the elliptic T-fold discussed in
section~\ref{subsec:nongeometricmonodromy}. The relation is in some sense similar to a
geometric transition in which brane sources are dissolved into
fluxes. 
In the T-fold picture, which is related to an asymmetric
$\mathbb{Z}_4$ orbifold construction~\cite{Condeescu:2013yma}, one can
argue for the presence of both $H$ and $Q$ fluxes (this can also be
inferred from a double field theory
approach~\cite{Hassler:2014sba}). This seems to be compatible with the
Dehn twist decomposition of the elliptic monodromy $M_{\rho} = UVU$, if we identify the
source of $H$-flux with NS5 branes and the source of Q-flux
with a $5^2_2$ branes. The charge of such branes should indeed be
identified with the parabolic monodromies $V$ and $U$. There is some
puzzle with this identification. The way to sum charges for
codimension-2 defects is closely related to braids, and Dehn
twists indeed satisfy the braid relation $UVU=VUV$. It is not clear
how this fact could be seen in the corresponding gauged
supergravity. Moreover, the elliptic asymmetric T-folds have fluxes
quantized in fractional units, so that the relation with the
corresponding brane is not obvious. It would be interesting to clarify
these issues. It would also be interesting to understand if there is a
way to understand directly the collision of NS5 and Q branes, namely
if there is an analogous of string junctions~\cite{Gaberdiel:1998mv} in the present context.

\subsubsection*{Hyperbolic $\rho$-fects}

The last example in order to exhaust all possible conjugacy classes
for the $\rho$ fibration is a hyperbolic monodromy. As we discussed
in the previous section, it is not possible to interpret such
monodromy as coming from a degeneration of elliptic curves. Let us
consider as an example the monodromy 
\begin{equation}
M_{\rho}= \begin{pmatrix} N & 1 \\- 1 & 0 \end{pmatrix}
\, , \quad N\geq 3 \, .
\end{equation}
We also show the corresponding $O(2,2,\mathbb{Z})$ element:
\begin{equation}
\Omega_{O(2,2)} =\begin{pmatrix}
  N &0 &0&1\\0&N&-1&0\\0&1&0&0\\-1&0&0&0\end{pmatrix} \, .
\end{equation}
The action on the B-field and the volume is then:
\begin{equation}\label{BVhyperbolictransf}
B \rightarrow - \frac{1}{\lambda}-\lambda -\frac{B}{B^2+ G} \, , \qquad \sqrt{G} \rightarrow
\frac{\sqrt{G}}{B^2+G} \, .
\end{equation}
We recall that $\lambda$ is the largest eigenvalue of $M_{\rho}$, and
$\lambda^{-1} +\lambda=N$.
The action~\eqref{BVhyperbolictransf} is similar to the elliptic case~\eqref{BVelliptictransf}, but
successive iterations of the hyperbolic transformation acts very
differently on the fields. From the
solution~\eqref{tauhyperbolic} it is possible to exhibit a
metric and flux with such monodromy. For simplicity, we show only the
solutions for the torus volume and the B-field:
\begin{align}
\sqrt{G} &= \frac{1}{\lambda}\frac{e^{\sigma}(\lambda^2-1)\left[e^{\sigma}-\lambda^{\theta/\pi}
    e^{\lambda^2\sigma}\cos
    \tilde r\right]}{e^{2\sigma}+\lambda^{2\theta/\pi}e^{2\sigma
    \lambda^2}-2\lambda^{\theta/\pi}e^{\sigma(\lambda^2+1)}\cos
    \tilde r}-\lambda \, , \\[10pt]
B & = -\frac{1}{\lambda}\frac{e^{\sigma(\lambda^2+1)}(\lambda^2-1)\sin
    \tilde r}{\lambda^{-\theta/\pi}e^{2\sigma}+\lambda^{\theta/\pi}e^{2\sigma\lambda^2}-2e^{\sigma(\lambda^2+1)}\cos
    \tilde r} \, ,
\end{align}
where we defined
\begin{equation}
\tilde r = \mu (\lambda^2-1) + \frac{1}{\pi} \log \lambda \log \left[\pi
r (\lambda^2-1) \right] \, .
\end{equation}
One can check that these solutions indeed
satisfy~\eqref{BVhyperbolictransf}. From the expression of $\sqrt{G}$ we see
that close to the origin there are infinite points at which $\sqrt{G}$ becomes
negative. As we discussed for the corresponding $\tau$-fects, this solution can at best approximates the geometry of
non-collapsed branes outside a small disk around the origin. Again,
the solution cannot be trusted for large $r$, as for the other
codimension two metrics.

\subsection{Colliding degenerations}

We finally discuss the most general case in which both $\tau$ and
$\rho$ vary along the base $\mathcal{B}$ and their degenerations
collide at the same point $z_0$, that we again take to be the origin. The semi-flat  metric ansatz
is now:
\begin{align}
ds^2 &= \eta_{\mu \nu} dx^{\mu}dx^{\nu} + e^{2\varphi_1}\tau_2\rho_2
dz d\bar z +\rho_2  G_{ab} dx^adx^b \, , \label{ansatztaurho}\\
B & = \rho_1 dx_8\wedge dx_9 \, , \nonumber
\qquad e^{2\Phi} = \rho_2 \, ,
\end{align}
with $G_{ab}$ given by~\eqref{Habtaubrane}. The conditions imposed by supersymmetry on
this ansatz have been studied in~\cite{Hellerman:2002ax,Becker:2009df}. This fixes again $\phi$, $\tau$ and $\rho$ to be
holomorphic functions on the punctured sphere. In
appendix~\ref{app:eom} we gives the expressions for the Ricci tensor
and Ricci scalar for this ansatz, by explicitly checking that it solves
the equations of motion.

In principle, by combining any of the solutions for $\rho(z)$ and
$\tau(z)$ derived in the previous section we can obtain an explicit
expression for the metric and the
B-field that have an arbitrary monodromy $(M_{\tau},
M_{\rho})$, i.e. an arbitrary $O(2,2,\mathbb{Z})$
element.\footnote{Except the exchange $\rho\rightarrow \tau$. This
  will be discussed in the next section.} A priori it is not obvious
that such degenerations should be accepted. However, as we will comment
in the next section, there is a concrete possibility to study such
objects from heterotic/F-theory duality. One could perhaps also think that
such geometries give an approximation of the geometry of non-collapsed
$\tau$ and $\rho$ T-fects.

\subsubsection*{Double parabolic T-fects}

A simple example is the case $M_{\tau}= V^N$, $M_{\rho}=V$ which should
represents a stack of $N$ NS5 branes on top 
of a Taub-NUT space. This is simple to check
from~\eqref{ansatztaurho}, which by using previous results reads:
\begin{align}
ds^2 &= \eta_{\mu \nu} dx^{\mu}dx^{\nu} + h_{5}(r) ds^2_{KKM}\, ,\label{NS5KKM}\\
B_2 & =  \frac{\theta}{2\pi}   dx_8\wedge dx_9 \, ,\nonumber
\qquad e^{2\Phi} =  h_{5}(r)  \, ,
\end{align}
where
\begin{equation}
h_{5}(r) = \frac{1}{2\pi}
\log\left(\frac{\mu}{r}\right) \, .
\end{equation}
This is a smeared approximation of the solution for an NS5 on a Taub-NUT
space, which has precisely the same form~\eqref{NS5KKM} by harmonic
superposition, with $h_{5}(r) \sim 1/r$.
Analogous results can be found for arbitrary $M_{\tau}$.
If $M_{\rho}$ is not a simple shift, we obtain non-geometric solutions that are not
T-dual to any geometric background. Here we limit ourselves to discuss
one example of this kind, obtained by considering an elliptic
monodromy for both the $\tau$ and $\rho$ modulus.

\subsubsection*{Double elliptic T-fects}

As an illustrative example, we consider the monodromy
\begin{equation}
M_{\tau} = M_{\rho} = UVU =\begin{pmatrix} 0&1\\ -1 &
  0 \end{pmatrix} \, ,
\end{equation}
which acts on $\tau$ and $\rho$ by
\begin{equation}\label{taurhodoubletfect}
 \tau\rightarrow -\frac{1}{\tau} \, ,\quad
\rho\rightarrow -\frac{1}{\rho} \, .
\end{equation}
This corresponds to the following $O(2,2)$ transformation:
\begin{equation}
\Omega_{O(2,2)} =\begin{pmatrix}
  0&0&-1&0\\0&0&0&-1\\-1&0&0&0\\0&-1&0&0\end{pmatrix} \, .
\end{equation}
From the solutions obtained above for the corresponding $\tau$ and
$\rho$-fects with the same monodromy we obtain a solution which is a
concrete realization of the fibration described in~\eqref{doubleellipticorder4}.
Here we show explicitly only the expressions for the metric and the B-field that
fiber over the two-dimensional base (note that we make a
particular choice for the integration constants, which is not the most
general one):
\begin{align}
G_{11} & = \tau_1^2+\tau_2^2 = \frac{r^2+\mu^2-2 r \mu \cos(\theta+
         2\sigma)}{\left[r+\mu+2\sqrt{r\mu}
         \cos\left(\theta/2+\sigma\right)\right]^2} \, ,\\[10pt]
G_{12} & =\tau_1 =
        \frac{\sin(\theta/2+\sigma)}{\cos(\theta/2+\sigma)+\cosh
         \left[\frac12 \log\left(\frac{r}{\mu}\right)\right]} \, , \\[10pt]
e^{2\Phi} &=\tau_2=  - \frac{\sinh\left[\frac12 \log\left(\frac{r}{\mu}\right)\right]}{\cos\left(\frac{
  \theta}{2}+\sigma\right)+
            \cosh\left[\frac12\log\left(\frac{r}{\mu}\right)\right]}
            \, ,\\
B & = \rho_1=\tau_1 =G_{12} \, .
\end{align}
The action of the elliptic transformation on the metric components is:
\begin{equation}
G_{11}(2\pi) =
               \frac{1}{G_{11}(0)}+\frac{G_{12}(0)^2}{G_{11}(0)^2}\left[G_{12}(0)^2-1\right]
               \, , \qquad
             G_{12}(2\pi)=
               -\frac{G_{12}(0)^2}{G_{11}(0)^2}
               \, .
\end{equation}
We can now repeat the discussion regarding the relation with the elliptic
T-folds described in section~\ref{subsec:nongeometricmonodromy} that
we did for the single $\rho$-fect. A
T-fold with moduli satisfying the double elliptic
monodromy~\eqref{taurhodoubletfect} has been studied
in~\cite{Condeescu:2012sp,Condeescu:2013yma,Hassler:2014sba}, where it was argued that such backgrounds contain both geometric and non-geometric $f$,
$H$ and $Q$ fluxes. It is interesting to ask if there is a relation
between these fluxes and the charges of the T-fects considered
here. A way to define the charge is via the monodromy that identifies a
given T-fect. For a NS5 branes, the monodromy $V_{\rho}$ indeed corresponds to
measuring one unit of $H$-flux around the source. The exotic $5^2_2$
brane is then identified by the parabolic monodromy $U_{\rho}$, and
one can declare this to correspond to one unit of non-geometric
$Q$-flux. For geometric fibrations, one could naively identify the
parabolic monodromies $V_{\tau}$ and $U_{\tau}$ that correspond to KK
monopoles with orthogonal special directions to the two kind of
``geometric $f$-fluxes'' that appear in the gauged supergravity. This
is little more then terminology, and there are of course subtleties in
making these relations concrete. For example, our discussion makes
very clear that the monodromy is closely related to the topology of
the space, so identifying a monodromy with a parameter of the effective
theory requires some care~\cite{McOrist:2012yc}. However, it is
interesting to see that the monodromy factorization in terms of Dehn
twists, which would correspond to adding the charges of the parabolic
T-fects, indeed contains the monodromies associated to NS5, $5^2_2$ and
KKM sources. It would be interesting to study further this relation.

\section{Heterotic T-fects from genus 2 fibrations}\label{sec:hyperellipticfibrations}

In this section we would like to generalize part of the previous discussion to
the study of monodromies and degenerations of genus two
fibrations. As we now explain, this is relevant in the context of
heterotic compactifications with a Wilson line. This setup has been
considered recently in~\cite{Malmendier:2014uka}, our discussion will be however more focused
on the description of local monodromies and their associated
T-fects. We mention that following the ``surface
diffeomorphism'' approach described in the previous section, we can
obtain a purely topological classification of genus two degenerations
that match the known analytic one~\cite{MatMont,Ashikaga_globaland}. We remark that similar models arise in
the context of U-duality in type II theories, and the resulting
U-folds, or better \emph{U-fects}, are relevant for holography~\cite{Martucci:2012jk,Braun:2013yla,Candelas:2014jma,Candelas:2014kma}.

Let us explain our motivation. We start with the eight dimensional
duality between heterotic string compactified on a $T^2$ and F-theory
compactified on an elliptically fibered K3 surface. We will consider
the case in which a single Wilson line expectation value is turned on. We then
adiabatically fiber the duality on a $\mathbb{P}^1$ to obtain a
six-dimensional duality. As explained in details
in~\cite{Malmendier:2014uka,Gu:2014ova}, one can map the data of the
fibration on the F-theory side to the data of a genus two fibered
surface $\mathcal{S}$. The data of this fibration will in turn
describe the fibration of the three 8d heterotic moduli $(\tau(z),
\rho(z), \beta(z))$ over a $\mathbb{P}^1$. If we allow monodromy in
$\tau$, namely large diffeomorphisms of the $T^2$, we  obtain the
duality between an elliptically fibered K3 
compactification 
and F-theory on a K3-fibered CY three-fold.  More generically, genus 2
monodromies
will determine patching conditions of the heterotic $T^2$ together with the
Wilson line bundle, that in general involve T-dualities, resulting
in a non-geometric heterotic compactification.

\subsection{Monodromy generators}

In order to understand the T-fects of these heterotic
non-geometric models, one can look at degenerations of genus 2
surfaces and try to understand the corresponding monodromy
factorization. This would give an understanding of a general T-fect
in terms of the charges of more fundamental objects.
Let us consider a genus 2 surface $\Sigma_2$. 
It is
useful to think about an element of
$\text{Aut}(H_1(\Sigma_2;\mathbb{Z}))=Sp(4,\mathbb{Z})$ as coming from an element of the mapping
class group $M(\Sigma_2)$, by the
surjective homeomorphism (see Appendix A for details):
\begin{equation}
\Phi: M(\Sigma_2) \rightarrow Sp(4,\mathbb{Z}) \, .
\end{equation}
This map has a large kernel, thus the homological monodromy does not
capture all the features of $M(\Sigma_2)$. However, we will use known results for
$M(\Sigma_2)$ to understand the trichotomy of mapping class elements
for the genus 2
case and the possible degenerations of $\Sigma_2$ fibrations.

\begin{figure}
\begin{center}
\includegraphics[scale=0.4]{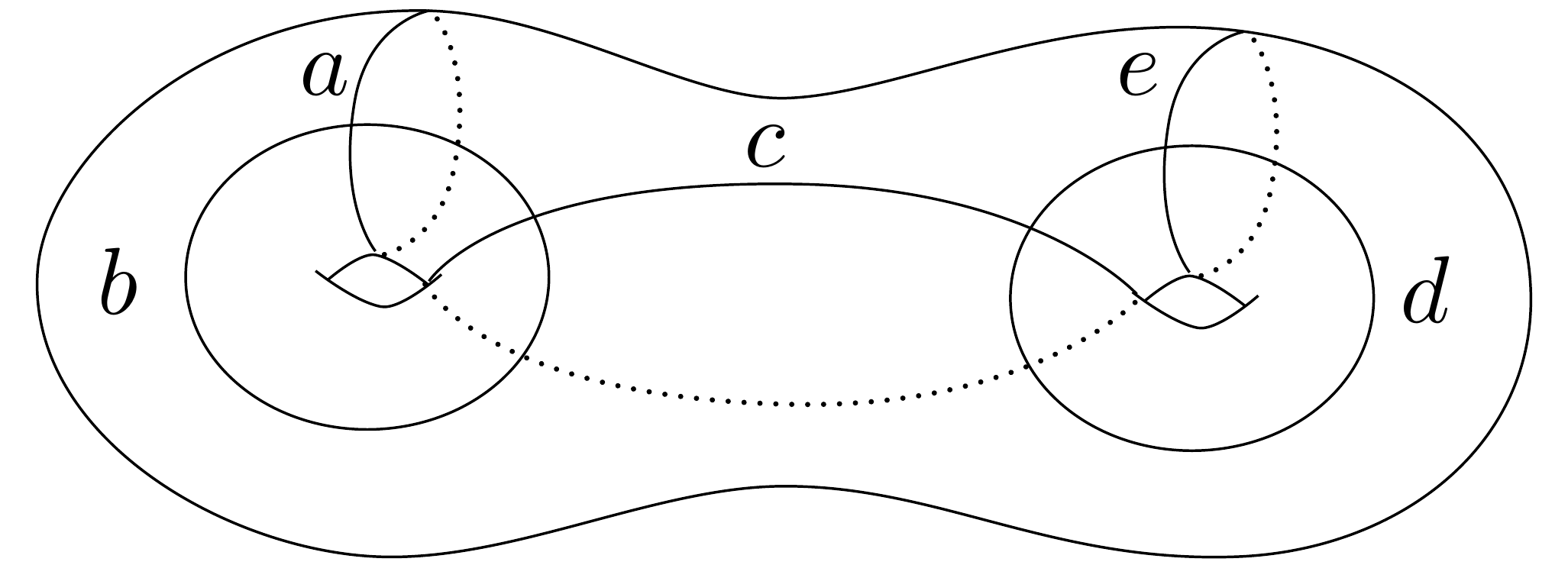}
\caption{The Humphries generators for the mapping class group of a
  genus 2 surface $\Sigma_2$: any monodromy is obtained from a
  combination of Dehn twits along the cycles $a,b,c,d,e$. We identify
  $a,b,e,d$ with the standard basis for the homology group.}
\label{Fig:genus2}
\end{center}
\end{figure}

We first give a presentation of $M(\Sigma_2)$ in terms of a minimal
number of right-handed Dehn twists, generalizing the discussion for
the torus in section~\ref{subsec:mappingtori}. This result is classical (see for
instance~\cite{BLTbook}). We consider the set of Humphries
generators shown in Figure~\ref{Fig:genus2}, and we denote a Dehn twist
around the cycle $a$ by $A$, ecc. We use $(a,b,d,e)$ as a basis for
$H_1(\Sigma_2;\mathbb{Z})$.
We then have:
\begin{align}
A &= \begin{pmatrix} 1 &1 & 0 &0 \\0 &1 & 0 &0 \\0 &0 & 1&0 \\0 &0 & 0
&  1  \end{pmatrix} \, , \quad
B = \begin{pmatrix} 1 &0 & 0 &0 \\-1 &1 & 0&0
  \\0 &0 & 1&0 \\0 &0 & 0 &1  \end{pmatrix} \, , \quad
C = \begin{pmatrix} 1& 1 & 0 &-1 \\0 &1 & 0 &0 \\0 &-1 &1 &1 \\0& 0& 0 &1  \end{pmatrix}
\, ,\\[10pt]
D &= \begin{pmatrix} 1 &0& 0 &0 \\0&1 & 0& 0 \\0 &0 & 1 &0 \\0&0&
  -1 &1  \end{pmatrix} \, , \quad  
E = \begin{pmatrix} 1 &0 & 0 &0 \\0 &1 & 0 &0 \\0 &0 & 1 &1 \\0 &0 & 0
  &1  \end{pmatrix} \, .\nonumber
\end{align}
One can prove that these generators are a presentation of the mapping
class group by showing that they satisfy the following relations:
\begin{align}
\text{(Disjointness)} & \qquad [ A,C ] = [A,D] = [A,E] = [B,D]= [B,E] = [C,E] = 0  \, ,
 \label{genus2presentation}\\
\text{(Braidness)} &\qquad ABA= BAB \, , BCB = CBC \, , CDC = DCD \,
,DED = EDE \, , \nonumber\\
\text{(3-chain)} & \qquad (ABC)^4 = E^2 \, , \nonumber\\
\text{(Hyperellitic)} & \qquad 
[H,A] = 0 \, , \quad H^2 = \mathbb{1} \, ,\nonumber
\end{align}
where $H = EDCBA^2BCDE$ is the hyperelliptic involution, namely a
rotation of $180$\textdegree~along a horizontal axis in
Figure~\ref{Fig:genus2}. Note that since the cycles $b$ and $d$ are
not linked, the sets $(A,B)$ and $(D,E)$ generate block diagonal
matrices
\begin{equation}\label{genus2block}
M_{\tau, \,\rho} = \begin{pmatrix}  M_{\tau} & 0 \\ 0 &
  M_{\rho} \end{pmatrix} \, ,
\end{equation}
where each block is of the kind discussed for torus
fibrations in~\ref{subsec:mappingtori}, so that $M_{\tau, \,\rho}$ corresponds to the $SL(2,\mathbb{Z})_{\tau}\times
SL(2,\mathbb{Z})_{\rho}$ T-duality monodromies. Note that the two
genus 1
components can be glued together with a non-trivial map, which is not
seen in the homology since it involves a separating curve on
$\Sigma_2$. Given a matrix of the form~\eqref{genus2block}, the action
on the moduli can be determined from the $Sp(4,\mathbb{Z})$ action on
Siegel upper half plane, as described for example
in~\cite{Martucci:2012jk}. In the limit $\beta \rightarrow 0$, this
action reduce to a ``double copy'' of the $SL(2,\mathbb{Z})$ action on
the upper half plane by M\"obius transformation, and we see that we
can embed in this way the double elliptic T-fects described in the
previous section. More interesting mapping group elements are
obtained by combining the $A, B, D, E$ twists with the action of
$C$. Interestingly, explicit Dehn twist factorizations of elliptic
elements containing the $C$ twist can be obtained. In particular, this
includes a monodromy that interchanges $\tau$ and $\rho$, thus geometrizing
the full T-duality group.\footnote{Such an element is for example given
  by the decomposition $ABCDEABCDABCABA$.} We list some
examples below:
\begin{equation}\label{elliptictwistgenus2}
M_3 = (ABCDE)^2 \, , \quad M_6 = ABCDE \, ,\quad M_8 = ABCDD \, ,\quad M_{10} =
ABCD \, ,
\end{equation}
 where the subscript indicates the order. One can prove that 10 is the
 maximum order for elliptic elements in $M(\Sigma_2)$.

\subsection{Degenerations}

While a detailed discussion of heterotic models is beyond the scope of
this paper, we would like to briefly outline how the discussion of the
previous sections generalizes to the genus 2 case. We can again
think of a non-geometric fibration of the heterotic T-duality group as
being geometrized by the mapping torus of a genus 2 surface
$\Sigma_2$. Specifically, if the base is $\mathbb{P}^1$, we can
consider the fibration of $\Sigma_2$ over the
boundary $S^1=\partial D^2$ of a small disk and give a classification of local
non-geometries on $D^2$ from
elements of the mapping class group $M(\Sigma_2)$ that specify the
gluing of the fibration on $S^1$.

From our previous discussion, we expect that monodromies of the
form~\eqref{genus2block}, were
$M_{\tau}$ and $M_{\rho}$ are parabolic or elliptic in
$SL(2,\mathbb{Z})$, or periodic monodromies such
as~\eqref{elliptictwistgenus2}, should arise around branch points of a
genus 2 degeneration. Indeed, this gives a rough derivation
of the local monodromies around degenerations of genus 2 surfaces. One
subtlety is that the homological monodromy is not a good invariant, so
different degenerations can have the same monodromy. 
A
more complete discussion should make use of an intrinsic
classification of mapping class elements and can be found in~\cite{Ashikaga_globaland},
while the algebraic classification is given
in~\cite{Namikawa:1973yq}. 

As in the torus case, one can consider more general monodromies that
might correspond to non-collapsed configurations of T-fects. These are
the analogous of hyperbolic conjugacy classes and are called
pseudo-Anosov maps. We refer to appendix~\ref{app:mcg} for a
definition. 
 For surfaces of genus $g>1$, compared to the torus, is more difficult
 to construct explicitly this kind of diffeomorphisms. If we
 restrict to the homology representation, there exists an explicit
 criterion (which is however not sufficient), and this is less simple then computing the trace as
 in the genus 1 case.
Here we limit ourselves to present one example in terms of a product of
Dehn twists and we refer to section 14.1.3 of~\cite{mcgprimerbook} for
a detailed discussion:
\begin{equation}
M_{pA} = ABCDE^{1-n} = \begin{pmatrix} 0 &1&0&0 \\ -1&0&-1& 1-n \\ 0&-1&0&1\\0&0&-1&1-n\end{pmatrix} \, .
\end{equation}
One can prove that for $|1-n|>1$ this map is pseudo-Anosov. Note that this is
the analogous to the hyperbolic maps~\eqref{hypmapsgenus1} in the genus
one case and this provides an example of a duality twist that do not
arise in the list of degenerating genus 2 surfaces.

We also mention that the geometric approach could be useful to
understand global models. As for the other codimension-2 defects, to
obtain a sensible global solution one
needs to consider configurations of T-fects that cancel the total
charge. Such configurations might be obtained, as in the torus case,
from known factorizations of the identity in terms of Dehn twists. For
example, in addition to the hyperelliptic relation
$(EDCBAABCDE)^2=\mathbb{1}$, there are additional relations, from~\eqref{elliptictwistgenus2}:
$(ABCDE)^6=\mathbb{1}$, $(ABCD)^{10}=\mathbb{1}$. These are configurations with
20, 30 and 40 elementary T-fects respectively (namely degenerations
associated to a vanishing cycle, described by a parabolic
monodromy).

It would be interesting to pursuit further this line of
investigations, in particular to study local solutions as we did in the previous sections and to
understand the relation with the T-folds studied
in~\cite{Blumenhagen:2014iua}. We note that left-right asymmetric heterotic T-folds have
also been studied from a CFT point of view
in~\cite{Bianchi:2012xz}. It would be interesting to see if their techniques
can be used to study some of the models outlined here, at least in some region of moduli
space.

\section{Conclusions}\label{sec:conclusions}

Let us summarize the finding of this paper. We developed a
correspondence between torus fibrations over circles and degenerating
torus fibrations over a two dimensional base, with focus on the
monodromy of the fibrations. We first applied this to string
theory compactified on a two torus $T^2$, and monodromies contained in
the group of large diffeomorphisms of the torus, i.e. the
$SL(2,\mathbb{Z})_{\tau}$ subgroup of the T-duality group. In this
case the trichotomy of conjugacy classes of $SL(2,\mathbb{Z})$ is
in correspondence with the possible geometries of the total space $\mathcal{N}$ of the fibration $T^2 \rightarrow
\mathcal{N} \rightarrow S^1$, namely Euclidean, Nil and Sol
geometries. We explicitly constructed a metric ons such manifolds for
each conjugacy classes of the $SL(2,\mathbb{Z})$ group. 

When the torus is fibered over a two dimensional base
$\mathcal{B}$, the possible degenerations (as classified by Kodaira)
are in correspondence with parabolic and elliptic conjugacy
classes, while the hyperbolic monodromies, as a result of a general theorem, do not arise as local
degenerations but might describe an expanded configuration of
non-collapsed defects. By using the knowledge of the metric for the three
manifolds arising from the fibration of the torus on a boundary of a
small disk $D^2$ containing a degeneration point $x_i$, we solved the corresponding
Cauchy-Riemann equations by separation of variables and we find, for
each conjugacy class in $SL(2,\mathbb{Z})$, the corresponding
holomorphic function on the punctured disk
$D^2\setminus\{x_i\}$ that determines the complex structure of the $T^2$, thus
obtaining the metric of all possible local semi-flat approximations to
the fibrations of $T^2$ over $D^2$. The simplest example is the
semi-flat approximation obtained from a Nil-geometry, which
approximates well a Taub-NUT space.

We then consider a general semi-flat ansatz that allows monodromies in
both the $SL(2,\mathbb{Z})_{\tau}$ and $SL(2,\mathbb{Z})_{\rho}$
factors of the T-duality group. When only $\rho$ varies, the local
solutions are trivially obtained by a fiberwise mirror symmetry from
the geometric fibrations. By applying this to all the conjugacy
classes of $SL(2,\mathbb{Z})$, we obtain a classification of the local
solutions. Since the monodromies now act non-trivially on the volume,
these solutions are non-geometric and should provide a
qualitative description of the corresponding T-fects. In order to
better understand these solutions, we need to supplement the local geometries
with a string description of the T-fect that sits at the
degeneration point, were the semi-flat approximation breaks
down. We could obtain this by noticing that in the heterotic theory
the fibration of the auxiliary two torus that geometrizes the $SL(2,\mathbb{Z})_{\rho}$
group can be studied
by using heterotic/F-theory duality, and in the F-theory side
some of such non-geometric solutions correspond to well known K3-fibered Calabi-Yau
compactifications. Since only bosonic degrees of freedom are excited
in the solutions, we should expect similar exotic objects in type IIA
and IIB as well. 

We finally briefly discussed the generalization of the geometric
fibration picture to surfaces of higher genus. Such situation for
example arises in non-geometric compactifications of the heterotic
theory with a Wilson line, by the identification of the T-duality
group $SO(2,3,\mathbb{Z})$ with the symplectic group
$Sp(4,\mathbb{Z})$ which arises as the action of the group of large diffeomorphisms
of a genus 2 surface on the homology. The geometric group picture
is then useful to understand the list of all possible
degenerations of such genus 2 surfaces, that was obtained by using
algebraic geometry techniques in~\cite{Namikawa:1973yq}. A detailed study of such
non-geometric backgrounds from the F-theory point of view will be
presented in~\cite{Ourpaper}.

\vspace{1cm}
\textbf{Acknowledgments:} We would like to thank Ralph Blumenhagen, Andreas Braun, 
Anamar\'ia Font, I\~naki Garc\'ia-Etxebarria, Daniel Junghans,
Christoph Mayrhofer, Ruben Minasian, Erik Plauschinn and
Federico Savelli for
useful discussions and correspondence and in particular Anamar\'ia Font, I\~naki
Garc\'ia Etxebarria and Christoph Mayrhofer for collaboration on
related topics. S.M. would like to thank the organizers of the ``Eighth Crete
Regional Meeting'' for hospitality while part of this work was completed.  
This work is supported by the ERC Advanced Grant 32004 -- Strings and Gravity.

\appendix

\section{Review of mapping class groups}\label{app:mcg}

In this section we give a short review of some classical results in the theory
of mapping class groups for surfaces, with the purpose of being
self-contained. A good source of material is the introductory
exposition in~\cite{mcgprimerbook}. 

\subsection{Generators and representations}

Let us consider a genus $g$ surface $\Sigma_g$. We will assume for
simplicity that $\partial \Sigma_g = 0$ and $g\ge 1$. The mapping class
group of $\Sigma_g$, that we denote by $M(\Sigma_g)$, is defined as
\begin{equation}
M(\Sigma_g) = \text{Diff}^+(\Sigma_g)/ \text{Diff}_0(\Sigma_g) \, ,
\end{equation}
where $\text{Diff}_0$ consists of elements isotopic to the identity
and $+$ indicate orientation-preserving. A fundamental result about
the generators of $M(\Sigma_g)$ is:
\begin{theorem}[Dehn] 
The mapping class group $M(\Sigma_g)$ is generated by finitely many
Dehn twists along non-separating simple closed curves in $\Sigma_g$.
\end{theorem}
The minimal number of twists needed to generate the whole group is
given by the following result:
\begin{theorem}[Humphries] 
Dehn twists along the $2g+1$ Humphries generators are enough to generate
the whole $M(\Sigma_g)$. Any set of generators contains at least $2g+1$ elements. 
\end{theorem}
The Humphries generators are showed in Figure~\ref{Fig:genus2} for a
genus 2 surface. A presentation of $M(\Sigma_1)$ and $M(\Sigma_2)$ by
using this set of Dehn twists is given respectively
in~\eqref{mcgtorus} and~\eqref{genus2presentation}. One tool to study
$M(\Sigma_g)$ is to consider its action on homology: $\Phi:
M(\Sigma_g) \rightarrow \text{Aut}(H_1(\Sigma_g,\mathbb{Z}))$. This
gives a representation of $M(\Sigma_g)$ on a symplectic vector space
$Sp(2g,\mathbb{Z})$:
\begin{equation}
\Phi: M(\Sigma_g) \rightarrow Sp(2g,\mathbb{Z}) \, .
\end{equation}
A $k$-power of a Dehn twist $B$ along a cycle $b$ act on the homology
in the obvious way: $\Phi(B^k)([a]) = [a] +k\cdot i(a,b) [b]$, where $i$ is the
intersection number. A fundamental result is that the map $\Phi$ is
surjective, so we can always see a monodromy in the homology as a
monodromy of a mapping torus. The converse is not true, and $\Phi$ has
a large kernel known as the Torelli group $T(\Sigma_g)$, giving an exact sequence
\begin{equation}
1 \longrightarrow T(\Sigma_g) \longrightarrow M(\Sigma_g)
\longrightarrow Sp(2g,\mathbb{Z})\longrightarrow 1 \, . 
\end{equation}
The Torelli group is empty for $g=1$,
while for $g=2$ a simple example is a Dehn twist along a separating
curve, whose action on the homology is the identity. Hence, for genus
two surfaces we can define two distinct notions of monodromy, one
coming from homology and one from surface diffeomorphisms. A better
representation is obtained on the fundamental group of $\Sigma_g$. We
have:
\begin{theorem}
There exists an isomorphism $\sigma: \widetilde{M(\Sigma_g)} \rightarrow
\text{Out}\big(\pi_1(\Sigma_g)\big)$.
\end{theorem}
Here the tilde means that orientation-reversing diffeomorphisms should
be included and Out is the outer automorphism group given by the quotient
Aut/Inn.

\subsection{Classification and hyperbolic maps}

An important result for the study of the geometry and topology of
3-manifolds is the classification of mapping class group elements that
generalizes the trichotomy of $M(T^2)$, which is given by the parabolic, elliptic and hyperbolic conjugacy
classes of $SL(2,\mathbb{Z})$. A geometric notion of such
classification is useful since gives a simple criterion to understand
which monodromies can arise from degenerations of surface fibrations
of arbitrary genus.

The fundamental result is a theorem of Thurston, related to some old
results by Nielsen:
\begin{theorem}[Nielsen-Thurston]
Let $\Sigma_g$ be a closed oriented 2-manifold. Then $f \in
M(\Sigma_g)$ is either periodic, reducible, or pseudo-Anosov. The
latter is neither periodic nor reducible.
\end{theorem}

We briefly summarize the definitions of the three kinds of mapping
class group elements. A map $\phi$ is \emph{periodic} if it exists a
positive integer $k$ such that $\phi^k = \mathbb{1}$ in
$M(\Sigma)$. One can prove a stronger result, that it is possible to
choose a representative of $\phi$ in $\text{Hom}^+(\Sigma)$ such that
  the $k$-th power of the map is precisely the identity, and not only
  isotopic to it. For the torus, periodic mapping classes are the
  elliptic conjugacy classes of $SL(2,\mathbb{Z})$.

A map is \emph{reducible} if it leaves invariant a non-empty set
$\mathcal{C}$ of
non intersecting simple closed curves in $\Sigma$. One can study such
maps by studying their action on $\Sigma \setminus \mathcal{C}$. On the torus,
these are precisely the parabolic conjugacy classes. For example, a
simple Dehn twist along a curve $c$ is reducible with $\mathcal{C} =\{
c\}$. In general periodic and reducible classes are not mutually
exclusive. A \emph{pseudo-periodic} map is either periodic or
reducible with all components periodic.

The remaining set is that of \emph{pseudo-Anosov} maps. The definition
is a generalization of the hyperbolic map of the torus and it relies
on the existence of a number $\lambda >1$ and so-called transverse measured foliations
$(\mathcal{F}, \mu), ( \mathcal{\tilde F}, \tilde \mu)$. A map $\phi$
is pseudo-Anosov if it acts on the two foliations by stretching and
contracting the measures: $\phi (\mu, \tilde \mu) = (\lambda \mu,
\lambda^{-1}\tilde \mu)$. This is a formalization of what we think an
hyperbolic map should do: on the torus a foliation is given by
repackaging all the parallels to a line in $\mathbb{R}^2$. For an
hyperbolic element $M\in SL(2,\mathbb{Z})$, the directions are given by the
expanding and contracting eigenspaces of $M$. 

\begin{figure}[t]
\begin{center}
\vspace{0.3cm}
\includegraphics[scale=0.4]{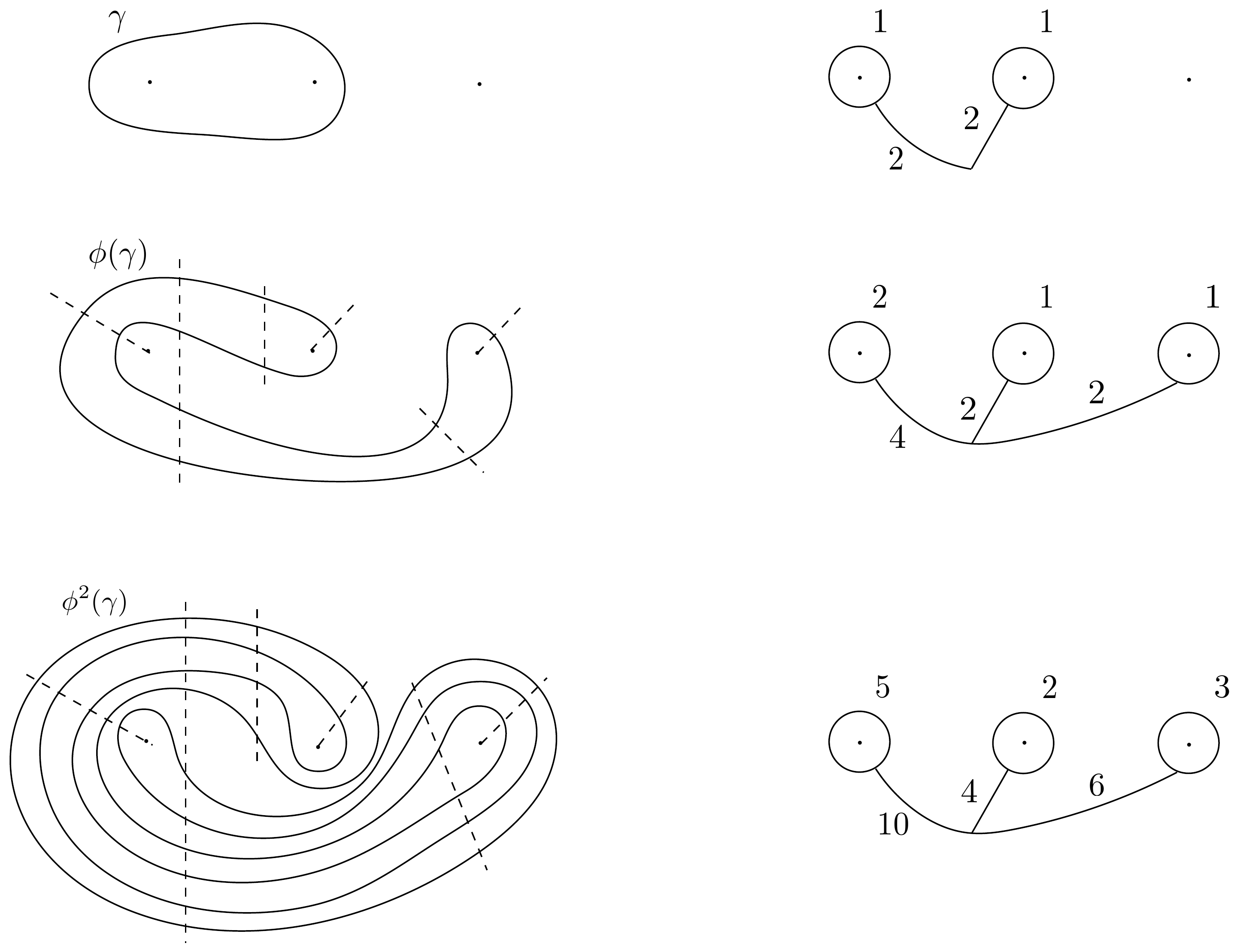}
\caption{Left: the action on a curve $\gamma$ of an hyperbolic
  diffeomorphism $\phi$ on a punctured sphere $\Sigma_{0,4}$ (one
  puncture at  infinity is not shown). Right: the corresponding
train tracks, which count the thickness of the foldings of $\gamma$
along the dotted line
under the iteration of the map.}
\label{Fig:traintrack}
\end{center}
\end{figure}

To give an intuition of the action of hyperbolic maps, and their Dehn
twist decomposition found in section~\ref{subsec:mappingtori}, we
briefly describe a classical example of a pseudo-Anosov mapping
class (see for example~\cite{mcgprimerbook}). We consider a sphere
with 4 punctures $1234$ and we consider two mapping class group
elements that implement half-twists, namely $T_{23}: 2
\leftrightarrow 3$, and $T_{12} : 1\leftrightarrow 2$. These maps
exchange two punctures clockwise. We then consider the map $M =T_{12}
T_{23}^{-1}$. Note that this map lifts to the closely related element in
$M(T^2)$ given by
\begin{equation}
M_{T^2} = V U^{-1} = \begin{pmatrix} 2 & 1 \\ 1&1 \end{pmatrix} \, ,
\end{equation}
that we encountered many times in the text, see for
instance~\eqref{hyperbolictaumonodromy}. We then study the action of
$M$ on a simple curve $\gamma$, shown in
Figure~\ref{Fig:traintrack}. This illustrates well the complicated
behavior of iterations of hyperbolic diffeomorphisms. In fact, we can
consider the thickness of the foldings of the images of $\gamma$,
which is usually depicted with a so-called train track shown on the
right of Figure~\ref{Fig:traintrack}. The number of foldings in our
example is governed by a Fibonacci sequence, and it grows
exponentially (see also the discussion near~\eqref{rhohyp}). One can
repeat this for different curves, and the result is a very complicated
dynamics, which share many similarities with chaotic systems.

\section{Hurwitz moves and ABC factorization}\label{app:braid}

In this appendix we give some more details about the relation between
our notation in section~\ref{subsec:monodromy} and the more familiar F-theory approach in terms of
moving branch cuts~\cite{Gaberdiel:1998mv}. First, we reproduce in
Table~\ref{tab:Kodaira} the well known Kodaira list for the
degeneration types of genus one curves.
\begin{table}[h!]\begin{center}
\begin{tabular}{|c|c|c|c|}
\hline
Order &  Singularity  & Type            &    Monodromy    \\
\hline
  1    &\text{dbl. point}&  $I_1$         &     $V$                    \\
  $n$    &     $A_{n-1}$    & $I_n$         &  $V^n$                 \\
   2    & \text{cusp}   & $II$             &    $U V$                  \\
  3     &     $A_{1}$      & $ III$          &    $UVU$                \\
  4     &     $A_{2}$    & $ IV $          &    $(UV)^2$                \\
  6     &      $D_4$      &  $I_0^{\ast}$ &    $(UV)^3$                \\
$ 6+n$ &   $D_{4+n}$     & $I_n^{\ast}$ &     $(UV)^3V^n = -V^n$ \\
  8    &        $E_6$       & $IV^{\ast} $ &     $(UV)^4  = (UV)^{-2}$ \\
  9    &        $E_7$      & $ III^{\ast} $ &    $(UV)^4U = (UVU)^{-1}$ \\
 10   &       $E_8$       & $ II^{\ast}$   &    $(UV)^5 =  (UV)^{-1}$ \\
\hline 
\end{tabular}\caption{Kodaira classification of degenerations of
  elliptic fibers.}\label{tab:Kodaira}\end{center}\end{table}
We give a factorization of the monodromy in terms of the two
generators $U$, $V$ of the mapping class group of the fiber torus, corresponding to a Dehn twist
along the $u$ cycle and a twist along the $v$ cycle, see Figure~\ref{Fig:twistT2}. Note that the
starred types have a monodromy which is the inverse of the
corresponding un-starred type. This follows trivially from the
relations which present $SL(2,\mathbb{Z})$, namely from $(UV)^6 = \mathbb{1}$
and from the braid relation $UVU = VUV$. We note that the
degenerations of the table precisely correspond to parabolic elements
(fibers $I_n$ and $I_n^{\ast}$)
and to the elliptic
conjugacy classes listed around~\eqref{ellipticsl2z}. More precisely, all the
mapping tori with parabolic and elliptic monodromies (where we
restrict to a given chirality) arise at the boundary of small disks encircling the given degenerations, in line with the general
theorem discussed in section~\ref{subsec:monodromy}.\footnote{Parabolic
elements such as $V^{-1}$ do not arise since they correspond to
left-handed twists. This offers a simple explanation of why
singularities of the kind $D_{N}$ with $N<4$ or type $\hat E_N$ (see
for instance~\cite{DeWolfe:1998pr})
represent non-collapsible singularities.}

\begin{figure}[h!]
\begin{center}
\includegraphics[scale=0.45]{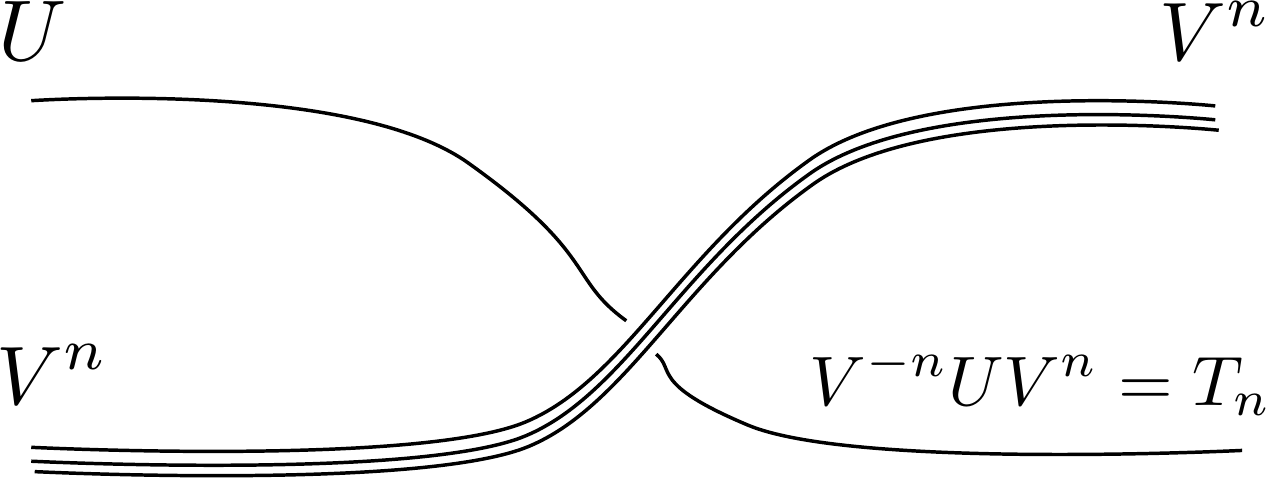}
\caption{Elementary move.}
\label{Fig:braidtn}
\end{center}
\end{figure}

Let us discuss now the Hurwitz moves on the above factorizations. We
define elementary moves that consist in shifting the $V$ components
one step from right to left. This gives $UV^n = V T_n$, where $T_n =
V^{-n} U V^{n}$ (see Figure~\ref{Fig:braidtn}). Analogously, we get $VU^n =U T_{(-n)}$.

\begin{figure}[t]
\begin{center}
\includegraphics[scale=0.4]{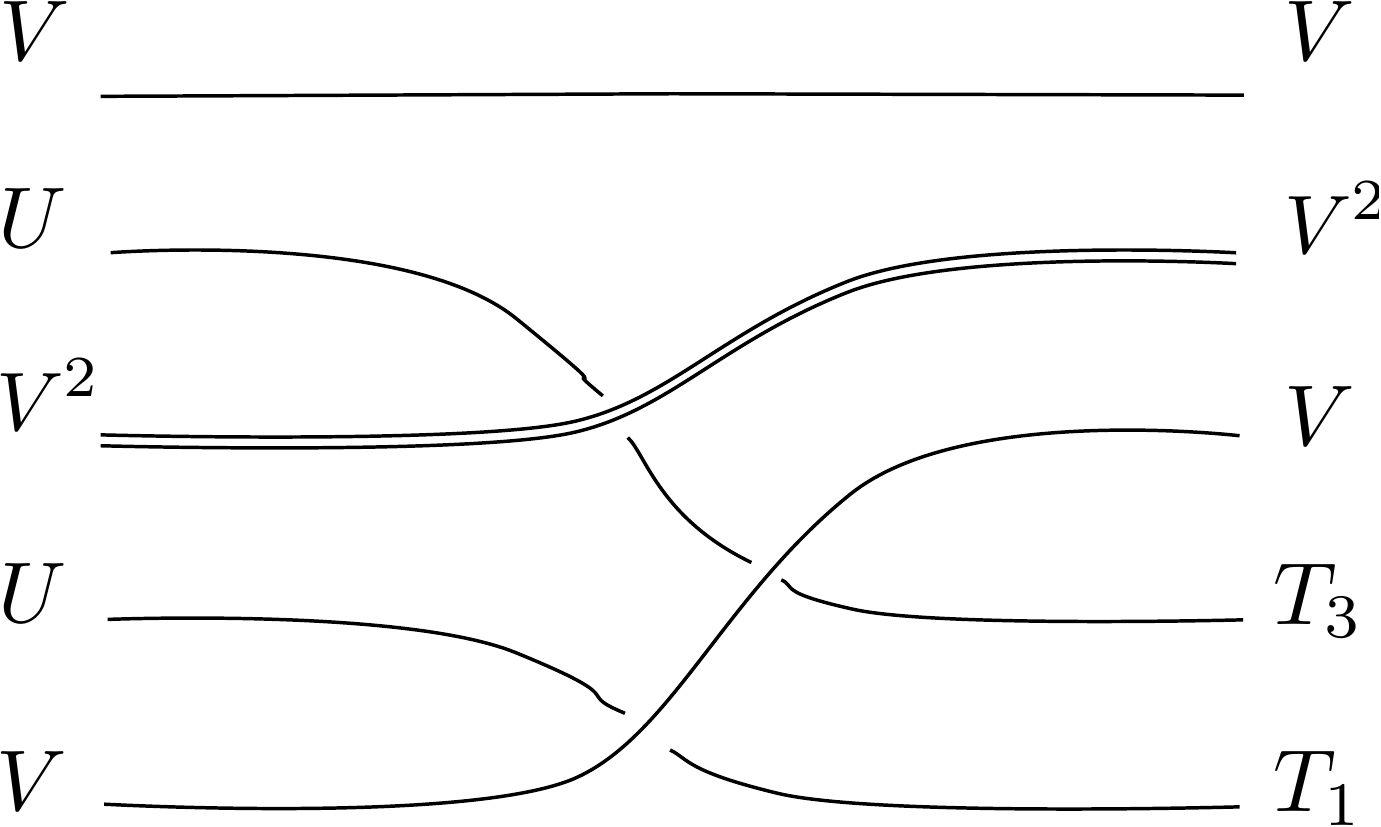}
\caption{Hurwitz moves on the $I_0^{\ast}$ factorization.}
\label{Fig:braidD4}
\end{center}
\end{figure}

\begin{figure}[h!]
\begin{center}
\includegraphics[scale=0.4]{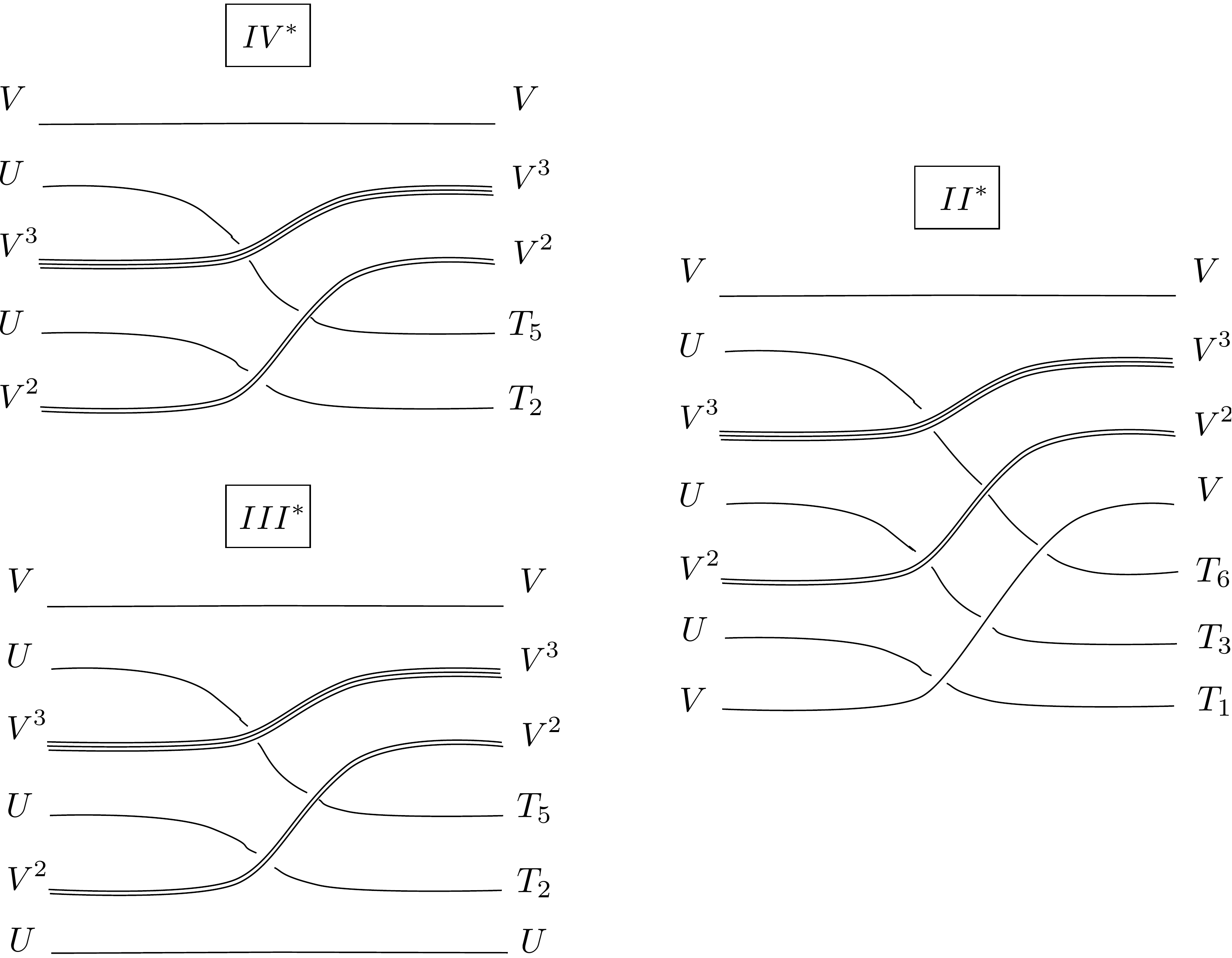}
\caption{Hurwitz moves on the monodromy factorizations for
  degenerations with exceptional singularities.}
\label{Fig:braidEn}
\end{center}
\end{figure}

Note that in terms of the more familiar ABC branes used in the
F-theory literature (see for instance~\cite{Johansen:1996am,Gaberdiel:1998mv,Blumenhagen:2010at}), we have $V=A$, $T_3=B$ and $T_{1} =
C$.\footnote{Sometimes $B$ and $C$ are defined as $B=T_1$,
  $C=T_{-1}$. For some factorizations this makes no
  difference since $T_{n+2}T_n$ is independent on $n$.} We can
use the elementary move on $V$ and the braid relation to bring the
UV factorization in Table~\ref{tab:Kodaira} to the corresponding ABC
factorization. As an example, we show in Figure~\ref{Fig:braidD4} the
case for the $D_4$ singularity. From the braid relation (applied to
the underlined twists) and the
Hurwitz moves we obtain
\begin{equation}
\underline{UVU}VUV = VUVVUV = A^4 BC \, ,
\end{equation}
 which is
the standard factorization for the $I_0^{\ast}$ monodromy. For the exceptional
singularities we have (see Figure~\ref{Fig:braidEn})
\begin{align}
\underline{UVU}V\underline{UVU}V &= VUVVVUVV = A^6 BC \, ,\nonumber \\
 \underline{UVU}V\underline{UVU}VU &= VUVVVUVVU = A^6 
                                                     BC^2 \, ,\nonumber\\
\underline{UVU}V\underline{UVU}VUV &= VUVVVUVVUV \approx A^7 BC \, ,
\end{align}
where the last relation follows from a global $SL(2,\mathbb{Z})$ conjugation.


The UV factorization is useful to understand colliding
singularities and the corresponding symmetry enhancements. For example, when two $III$ fibers collide we get a
monodromy
\begin{equation}
III+III = UVU\, \underline{UVU} = UVU\, VUV = (UV)^3 = I_0^{\ast} \, ,
\end{equation} 
giving an $SO(8)$ gauge group. Another example is the branch II at
constant $\tau$. There are generically
twelve type $II$ degenerations, each with monodromy $UV$. To understand enhanced symmetries on
this branch, we see that when $n$ singularities coincide, $n=(2,3,4,5)$,
we get a monodromy $(UV)^n$, corresponding to 
the gauge groups $SU(3)$, $SO(8)$, $E_6$, $E_8$. This reproduces the
list in~\cite{Gaberdiel:1998mv}.

\section{Geometry of torus fibrations}\label{app:eom}

We summarize some details about the geometry of the
ansatz~\eqref{ansatztaurho}. The relevant equations of motion are:
\begin{align}
& R-4 (d\Phi)^2+ 4\nabla^2\Phi -\frac12 H^2 = 0 \, ,\\
& R_{MN} + 2\nabla_M\nabla_N \Phi -\frac12\iota_MH\iota_N H = 0 \, ,\\
& d(e^{-2\Phi}\star_{10} H) = 0 \, .
\end{align}
It is easy to check that the Cauchy-Riemann equations for the
functions $\tau$, $\rho$ and $\phi$ imply that the
ansatz~\eqref{ansatztaurho} solves the equations of motion. The Ricci
scalar (in the string frame) for such ansatz is:
\begin{equation}
R =
\frac{3e^{-2\varphi_1(r,\theta)}}{2r^2[\rho_2(r,\theta)]^3\tau_2(r,\theta)}\left[
  r^2 \left(\frac{\partial
      \rho_2(r,\theta)}{\partial r}\right)^2+\left(\frac{\partial
      \rho_2(r,\theta)}{\partial \theta}\right)^2\right] \, .
\end{equation}
For completeness we also give the components of the Ricci tensor along the torus and
the base directions:
\begin{align}
R_{r,r} &=
\frac{(1+\partial_{\theta}^2\varphi_2)\partial_r\rho_2+\partial_{\theta}\rho_2\partial_r\varphi_2}{r\rho_2}+
\frac{r^2\partial_r\rho_2\partial_r\tau_2-\partial_{\theta}\rho_2\partial_{\theta}\tau_2}{2r^2\rho_2\tau_2}+
          \frac{\partial_{\theta}^2\rho_2}{r^2\rho_2}+\frac{3(\partial_r\rho_2)^2}{2\rho_2^2}
  \, , \nonumber\\
R_{r,\theta} & =
               \frac{\partial_{\theta}\rho_2(1+\partial_{\theta}\varphi_2)}{r\rho_2}+\frac{3\partial_{\theta}\rho_2\partial_r\rho_2}{2\rho_2^2}+\frac{\partial_{\theta}\rho_2\partial_r\tau_2+\partial_{\theta}\tau_2\partial_r\rho_2}{2\rho_2\tau_2}-\frac{r\partial_r\rho_2\partial_r\varphi_2+\partial_r\partial_{\theta}\rho_2}{\rho_2}\,
               ,\nonumber\\
R_{\theta,\theta}& =
                   \frac{3(\partial_{\theta}\rho_2)^2}{2\rho_2^2}-\frac{\partial_{\theta}^2\rho_2+
                   r(1+\partial_{\theta}\varphi_2)\partial_r\rho_2}{\rho_2}+\frac{\partial_{\theta}\rho_2\partial_{\theta}\tau_2-r^2\partial_r\rho_2\partial_r\tau_2}{2\rho_2\tau_2}-\frac{r\partial_{\theta}\rho_2\partial_r\varphi_2}{\rho_2}\,
                   ,\nonumber\\
R_{8,8}& =
         \frac{e^{-2\varphi_1}\tau_1}{r\rho_2\tau_2^2}\left[\partial_{\theta}\rho_2\partial_r\tau_2-\partial_r\rho_2\partial_{\theta}\tau_2\right]+\frac{e^{-2\varphi_1}(\tau_1^2-\tau_2^2)}{2r^2\rho_2\tau_2^3}\left[\partial_{\theta}\rho_2\partial_{\theta}\tau_2+r^2\partial_r\rho_2\partial_r\tau_2\right]\,
         ,\nonumber\\
R_{8,9} & =
          \frac{e^{-2\varphi_1}}{2r\rho_2\tau_2^2}\left[\partial_{\theta}\rho_2\partial_r\tau_2-\partial_r\rho_2\partial_{\theta}\tau_2\right]+
          \frac{e^{-2\varphi_1}\tau_1}{2r^2\rho_2\tau_2^3}\left[\partial_{\theta}\rho_2\partial_{\theta}\tau_2+
          r^2\partial_r\rho_2\partial_{r}\tau_2\right]\, , \nonumber\\
R_{9,9}&=
         \frac{e^{-2\varphi_1}}{2r^2\rho_2\tau_2^3}\left[\partial_{\theta}\rho_2\partial_{\theta}\tau_2+
         r^2 \partial_r\rho_2\partial_r\tau_2\right] \, . 
\end{align}

\section{T-duality}\label{app:T-duality}

We recall well known facts about the action of an element of the
T-duality group
$O(2,2,\mathbb{Z})$ (see for example~\cite{Giveon:1994fu}). If we define the background matrix
\begin{equation}
E = G + B \, ,
\end{equation}
the action of an element
\begin{equation}
\Omega_{O(d,d)} = \begin{pmatrix} \mathbf{A} & \mathbf{B}\\ \mathbf{C} & \mathbf{D}\end{pmatrix} 
\end{equation}
on $E$ is
\begin{equation}
E \rightarrow \frac{\mathbf{A} E + \mathbf{B}}{\mathbf{C} E+ \mathbf{D}} \, ,
\end{equation}
where the following relations hold:
\begin{equation}
 \mathbf{A}^T
\mathbf{C}+ \mathbf{C}^T\mathbf{A} =
\mathbf{B}^T\mathbf{D}+\mathbf{D}^T\mathbf{B}=0 \, , \quad \mathbf{A}^T\mathbf{D} + \mathbf{C}^T\mathbf{B}=\mathbb{1} \, .
\end{equation}
If we introduce the generalized matrix
\begin{equation}
M_E = \begin{pmatrix} G- B G^{-1}B& B G^{-1} \\ -G^{-1} B & G^{-1} \end{pmatrix} \, ,
\end{equation}
the action is $M_E\rightarrow \Omega M_E \Omega^T$.
In terms of the $SL(2,\mathbb{Z})$ action on $\tau$ and $\rho$,
defined as
\begin{equation}
\tau \rightarrow \frac{a\tau + b}{c\tau + d} \, , \qquad \rho
\rightarrow \frac{\tilde a \rho + \tilde b}{\tilde c\rho + \tilde d}
\, , \quad ad-bc=\tilde a \tilde d - \tilde b \tilde c =1 \, ,
\end{equation}
the element $\Omega_{O(2,2)}$ is determined as follows:
\begin{equation}
\mathbf{A} = \tilde a \begin{pmatrix} a & b\\ c & d\end{pmatrix} \, ,\quad
 \mathbf{B} =\tilde b \begin{pmatrix} -b & a\\ -d & c\end{pmatrix} \, ,\quad 
\mathbf{C} = \tilde c \begin{pmatrix} -c & -d\\ a & b\end{pmatrix} \, ,\quad 
\mathbf{D} = \tilde d \begin{pmatrix} d & -c\\ -b & a\end{pmatrix} \, .
\end{equation}
In particular, the action of $SL(2,\mathbb{Z})_{\tau}$ is given by
$O(2,2,\mathbb{Z})$ matrices of the form:
\begin{equation}
\Omega_{O(2,2),\tau}=\begin{pmatrix} M_{\tau} & 0 \\ 0 & (M_{\tau}^T)^{-1} \end{pmatrix} \, ,
\end{equation}
which is just a basis change $E\rightarrow M_{\tau} E M_{\tau}^T$.
The action of $SL(2,\mathbb{Z})_{\rho}$ is embedded as
\begin{equation}
\Omega_{O(2,2),\rho}=\begin{pmatrix} \tilde a & \tilde b\, \omega\\ -\tilde c\,\omega &\tilde d \end{pmatrix} \, ,
\end{equation}
where $\omega = i \sigma_2 = \bigl(\begin{smallmatrix} 0 & 1\\
  -1&0 \end{smallmatrix} \bigr)$. Perturbative monodromies in $\rho$
given by the $SL(2,\mathbb{Z})$ element $V^{\tilde b}$ in~\eqref{UVtwists},
are identified with the upper right corner.
We see then that upper triangular matrices represent
a geometric subgroup $G_{geom} \subset O(2,2,\mathbb{Z})$ which is a
product of diffeomorphisms and B-transformations of the form:
\begin{equation}
\Omega_{geom} = \begin{pmatrix}  M_{\tau} & \tilde b\, M_{\tau}\cdot\omega  \\ 0& (M_{\tau}^T)^{-1}\end{pmatrix} \, .
\end{equation}
A diffeomorphism combined with a $(0,1)$ element of the parabolic conjugacy
classes in $SL(2,\mathbb{Z})_{\rho}$, namely $U^{\tilde
  c}$~\eqref{UVtwists} which acts as:
\begin{equation}\label{APP:Urho}
\rho \rightarrow \frac{\rho}{\tilde c \rho + 1} \, ,
\end{equation}
is represented by a lower triangular matrix
\begin{equation}
\Omega_{\beta} = \begin{pmatrix}  M_{\tau} & 0
  \\ \tilde c\, \omega \cdot M_{\tau}& (M_{\tau}^T)^{-1} \end{pmatrix} \, .
\end{equation}
The action given by the transformation~\eqref{APP:Urho} forms an Abelian
subgroup of $O(2,2,\mathbb{Z})$ usually referred
to as $\beta$-transforms~\cite{Gualtieri:2004kx} since it acts
naturally on bivectors $\beta \in \bigwedge^2 TM$. Such
elements implement TsT transformations and are very useful in
holography to construct gravity duals of marginal deformations, such
as $\beta$-deformations of $\mathcal{N}=4$
SYM~\cite{Lunin:2005jy,Minasian:2006hv}. 

Other $SL(2,\mathbb{Z})_{\rho}$ transformations, such as the parabolic
$(p,q)$ monodromy, will result in $O(2,2,\mathbb{Z})$ elements in which
both B- and $\beta$-transformations are present at the same
time.

%%%%%%%%%%%%%%%%%%%%%%%%%%%%%%%%%%%%%%%%%%%%%%%%%%%%%%%%%%%%%%%%%%%%%%%%%%%
%\providecommand{\href}[2]{#2}\begingroup\raggedright\begin{thebibliography}{10}

%\bibliographystyle{utphys}
%\bibliography{biblioexotic}
\providecommand{\href}[2]{#2}\begingroup\raggedright\endgroup

%%%%%%%%%%%%%%%%%%%%%%%%%%%%%%%%%%%%
%\end{thebibliography}
\end{document}